\documentclass[english]{article}
\usepackage[T1]{fontenc}
\usepackage{geometry}
\usepackage{babel}
\usepackage{array}
\usepackage{units}
\usepackage{bm}
\usepackage{multirow}
\usepackage{amsmath}
\usepackage{amssymb}
\usepackage{stackrel}
\usepackage{graphicx}
\usepackage[authoryear]{natbib}
\usepackage[unicode=true,pdfusetitle,
 bookmarks=true,bookmarksnumbered=false,bookmarksopen=false,
 breaklinks=false,pdfborder={0 0 1},backref=false,colorlinks=false]
 {hyperref}
\hypersetup{
 hypertexnames=false}

\makeatletter

\newcommand{\noun}[1]{\textsc{#1}}
\providecommand{\tabularnewline}{\\}
\usepackage{tikz}
\usetikzlibrary{calc}

\usepackage[font=footnotesize]{caption}
\usepackage{xcolor}

\usepackage[mathlines]{lineno}

\newif\ifComments
\Commentsfalse

\date{}

\makeatother

\begin{document}
\title{Events in Noise-Driven Oscillators: Markov Renewal Processes and the
``Unruly'' Breakdown of Phase-Reduction Theory}
\author{Avinash J. Karamchandani}
\date{}
\maketitle
\begin{abstract}
We introduce an extension to the standard reduction of oscillatory
systems to a single phase variable. The standard reduction is often
insufficient, particularly when the oscillations have variable amplitude
and the magnitude of each oscillatory excursion plays a defining role
in the impact of that oscillator on other systems, i.e. on its output.
For instance, large excursions in bursting or mixed-mode neural oscillators
may constitute events like action potentials, which trigger output
to other neurons, while smaller, sub-threshold oscillations generate
no output and therefore induce no coupling between neurons. Noise
induces diffusion-like dynamics of the oscillator phase on top of
its otherwise constant rate-of-change, resulting in the irregular
occurrence of these output events. We model the events as corresponding
to distinguished crossings of a Poincare section. Using a linearization
of the noisy Poincare map and its description under phase-isostable
coordinates, we determine the diffusion coefficient for the occurrence
and timing of the events using Markov renewal theory. We show that
for many oscillator models the corresponding point process can exhibit
``unruly'' diffusion: with increasing input noise strength the diffusion
coefficient vastly increases compared to the standard phase reduction
analysis, and, strikingly, it also \emph{decreases} when the input
noise strength is increased further. We provide a thorough analysis
in the case of planar oscillators, which exhibit unruliness in a finite
region of the natural parameter space. Our results in part explain
the surprising synchronization behavior obtained in pulse-coupled,
mixed-mode oscillators as they arise, e.g., in neural systems.
\end{abstract}

\section{Introduction}

\subsection{Background: Phase reduction, Population Dynamics of Coupled and Noise-driven
Oscillators, and Limitations}

Phase reduction is a well-established dimension-reduction technique
for dynamical systems that exhibit limit cycle oscillations \citep{winfree_geometry_1980,kuramoto_chemical_1984,ermentrout_mathematical_2010,schultheiss_phase_2012}.
In appropriate limits, such as weak perturbative forcing or strong
contraction to the limit cycle, it replaces the possibly high-dimensional
oscillator state by a single phase variable $\phi$, which represents
the oscillator's `internal clock', and allows one to describe the
long-time dynamics of the oscillator by a single differential equation
for $\phi$. Since this achieves a substantial reduction in the complexity
of the description, phase reduction is a powerful tool for the analysis
of individual and coupled oscillators \citep{kuramoto_chemical_1984}
and their applications in many areas, including neuroscience \citep{stiefel_neurons_2016}
and circadian rhythms \citep{gunawan_isochron-based_2006}.

In the presence of noise and in the limit of infinitely many oscillators,
globally coupled oscillators can be described by a nonlinear, non-local
Fokker-Planck equation for the population distribution $\rho(\phi,t)$
of the phase \citep{shinomoto_phase_1986,strogatz_stability_1991,kilpatrick_sparse_2011,karamchandani_pulse-coupled_2018},
\begin{equation}
\frac{\partial\rho(\phi,t)}{\partial t}=D_{\textit{phase}}\frac{\partial^{2}\rho(\phi,t)}{\partial\phi^{2}}-\frac{\partial}{\partial\phi}\left[f_{0}\rho(\phi,t)+\gamma\left(\mbox{nonlinear, nonlocal interaction term}\right)\right].\label{eq:Fokker-Planck}
\end{equation}
In this framework, a key role in determining the tendency of the oscillators
to synchronize is played by the diffusion coefficient of the phase,
$D_{\textit{phase}}$. It represents the leading impact of the noise on the
oscillators, which tends to distribute their phases uniformly. (\ref{eq:Fokker-Planck})
quantifies the intuition that, as long as the diffusion is small in
comparison with the strength $\gamma$ of the interactions between
the oscillators, coherent population-level dynamics, like synchronous
states, will emerge. In particular, (\ref{eq:Fokker-Planck}) predicts
that coherent states arise for coupling strengths $\gamma$ above
a critical value $\gamma_{\textit{crit}}$ that is proportional to $D_{\textit{phase}}$:
$\gamma>\gamma_{\textit{crit}}\propto D_{\textit{phase}}$ (Figure \ref{fig: MMO_DEff_vs_DPhase}c,
dashed line). This prediction should be accurate in the limit of weak
interactions and weak noise.

In previous work \citep{karamchandani_pulse-coupled_2018}, we applied
the Fokker-Planck framework to mixed-mode oscillations comprised dichotomously
of large- and small-amplitude oscillations (Figure \ref{fig: MMO_DEff_vs_DPhase}a).
This model arose in a neuronal context where only the large oscillations,
which represent action potentials, lead to any output to other neurons
via chemical synapses. Therefore, from the perspective of the other
neurons in the network only these large-amplitude oscillations are
relevant \emph{events}, and it is their timing that determines the
population-level dynamics. Numerical simulations reveal transitions
from asynchrony to synchrony and more exotic coherent states, which
depend on the strength of the noise $D_{\textit{in}}$. Standard phase reduction
provides a direct linear connection between $D_{\textit{in}}$ and $D_{\textit{phase}}$,
and thus the Fokker-Planck theory (\ref{eq:Fokker-Planck}) predicts
the onset $\gamma_{\textit{crit}}$ for these coherent states as a function
of the input noise strength $D_{\textit{in}}$. For the mixed-mode oscillations
investigated in \citet{karamchandani_pulse-coupled_2018}, that prediction
fails spectacularly when compared to simulation of the full system,
even when the noise and the interactions are relatively weak (compare
the dashed line and blue dots in Figure \ref{fig: MMO_DEff_vs_DPhase}c).
To wit, it vastly underestimates the strength of the oscillator interactions
needed to overcome the independent noise and produce coherent dynamics
states. In our previous work, we found that the amount of diffusion
is substantially underestimated by the phase reduction theory, since
$D_{\textit{phase}}$ does not take into account the fact that the oscillators
interact only via the amplitude-dependent events. Standard phase reduction,
it turns out, fails because it discards key amplitude information.

\subsection{Going Beyond Standard Phase Reduction: Events and Effective Phase}

The need to go beyond the standard phase reduction has been recognized
previously in various other contexts, e.g. in shear-induced chaos
\citep{lin_shear-induced_2008} and systems involving multiple time
scales \citep{monga_augmented_2021,lin_limitations_2013}. Recent
works have considered augmenting the phase variable with one or more
amplitude-like variables to increase the fidelity of the reduction
(see e.g. \citet{wedgwood_phase-amplitude_2013,castejon_phase-amplitude_2013}).
Of particular interest is the phase-isostable reduction \citep{wilson_isostable_2016},
offering a coordinate system in which the unperturbed oscillator has
especially simple dynamics.

Here we offer an \emph{event-centric} extension to the standard phase
reduction. In particular, we address the influence of noise on oscillators
in which the oscillatory excursions can be divided into two classes:
excursions that qualify as events and excursions that do not. More
concretely, describing each oscillatory excursion as the crossing
of a Poincare section, events correspond to a crossing within a limited
domain of the Poincare section; crossings outside that domain constitute
non-events. In the neuronal mixed-mode oscillators, for example, only
the large-amplitude peaks in the voltage are events (action potentials),
whereas small-amplitude extrema are non-events.

In noise-driven oscillators, the events occur in an irregular fashion
and define a stochastic point process. We are particularly interested
in the long-term statistics of the events that reflect the degree
of diffusion in the stochastic process. The quantity that is of central
interest for the description of the event irregularity is the growth
rate of the variance of the total number of events with time, which
we propose to call the ``temporal variance growth rate'' (TVGR)\footnote{This quantity has also been referred to as ``the slope of the variance-time
curve'' \citep{cox_statistical_1966}.} and represent by the symbol $\mathcal{V}_{E}^{\left(t\right)}$.
As it turns out, this quantity has a direct relationship to the phase
diffusion coefficient $D_{\textit{phase}}$ in the limit of weak noise. Namely,
$\mathcal{V}_{E}^{\left(t\right)}\sim2D_{\textit{phase}}$ as $D_{\textit{in}}\rightarrow0$.
This connection between the event-based TVGR and the phase diffusion
motivates us to introduce an \emph{effective} phase that increases
by 1 between events, whose diffusion coefficient is given by an \emph{effective}
phase diffusion $D_{\textit{eff}}\equiv\frac{1}{2}\mathcal{V}_{E}^{\left(t\right)}$.

\begin{figure}
\noindent \begin{centering}
\includegraphics[width=0.9\textwidth]{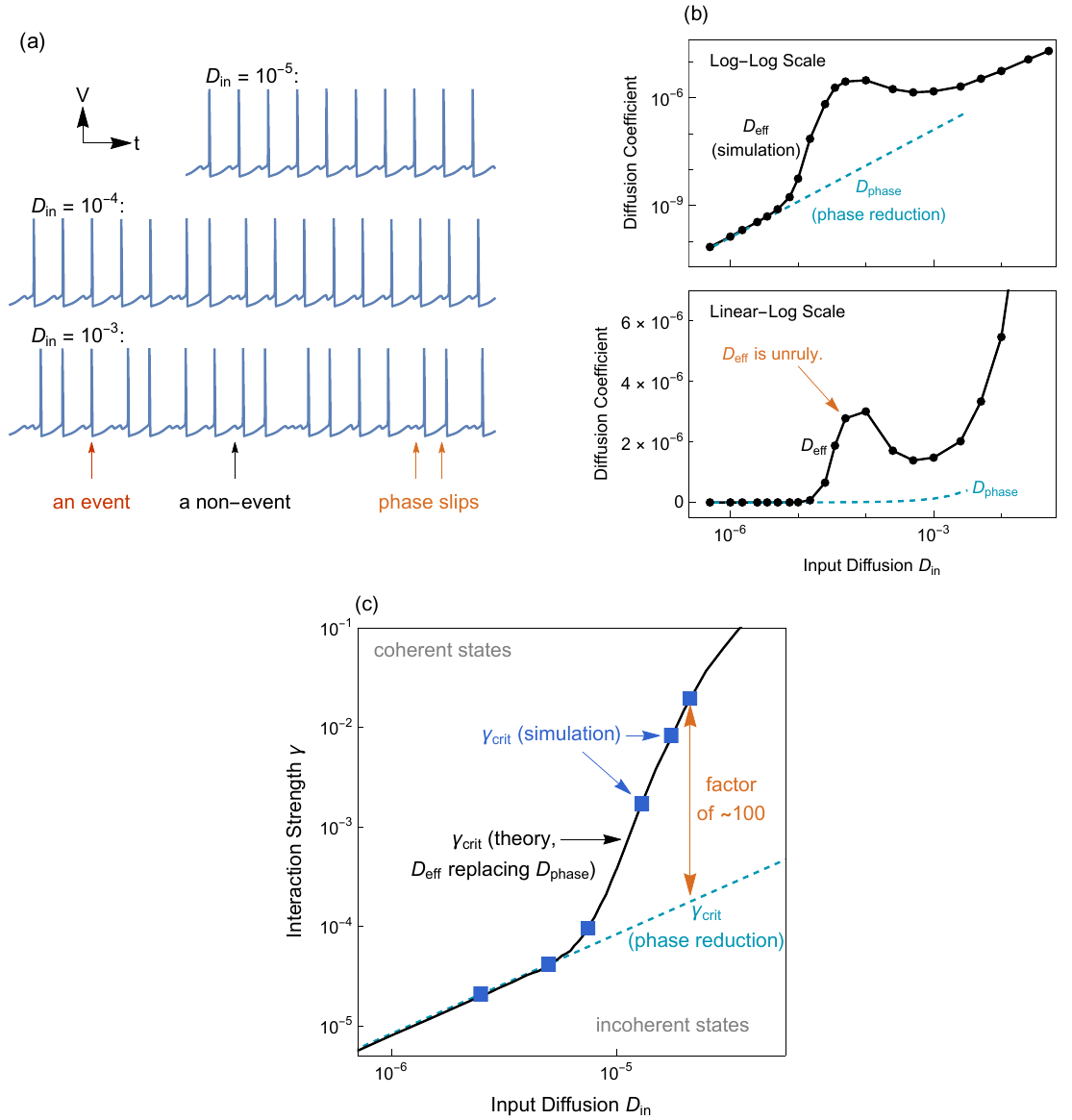}
\par\end{centering}
\caption{\textbf{Effective Diffusion in a Mixed-Mode Oscillator} \citep{karamchandani_pulse-coupled_2018}\textbf{.
}(a): Sample voltage traces of a neural mixed-mode oscillator driven
by noise. Only the large-amplitude excursions are events in which
the oscillator produces \textquotedblleft output\textquotedblright ;
the small-amplitude excursions are non-events. In the absence of noise,
a fixed, periodic pattern of alternating large-amplitude and small-amplitude
oscillations is produced. With noise $D_{\textit{in}}$, \textquotedblleft phase
slips\textquotedblright{} perturb the pattern, in which events are
added or omitted. (b): The phase diffusion $D_{\textit{phase}}$ from phase
reduction theory and the effective diffusion $D_{\textit{eff}}$ (same data
shown on two different scales), treating large-amplitude oscillations
as events. As a function of the input diffusion strength, $D_{\textit{eff}}$
initially agrees with $D_{\textit{phase}}$. But it becomes greatly amplified
for moderate noise strengths and then eventually decreases as the
noise strength increases. We call this qualitative non-monotonic behavior
\textquotedblleft unruly\textquotedblright . (c): A phase diagram
for population states in globally coupled oscillators. Standard phase
reduction with the Fokker-Planck theory, (\ref{eq:Fokker-Planck}),
predicts that the boundary between stable coherent states and stable
incoherence is linear. The theory only agrees with the full, coupled-oscillator
simulation once $D_{\textit{phase}}$ is replaced with $D_{\textit{eff}}$. \label{fig: MMO_DEff_vs_DPhase}}
\end{figure}

In our previous work \citep{karamchandani_pulse-coupled_2018}, we
suggested that $D_{\textit{phase}}$ be replaced by $D_{\textit{eff}}$ in (\ref{eq:Fokker-Planck})
as a way to remedy the deficiencies in the Fokker-Planck theory. We
determined $D_{\textit{eff}}$ computationally via simulations of a single,
uncoupled oscillator and found that it indeed compensates for the
discrepancy in the predictions and recovers the actual onset of coherent
states (Figure \ref{fig: MMO_DEff_vs_DPhase}b,c). The effective phase
thus not only provides an intuitive picture of the long-time dynamics
of the events for an isolated oscillator, but also captures quantitatively
the synchronization behavior of a large ensemble of such oscillators
when their interaction is contingent on the events. Comparing the
effective diffusion with the phase diffusion shows just how poorly
the standard phase reduction performs: $D_{\textit{eff}}$, which accounts
for the fact that the oscillators interact only via the amplitude-dependent
events, is orders of magnitude larger than $D_{\textit{phase}}$ (Figure \ref{fig: MMO_DEff_vs_DPhase}b).

Strikingly, we find that the effective diffusion coefficient, exhibits
``unruly'' behavior for a wide range of oscillators: while in the
limit of weak noise the effective diffusion coefficient $D_{\textit{eff}}$
converges to the diffusion coefficient $D_{\textit{phase}}$ obtained from
the standard phase reduction, already for surprisingly small values
of the noise it can become orders of magnitude larger than $D_{\textit{phase}}$,
only to \emph{decrease} when noise is increased further to larger
values (see Figure \ref{fig: MMO_DEff_vs_DPhase}b for an example).
For coupled such oscillators, this implies strongly enhanced sensitivity
to desynchronization for intermediate noise values and reduced sensitivity
for strong noise \citep{karamchandani_pulse-coupled_2018}.

In the present work, we address the source and abundance of the unruliness
in the TVGR and thus the effective diffusion coefficient. In contrast
with our prior, purely computational approach, we obtain explicit
expressions for the TVGR by considering the point process that arises
from a linearized Poincare map. The paradigm of point processes has
already previously been used in the analysis of oscillatory and excitable
systems. In neuroscience, for example, the times $T_{n}$ at which
action potentials occur in a neuron are often considered to arise
from a point process. Indeed, the distribution of so-called inter-spike-intervals,
$T_{n+1}-T_{n}$, is a subject of great interest (see e.g. \citep{gabbiani_mathematics_2010}).
More broadly, point-process theory offers various measures of the
temporal variability in oscillatory and excitable dynamics (Table
\ref{tab:Point-process-statistics}). The Fano factor, for instance,
has been used to measure the similarity of a given point process to
a Poisson process, which has a Fano factor of $1$. For noise-driven,
nonlinear oscillatory and excitable systems, the Fano factor and effective
diffusion coefficient have each been used to identify a variety of
``resonances'' that appear as a function of input noise strength:
coherence resonance and incoherence maximization \citep{pikovsky_coherence_1997,lindner_maximizing_2002}.

The analysis of point processes is often limited to renewal processes,
i.e. to point processes in which the intervals between events are
independent and identically distributed. Noise-driven limit-cycle
oscillators will, however, in general maintain correlations from one
cycle to the next and therefore do not fit the framework of renewal
processes. Indeed, non-renewal dynamics are seen in real world oscillatory
systems \citep{farkhooi_serial_2009,gabbiani_mathematics_2010,avila-akerberg_nonrenewal_2011}.
These correlations do not preclude a description of the oscillators
within the framework of point processes. In this work, we therefore
include correlations in the time intervals between events by considering
the Markov renewal process associated with the stochastic Poincare
map.

\subsection{Organization of the Paper}

The paper is organized as follows. In Section \ref{sec:effective phase diffusion}
we introduce the effective phase via the events and their variance
statistics and outline in more detail our approach for its analysis.
In the somewhat technical Section \ref{sec:Fano-Variance-and-Function-Variance}
and the corresponding Supplementary Information (SI) Section \ref{sec: SI - Event TVGR Formula},
we review the theory for variance statistics of Markov and Markov
renewal reward processes, offering a novel formula for the TVGR. We
use that formula in Section \ref{sec:Fano-Variance-for-Toy-Model}
to demonstrate in a simple toy model how unruliness can arise in $\mathcal{V}_{E}^{\left(t\right)}$,
offering an simple explanation of the eventual decrease of the TVGR
with increasing noise strength. We then apply the general theory to
the linearization of Poincare map dynamics for limit cycle oscillators
in the Section \ref{sec:Fano-Variance-for-Limit-Cycle} and the corresponding
SI Section \ref{sec: SI - TVGR Oscillator Derivation}. In Section
\ref{sec: Main Results - Planar Oscillators} we narrow our focus
to two-dimensional oscillators and make an argument that unruliness
in $\mathcal{V}_{E}^{\left(t\right)}$ is a common occurrence. We
conclude with plausible extensions to this work in Section \ref{sec:Extensions}
and a broader discussion in Section \ref{sec:Discussion}.

\section{Phase Diffusion and The Temporal Variance Growth Rate}

\label{sec:effective phase diffusion}

\subsection{Phase Reduction and Its Breakdown}

The specific subjects of this work are noise-driven oscillatory systems
and the statistical characteristics of the events they produce. Motivated
by systems in which output is only generated by events, we will take
the event timings to be the relevant read-out of the system's dynamics,
and we will be particularly interested in quantifying the long-time-scale,
noise-driven dispersion of events. In this section we connect the
statistics of the events with the diffusive dynamics of the phase
and introduce an effective phase, which becomes relevant when the
noise induces ``phase slips'' during which events are skipped or
extra ones are produced (as in, for example, Figure \ref{fig: MMO_DEff_vs_DPhase}a).

We investigate systems in $d+1$ dimensions of the form
\begin{equation}
d\vec{y}=\vec{F}(\vec{y})\,dt+\sqrt{2D_{\textit{in}}}\mathbf{G}\left(\vec{y}\right)\,d\vec{W}\label{eq: system}
\end{equation}
that have a stable limit cycle when $D_{\textit{in}}=0$, where $\vec{W}$
is the standard Wiener process of dimension $d+1$ and the matrix
$\mathbf{G}(\vec{y})$ characterizes the noise correlations. To define
events we choose a set $E$ in phase space, such that an event occurs
when $\vec{y}\left(t\right)$ intersects $E$. For instance, in neuroscience
applications the event often corresponds to an action potential, where
the voltage variable has a large ``spike''.

For weakly perturbed limit cycles, the system description can be substantially
simplified by a phase reduction, in which the state $\vec{y}\in\mathbb{R}^{d+1}$
is approximately represented by a single phase variable $\phi$ parameterizing
the limit cycle $\vec{y}_{\textit{LC}}\left(\phi\right)$. The dynamics of
(\ref{eq: system}) are thus captured via a single equation \citep{kuramoto_chemical_1984,aminzare_phase_2019},
\[
d\phi=f_{0}\,dt+\sqrt{2D_{\textit{in}}}\,Z^{T}\left(\phi\right)\mathbf{G}\left(\vec{y}_{\textit{LC}}\left(\phi\right)\right)\,d\vec{W}+\mathcal{O}\left(D_{\textit{in}}\right),
\]
in which $\phi\in\left[0,1\right);$ upon reaching $1$, $\phi$ is
reset to $0$. Subsequently applying the method of averaging (see
e.g. Section 6.2 of \citet{schwemmer_theory_2012}) yields 
\begin{equation}
d\phi\sim f_{0}\,dt+\sqrt{2D_{\textit{phase}}}\,dW,\label{eq: phase}
\end{equation}
where
\begin{equation}
D_{\textit{phase}}=\left[f_{0}\int_{0}^{\nicefrac{1}{f_{0}}}Z^{T}\left(\phi\right)\mathbf{G}\left(\vec{y}_{\textit{LC}}\left(\phi\right)\right)\mathbf{G}^{T}\left(\vec{y}_{\textit{LC}}\left(\phi\right)\right)Z\left(\phi\right)d\phi\right]D_{\textit{in}}.\label{eq: linear phase diffusion}
\end{equation}
The reduction thus offers a prediction for the amount of diffusion
$D_{\textit{phase}}$ in the phase variable. In particular, from (\ref{eq: linear phase diffusion}),
it is proportional to the input diffusion coefficient. The dynamics
of the reduced phase model are localized to the limit cycle, and so
the events produced by (\ref{eq: phase}) are governed by the intersection
of the limit cycle with $E$. In this work, unless otherwise indicated,
we will consider the simplest scenario, in which the noiseless limit
cycle intersects $E$ once and does so transversely. Without loss
of generality, we take the phase on the limit cycle at the intersection
to be $\phi=0$.

We can compare the full dynamics (\ref{eq: system}) with that of
the phase oscillator (\ref{eq: phase}) via the corresponding point
processes for the events. Define the point process $T_{n}$ for the
full system (\ref{eq: system}) as the time $t$ at which the $n^{\textit{th}}$
event occurs. In contrast with the point process for the phase oscillator,
given by the times at which $\phi=0$, the process $T_{n}$ will in
general be non-renewal. Indeed, a noise-driven limit-cycle oscillator
will support correlations in its dynamics from one intersection of
$E$ to the next. As the noise strength is increased from infinitesimal
values, $\vec{y}$ will deviate from $\vec{y}_{\textit{LC}}$ and the time
interval, $\Delta T_{n}=T_{n+1}-T_{n}$, between events will deviate
from the period of the unperturbed limit cycle. More drastic changes
occur with increasing noise strength if the event surface $E$ extends
only a finite distance away from the limit cycle. In that case non-infinitesimal
noise may cause the trajectory to miss $E$ in one cycle and return
to it only in a subsequent cycle, inserting an additional non-event
crossing. The phase $\phi$ will then not capture the event point
process and - in terms of describing the events - the phase equation
(\ref{eq: phase}) breaks down.

\subsection{Effective Phase Diffusion for Events}

In order to obtain a description that is able to deal with phase slips
we introduce an alternative, \emph{effective} phase oscillator, one
whose \emph{event }statistics by definition match that of the full
oscillator in the \emph{long-time limit}. It will be useful to characterize
the point process by the function $N_{t}$, the number of events that
have occurred by time $t$. Take
\begin{equation}
d\phi_{\textit{eff}}=f_{\textit{eff}}\,dt+\sqrt{2D_{\textit{eff}}}\,dW,\label{eq: effective phase}
\end{equation}
where each reset from $\phi_{\textit{eff}}=1$ to $0$ constitutes an event
and the effective oscillation frequency,
\begin{equation}
f_{\textit{eff}}\equiv\frac{1}{\mu}=\lim_{t\rightarrow\infty}\frac{1}{t}\mathrm{E}\left\{ N_{t}\right\} ,\label{eq: f_eff}
\end{equation}
and the effective phase diffusion coefficient,
\begin{equation}
D_{\textit{eff}}\equiv\frac{1}{2}\mathcal{V}_{E}^{\left(t\right)}=\frac{1}{2}\lim_{t\rightarrow\infty}\frac{1}{t}\mathrm{var}\left\{ N_{t}\right\} ,\label{eq: D_eff}
\end{equation}
are defined by the event point process $N_{t}$ of the full oscillator.
We expect that in the weak-noise limit, where phase reduction applies,
$f_{\textit{eff}}\rightarrow f_{0}$, $D_{\textit{eff}}\rightarrow D_{\textit{phase}}$, and
so $\phi_{\textit{eff}}\rightarrow\phi$. In general, and in contrast with
$\phi$, however, we do not think of a given value of $\phi_{\textit{eff}}$
as referring to a particular point or set of points in phase space.
Instead, $\phi_{\textit{eff}}$ is analogous to an averaged dynamical variable
whose dynamics only reflect those of the original system over long
time scales.

Note that because each passage from $\phi_{\textit{eff}}=0$ to $\phi_{\textit{eff}}=1$
governed by (\ref{eq: effective phase}) is independent from all others,
the corresponding point process $N_{t}^{\textit{eff}}$ (with events defined
by $\phi_{\textit{eff}}=1$) is a renewal process, and its statistics are fully
determined by the first passage time distribution. Ignoring the reset,
(\ref{eq: effective phase}) produces Brownian motion with a drift,
which has a first passage time density given by an inverse Gaussian
distribution with mean $\mu$ and variance $\mu^{3}\mathcal{V}_{E}^{\left(t\right)}$
(see e.g. \citet{folks_inverse_1978,aminzare_phase_2019}). We note
that because (\ref{eq: effective phase}) incorporates each a mean-
and variance- type statistic, $N_{t}^{\textit{eff}}$ will correctly reproduce
all of the long-time statistics of $N_{t}$ that appear in Table \ref{tab:Point-process-statistics},
which are inter-related \citep{cox_statistical_1966}. To facilitate
the comparison of the standard phase reduction (\ref{eq: phase})
with the effective phase oscillator (\ref{eq: effective phase}),
here we specifically focus on the effective diffusion $D_{\textit{eff}}=\frac{1}{2}\mathcal{V}_{E}^{\left(t\right)}$.
In this context, a ``break-down'' in the phase reduction theory
occurs when $D_{\textit{eff}}$ deviates greatly from the linear prediction,
$D_{\textit{phase}}$.

\begin{table}
\begin{tabular}{|c|c|c|c|}
\hline 
\multirow{2}{*}{} & \multirow{2}{*}{definition} & \multirow{2}{*}{... in the long-time limit} & .... in the effective phase oscillator\tabularnewline
 &  &  & or effective renewal process\tabularnewline
\hline 
\hline 
Event Rate & $\frac{1}{t}\mathrm{E}\left\{ N_{t}\right\} $ & $1/\mu$ & $f_{\textit{eff}}=\frac{1}{\mathrm{E}\left\{ \Delta\tau\right\} }$ \tabularnewline
\hline 
\multirow{2}{*}{Mean Inter-event Interval} & \multirow{2}{*}{$\frac{1}{n}\mathrm{E}\left\{ T_{n}\right\} $} & \multirow{2}{*}{$\mu$} & mean first passage time\tabularnewline
 &  &  & $=\mathrm{E}\left\{ \Delta\tau\right\} $\tabularnewline
\hline 
Temporal Variance & \multirow{2}{*}{$\frac{1}{t}\mathrm{var}\left\{ N_{t}\right\} $} & \multirow{2}{*}{$\mathcal{V}_{E}^{\left(t\right)}$} & \multirow{2}{*}{$2D_{\textit{eff}}=\frac{\mathrm{var}\left\{ \Delta\tau\right\} }{\mathrm{E}\left\{ \Delta\tau\right\} ^{3}}$}\tabularnewline
Growth Rate &  &  & \tabularnewline
\hline 
\multirow{2}{*}{Event Dispersion Rate} & \multirow{2}{*}{$\frac{1}{n}\mathrm{var}\left\{ T_{n}\right\} $} & \multirow{2}{*}{$\mu^{3}\mathcal{V}_{E}^{\left(t\right)}$} & variance in the first passage time \tabularnewline
 &  &  & $=\mathrm{var}\left\{ \Delta\tau\right\} $\tabularnewline
\hline 
Fano Factor & \multirow{2}{*}{$\frac{\mathrm{var}\left\{ N_{t}\right\} }{\mathrm{E}\left\{ N_{t}\right\} }$} & \multirow{2}{*}{$\mu\mathcal{V}_{E}^{\left(t\right)}$} & \multirow{2}{*}{$\frac{\mathrm{var}\left\{ \Delta\tau\right\} }{\mathrm{E}\left\{ \Delta\tau\right\} ^{2}}$}\tabularnewline
(or Index of Dispersion) &  &  & \tabularnewline
\hline 
\end{tabular}

\caption{Typical point process statistics, their inter-relationships in the
long-time limit ($n\rightarrow\infty$ or $t\rightarrow\infty$),
and their connection to the effective phase oscillator (\ref{eq: effective phase}).
Note that as long as these long-time-scale statistics exist for the
general (non-renewal) point process, they can be reproduced by a renewal
process: choose the random renewal intervals $\Delta\tau$ such that
$\mu=\mathrm{E}\left\{ \Delta\tau\right\} $ and $\mathcal{V}_{E}^{\left(t\right)}=\frac{\mathrm{var}\left\{ \Delta\tau\right\} }{\mathrm{E}\left\{ \Delta\tau\right\} ^{3}}$.
(\ref{eq: effective phase}) is just one example of such a renewal
process. \label{tab:Point-process-statistics}}
\end{table}

The break-down can be quite dramatic. Indeed, for the neuronal mixed-mode
oscillators \citep{karamchandani_pulse-coupled_2018}, as a function
of the input noise strength $D_{\textit{in}}$, the effective diffusion coefficient
$D_{\textit{eff}}$ grows orders of magnitude above the linear phase diffusion
before eventually decreasing to comparable levels for large input
noise (Figure \ref{fig: MMO_DEff_vs_DPhase}b). We call this behavior
``unruly'' because its graph is highly nonlinear and non-monotonic
in stark contrast with the linear prediction from phase reduction,
(\ref{eq: linear phase diffusion}). Similarly dramatic effects have
been noted in the literature: non-monotonicity in the Fano factor
has been called incoherence maximization \citep{lindner_maximizing_2002}
when a maximum appears and coherence resonance \citep{pikovsky_coherence_1997}
when there is a minimum.

\subsection{Theoretical Approach}

\begin{figure}
\noindent \begin{centering}
\includegraphics[width=0.9\textwidth]{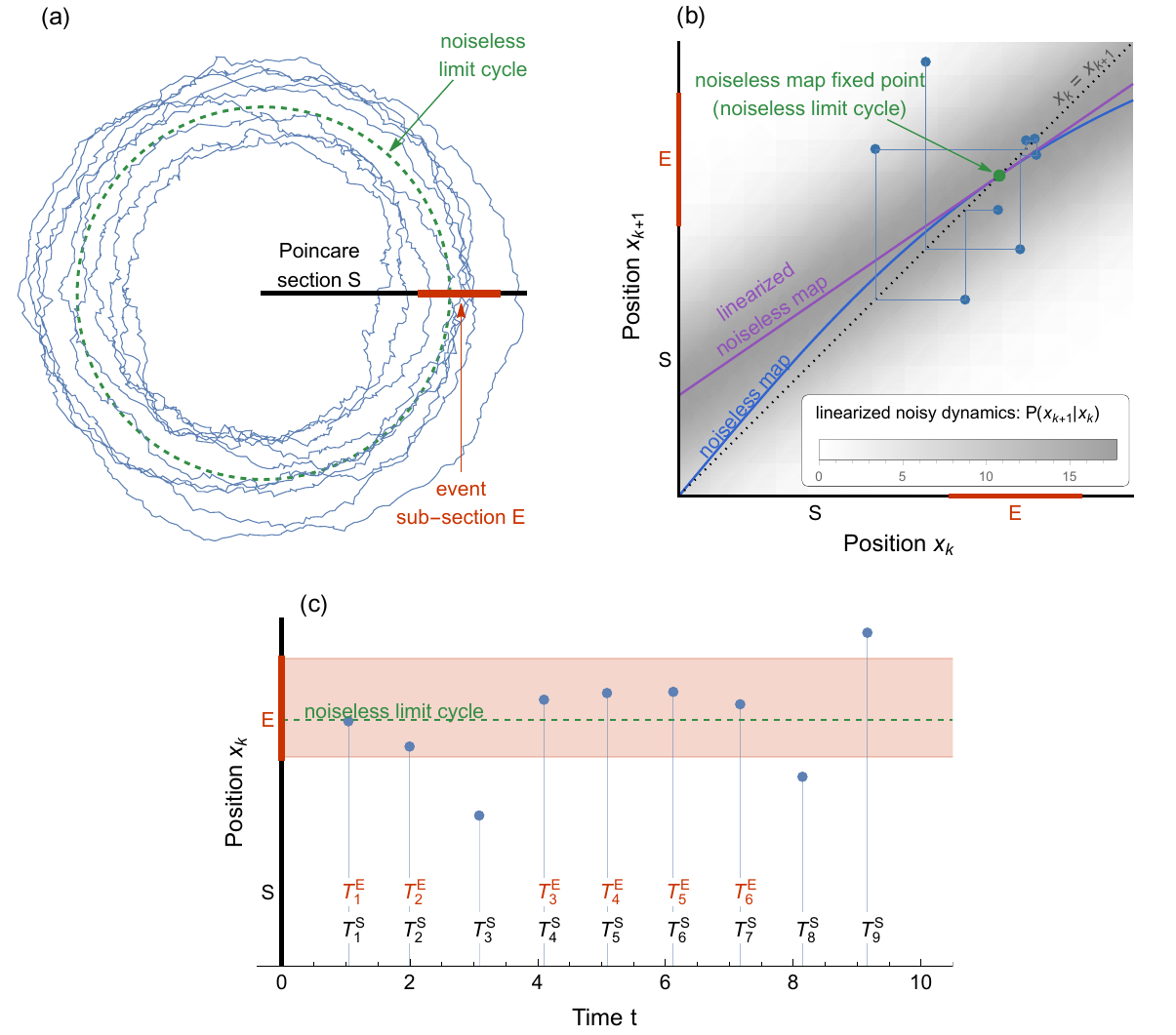}
\par\end{centering}
\caption{\textbf{Events in a Noise-driven Limit-cycle Oscillator. }(a): Noisy
trajectory crossing a Poincare section $S$ and an event sub-section
$E$. (b): The linearized, stochastic, discrete-time dynamics for
$x_{k}$ on the Poincare section $S$. The shading gives the probability
of $x_{k+1}$ conditioned on $x_{k}$. (c): The designation $1_{E}\left(x_{k}\right)$
of each $x_{k}$ as an event or non-event and the two associated point
processes, $T_{k}^{S}$ and $T_{k}^{E}$. \label{fig: schematic_PoincareS+linearizedDynamics+eventLabel}}
\end{figure}

The main contribution of this work is an asymptotic expression for
the effective diffusion coefficient (\ref{eq: D_eff}) that applies
to a wide class of limit-cycle oscillators and that is readily analyzed
and interpreted. In deriving that formula, we make the following over-arching
choices (see also Figure \ref{fig: schematic_PoincareS+linearizedDynamics+eventLabel}):
\begin{enumerate}
\item We define an event as occurring when the trajectory $\vec{y}$ of
the system (\ref{eq: system}) crosses a subset $E$ of a codimension-$1$
surface $S$ that is transverse to the limit cycle.
\item Of the variance-like statistics listed in Table \ref{tab:Point-process-statistics},
we compute in particular the asymptotic temporal variance growth rate,
$\mathcal{V}_{E}^{\left(t\right)}=2D_{\textit{eff}}$.
\item We approximate the dynamics in the vicinity of the limit cycle to
linear order, but - in contrast with the phase reduction (\ref{eq: phase})
- we explicitly include dynamics off of the limit cycle.
\end{enumerate}
Choice 1 defines a point process, which arises from crossings of a
Poincare section $S$. It has the structure of a Markov renewal process,
where the ``Markov'' aspect governs the positions on the Poincare
section and the ``renewal'' aspect reflects the randomly varying
times between crossings. As compared with the renewal process generated
by (\ref{eq: phase}), including the state space of the Poincare section
enriches the dynamics. The dynamics can be non-renewal, since the
position on the section and the time intervals between crossings will
in general be correlated. We make the further choice to analyze the
system by including all crossings of $S$ as steps in the process
but distinguishing those in $E$ as events. There are therefore two
related point processes that we consider, given by $T_{n}^{S}$ (or
$N_{t}^{S}$) for the process on $S$ and $T_{n}^{E}$ (or $N_{t}^{E}$)
for the one limited to $E$. While in principle it is also an option
to simply limit the point process to $E$ in the first place by taking
$S=E$, our choice makes calculation feasible. Unless otherwise indicated,
we restrict our consideration to sections that intersect the noiseless
limit cycle once with the intersection occurring in $E$. Then, in
the absence of noise, $T_{n}^{S}=T_{n}^{E}$, which is also the point
process produced by the phase-reduced model, (\ref{eq: phase}).

The inclusion of $E$ informs our Choice 2, since the asymptotic temporal
variance growth rate can be calculated as 
\begin{equation}
\mathcal{V}_{E}^{\left(t\right)}=\lim_{t\rightarrow\infty}\frac{1}{t}\mathrm{var}\left\{ N_{t}^{E}\right\} =\lim_{t\rightarrow\infty}\frac{1}{t}\mathrm{var}\left\{ \sum_{k=1}^{N_{t}^{S}}1_{E}\left(x_{k}\right)\right\} ,\label{eq:Fano variance}
\end{equation}
where $1_{E}$ is the indicator function for $E$ on $S$ and $x_{k}\in S$
is the position of the $k^{\textit{th}}$ crossing on the section - the $k^{\textit{th}}$
step in the Markov process, which occurs at time $T_{k}^{S}$. We
note that with this specification of the function $1_{E}$ on the
state space, this process could be considered a Markov renewal reward
process in which each event ($x_{k}\in E$) produces a reward of $1$
and each non-event crossing ($x_{k}\in S$, but $x_{k}\notin E$)
a reward of $0$. In our non-rigorous analysis, we will assume that
the Markov processes are sufficiently ``nice''; at the least, we
require that the processes have an invariant probability density $\pi\left(x\right)$
which is nonzero for all $x\in S$ and has the necessary recurrence
properties to render $\mathcal{V}_{E}^{\left(t\right)}$ independent
of initial conditions.

Finally, Choice 3 means that, while our calculation of the effective
diffusion coefficient $D_{\textit{eff}}=\frac{1}{2}\mathcal{V}_{E}^{\left(t\right)}$
will be an approximation of (\ref{eq: D_eff}), it will be a correction
to $D_{\textit{phase}}$. In contrast with our previous study, \citet{karamchandani_pulse-coupled_2018},
where we found $D_{\textit{eff}}$ computationally, we here offer an analytical
treatment that is made possible by the linearization. Our analysis
combines the ideas presented by \citet{hitczenko_poincare_2013} and
\citet{wilson_isostable_2016}. The former make use of an orthonormal
moving coordinate frame to describe the Poincare map for the position
$x_{k}$ on the Poincare section, approximating the dynamics to linear
order in $x_{k}$. We expect this approximation to be accurate when
the Poincare map is roughly linear in and near the subset $E$, e.g.
if $E$ is sufficiently small. We perform this analysis using phase-isostable
coordinates \citep{wilson_isostable_2016} under which the linearized
dynamics are particularly simple.

We note that \citet{wilson_operational_2018} have previously introduced
a notion of phase that also addresses features present in output signals,
like neuronal voltage spikes. In light of the work we present here,
they, too, consider output features that correspond to the crossing
of a Poincare section $S$, and, under their ``operational'' definition
of phase, the oscillator completes one cycle with every crossing of
$S$. Using phase-isostable coordinates, they derive a representation
of the transient response of the operational phase to forcing. But,
since their analysis is perturbative, it is not clear that the operational
phase can be immediately generalized to account for amplitude-dependent
events (and thus phase slips) defined via an event subsection $E$.
The amplitude-dependent events, however, underlie the unruly noise
response of oscillators that is of interest here. In our approach,
in (\ref{eq:Fano variance}) in particular, we capture the distinguished
crossings of the Poincare section within the Markov renewal framework
via an additional element: the reward function $1_{E}\left(x\right)$.

We also note that \citet{schwalger_theory_2010} have previously considered
Markov renewal processes on finite state spaces where, likewise, only
some transitions are distinguished as events, and they provide a broad
analysis of such models. As an application, they consider the negative
correlations between subsequent inter-spike-intervals that are seen
in some neurons, and they offer ad-hoc, minimal models with a small
number of discrete states that qualitatively reproduce the correlation.
In this work, we have a similar goal: capturing a statistical property
of a non-renewal point process, here the unruliness in the temporal
variance growth rate for events, in a reduced model. Rather than using
an ad-hoc model, however, we make use of a systematic reduction that
connects the full system (\ref{eq: system}) directly to a linear
model on the continuous-state-space Poincare section.

\section{Asymptotic Variance Growth Rates for Markov and Markov Renewal Processes}

\label{sec:Fano-Variance-and-Function-Variance}Our immediate goal
is an understanding of the (asymptotic)\footnote{``Asymptotic'' here reflects the fact that we are taking the limit
$t\rightarrow\infty$ or $n\rightarrow\infty$. Since, throughout
this work, we are only considering long-time statistics - i.e. this
type of limit, we will drop this modifier in future references to
this and related quantities.} temporal variance growth rate (\ref{eq:Fano variance}). In this
section and the corresponding SI Section \ref{sec: SI - Event TVGR Formula},
we derive an expression for it, (\ref{eq:FanoVar decomposed}), that
is amiable to the analysis that appears in the following sections.
We find (\ref{eq:FanoVar decomposed}) in two steps: we first write
the temporal variance growth rate in terms of a \emph{sequential }variance
growth rate (Section \ref{subsec:TVGR as SVGR} and SI Sections \ref{subsec: SI - TVGR Formula by Analogy}
and \ref{subsec: SI - TVGRFormula Discrete}), and then we decompose
the resulting formula into components that will show qualitatively
different behaviors when applied to the Markov renewal processes arising
from oscillatory systems (Section \ref{subsec:TVGR decomposed}).

\subsection{Sequential Variance Growth Rates\label{subsec: SVGR}}

The TVGR $\mathcal{V}_{\mathrm{E}}^{\left(t\right)}$, (\ref{eq:Fano variance}),
is a statistical property of the Markov renewal process $T_{k}^{S}$
on the Poincare section $S$. Recall that a Markov renewal process
is a Markov process in which each step $k$ occurs at a particular
time $T_{k}$. The time interval $\Delta T_{k}\equiv T_{k+1}-T_{k}$
between steps is a random variable that may depend on the preceding
and subsequent states, $x_{k}$ and $x_{k+1}$, and the $\Delta T_{k}$
are independent of each other once conditioned on the states $x_{k}$
and $x_{k+1}$ (see e.g. Chapter 6 of \citet{kao_introduction_2019}
for a pedagogical introduction). Since $\mathcal{V}_{\mathrm{E}}^{\left(t\right)}$
is defined as a growth rate over time, it incorporates that temporal
information. By contrast, we define the sequential variance growth
rate $\mathcal{V}_{x\mapsto f\left(x\right)}^{\left(n\right)}$ (SVGR)
of a function $f$ on a Markov process with states $x\in S$ as
\begin{equation}
\mathcal{V}_{x\mapsto f\left(x\right)}^{\left(n\right)}\equiv\mathcal{V}^{\left(n\right)}\left[x\mapsto f\left(x\right)\right]\equiv\lim_{n\rightarrow\infty}\frac{1}{n}\mathrm{var}\left\{ \sum_{k=1}^{n}f\left(x_{k}\right)\right\} .\label{eq:function variance}
\end{equation}
In our notation, the subscript $x\mapsto f\left(x\right)$ of $\mathcal{V}$
identifies how one maps the Markov states $x$ to the value $f\left(x\right)$
that is then accumulated from one step to the next. In some instances,
for readability, we will alternatively write the expression $x\mapsto f\left(x\right)$
in square brackets. As in the TVGR, the variance in the SVGR is taken
over sample sequences given by $x_{k}$. However, unlike the TVGR,
the SVGR $\mathcal{V}_{x\mapsto f\left(x\right)}^{\left(n\right)}$
does not take into account any information about the time between
steps. In the case that $f$ is an indicator function for those $x$
that correspond to events, the function variance is a simplified version
of the TVGR wherein the randomness of the time intervals in the Markov
renewal process is ignored and all intervals are set to $1$.

Note that, like $x_{k}$, $f_{k}=f\left(x_{k}\right)$ is itself a
sequence of correlated random variables. When the variance of the
normalized sum $\mathcal{V}_{x\mapsto f\left(x\right)}^{\left(n\right)}$
exists, it follows the standard formula for series of random variables,

\begin{equation}
\mathcal{V}_{x\mapsto f\left(x\right)}^{\left(n\right)}=\mathrm{var}\left\{ f\left(x_{0}\right)\right\} +\sum_{k=1}^{\infty}\mathrm{cov}\left\{ f\left(x_{0}\right),f\left(x_{k}\right)\right\} +\sum_{k=1}^{\infty}\mathrm{cov}\left\{ f\left(x_{k}\right),f\left(x_{0}\right)\right\} .\label{eq:FV sum of variances}
\end{equation}
Here, we imagine choosing the initial state $x_{0}$ randomly according
to the invariant density of the Markov process. In that case, by the
stationarity of the Markov process, $x_{0}$ is as good as any arbitrary
element of the sequence. We write the two sums in (\ref{eq:FV sum of variances})
separately to allow $f$ to be vector-valued, in which case $\mathrm{cov}\left(\cdot,\cdot\right)$
is in general a non-symmetric cross-covariance matrix. Note that the
definition of the SVGR (\ref{eq:function variance}) and the formula
(\ref{eq:FV sum of variances}) also allow for situations where the
function $f\left(x\right)$ itself is not deterministic, but takes
on random values that may depend on $x$. This is a feature we will
make use of in the following formulae.

We will also make use of a generalization of the SVGR, allowing for
functions $f$ that depend on adjacent points $x_{k}$ and $x_{k+1}$
of the Markov process:
\begin{equation}
\mathcal{V}_{\left(x\rightarrow x^{\prime}\right)\mapsto f\left(x,x^{\prime}\right)}^{\left(n\right)}\equiv\lim_{n\rightarrow\infty}\frac{1}{n}\mathrm{var}\left\{ \sum_{k=1}^{n}f\left(x_{k},x_{k+1}\right)\right\} .\label{eq: genearlized function variance}
\end{equation}
For that we find it useful to define the Markov process on the ``edges''
of the original process. We notate the sequence of states of the new
process as $z_{k}\equiv\left(x_{k}\rightarrow x_{k+1}\right)$. The
process's dynamics are wholly induced by that of the original process:
the probability to transition from $\left(a\rightarrow b\right)$
to $\left(c\rightarrow d\right)$ under the new process is the probability
to transition from $b$ to $d$ under the original process if $b=c$
and is $0$ otherwise. The generalized SVGR on the original process
is then just a standard SVGR on the new process, $\mathcal{V}_{\left(x\rightarrow x^{\prime}\right)\mapsto f\left(x,x^{\prime}\right)}^{\left(n\right)}=\mathcal{V}_{z\mapsto f\left(z\right)}^{\left(n\right)}$,
and so it can be computed via (\ref{eq:FV sum of variances}).

\subsection{The Event Temporal Variance Growth Rate as a Sequential Variance
Growth Rate\label{subsec:TVGR as SVGR}}

In SI Section \ref{sec: SI - Event TVGR Formula}, we use a result
from reward-renewal theory \citep{smith_regenerative_1955} to make
the non-rigorous argument that the TVGR for events $\mathcal{V}_{E}^{\left(t\right)}$
can be written in terms of the generalized SVGR of an (ordinary, non-renewal)
Markov process. We find
\begin{equation}
\mathcal{V}_{E}^{\left(t\right)}=\frac{1}{\mathrm{E}\left\{ \Delta T\right\} }\mathcal{V}^{\left(n\right)}\left[\left(x\rightarrow x^{\prime}\right)\mapsto\left(1_{E}\left(x\right)-\mathrm{E}\left\{ 1_{E}\left(x\right)\right\} \frac{\Delta t\left(x,x^{\prime}\right)}{\mathrm{E}\left\{ \Delta T\right\} }\right)\right],\label{eq:FanoVar from funcVar}
\end{equation}
where $\Delta t\left(x,x^{\prime}\right)$ is a random-valued function
with the property that $\Delta t\left(x_{k},x_{k+1}\right)$ has the
same distribution as $\Delta T_{k}$ when conditioned on $x_{k}$
and $x_{k+1}$. Below and in the following sections, we will rescale
time so that the expected time interval is $1$, and, because we will
use it often, introduce $\mathcal{E}$ as the event probability; thus
\begin{eqnarray}
\mathrm{E}\left\{ \Delta T\right\}  & = & 1\label{eq:expectation_Deltat}\\
\mathrm{E}\left\{ 1_{E}\left(x\right)\right\}  & = & \mathcal{E}.\label{eq:expectation_event}
\end{eqnarray}
With these simplifying choices, (\ref{eq:FanoVar from funcVar}) becomes
\begin{equation}
\mathcal{V}_{E}^{\left(t\right)}=\mathcal{V}_{\left(x\rightarrow x^{\prime}\right)\mapsto\left(1_{E}\left(x\right)-\mathcal{E}\Delta t\left(x,x^{\prime}\right)\right)}^{\left(n\right)},\label{eq:FanoVar from funcVar simple}
\end{equation}
which can in principle be computed via (\ref{eq:FV sum of variances}).

Before moving on, we make the following disclaimer. In SI Section
\ref{subsec: SI - TVGR Formula by Analogy}, we only give a plausibility
argument that the TVGR can by computed via (\ref{eq:FanoVar from funcVar simple}).
We offer a more detailed derivation in SI Section \ref{subsec: SI - TVGRFormula Discrete}
resulting in the same formula, but that derivation is only valid for
Markov chains with finitely-many states. Even so, (\ref{eq:FanoVar from funcVar simple})
has a clear interpretation for more general state spaces. We assume
that the result holds for the Markov renewal processes we consider
in this work, which, as we discuss in Section \ref{subsec:Linearized-Poincare-Map},
are innocuous Gaussian processes. We provide some empirical validation
of (\ref{eq:FanoVar from funcVar simple}) in Section \ref{subsec:comparisonTheoryNumericsSimulation}.

\subsection{Decomposition of the Temporal Variance Growth Rate\label{subsec:TVGR decomposed}}

The TVGR (\ref{eq:FanoVar from funcVar simple}) is a (type of) variance
of the sum of the correlated quantities $1_{E}\left(x^{\prime}\right)$
and $-\mathcal{E}\Delta t\left(x,x^{\prime}\right)$, and it would
be reasonable to guess that it can be decomposed as the variance of
each term plus a type of covariance $\mathcal{CV}$, 
\begin{equation}
\mathcal{V}_{E}^{\left(t\right)}=\mathcal{V}_{x\mapsto1_{E}\left(x\right)}^{\left(n\right)}-2\,\mathcal{E}\,\mathcal{CV}_{\left(x\rightarrow x^{\prime}\right)\mapsto1_{E}\left(x\right),\left(x\rightarrow x^{\prime}\right)\mapsto\Delta t\left(x,x^{\prime}\right)}^{\left(n\right)}+\mathcal{E}^{2}\,\mathcal{V}_{\left(x\rightarrow x^{\prime}\right)\mapsto\Delta t\left(x,x^{\prime}\right)}^{\left(n\right)}.\label{eq:FanoVar decomposed}
\end{equation}
This is indeed the case, as we show in SI Section \ref{subsec: SI - TVGR Decomposition},
where we expand (\ref{eq:FanoVar from funcVar simple}) via (\ref{eq:FV sum of variances})
and reorganize the resulting terms. $\mathcal{CV}_{x\mapsto f\left(x\right),x\mapsto g\left(x\right)}^{\left(n\right)}$
is defined via a natural extension of (\ref{eq:function variance})
and (\ref{eq:FV sum of variances}),
\begin{eqnarray}
\mathcal{CV}_{x\mapsto f\left(x\right),x\mapsto g\left(x\right)}^{\left(n\right)} & \equiv & \lim_{n\rightarrow\infty}\frac{1}{n}\mathrm{cov}\left\{ \sum_{k=0}^{n-1}f\left(x_{k}\right),\sum_{k=0}^{n-1}g\left(x_{k}\right)\right\} \label{eq: funcCV}\\
 & = & \mathrm{cov}\left\{ f\left(x_{0}\right),g\left(x_{0}\right)\right\} +\sum_{k=1}^{\infty}\mathrm{cov}\left\{ f\left(x_{0}\right),g\left(x_{k}\right)\right\} +\sum_{k=1}^{\infty}\mathrm{cov}\left\{ f\left(x_{k}\right),g\left(x_{0}\right)\right\} .\label{eq:funcCV sum of variances}
\end{eqnarray}
Note that if $f$ and $g$ are vector-valued, $\mathcal{CV}_{x\mapsto f\left(x\right),x\mapsto g\left(x\right)}^{\left(n\right)}$
is analogous to a cross-covariance matrix and will in general be non-symmetric.

The first term of (\ref{eq:FanoVar decomposed}) is just the sequential
variance that arises by ignoring the temporal aspect of the process,
i.e. by setting $\Delta T=1$. We call it the ``Markov-only'' term.
Similarly, if one ignores the partitioning of the state-space by $E$
by, say, setting $E=S$, there is no variance in $1_{E}\left(x\right)$.
In that case, only the last term of (\ref{eq:FanoVar decomposed}),
$\mathcal{V}_{\left(x\rightarrow x^{\prime}\right)\mapsto\Delta t\left(x,x^{\prime}\right)}^{\left(n\right)}$,
is non-zero. We call this the ``temporal'' term. We call the remaining
covariance term ``mixed'', since it only appears when both the temporal
and partitioning aspects of the process are present and are inter-dependent.

\section{Temporal Variance Growth Rate for a Toy Model}

\label{sec:Fano-Variance-for-Toy-Model}

Before considering the TVGR for stochastic limit-cycle oscillators,
we introduce a simple, discrete $2$-state toy model that reveals
the quality of each the Markov, the mixed, and the temporal components
in (\ref{eq:FanoVar decomposed}) and how they interact to generate
the non-monotonic behavior of $\mathcal{V}_{E}^{\left(t\right)}$.
We take $x\in\left\{ 0,1\right\} =S$ and $E=\left\{ 1\right\} $.
In order to fully specify the Markov-renewal process, we must prescribe
the joint, conditional distribution for $x_{k+1}$ and $\Delta T_{k}^{S}=T_{k+1}^{S}-T_{k}^{S}$
given $x_{k}$. For simplicity, we assume that $\Delta T_{k}^{S}$
is independent of $x_{k}$ and that $x_{k+1}$ is completely determined
by $\Delta T_{k}^{S}$. Equivalently, there is a subset $E_{\Delta T}$
of the positive real line, so that $x_{k+1}$ corresponds to an event
if and only if $\Delta T_{k}^{S}\in E_{\Delta T}$. Thus, 
\begin{equation}
x_{k+1}=1_{E_{\Delta T}}\left(\Delta T_{k}^{S}\right).\label{eq:toy_model}
\end{equation}
This very roughly captures what we expect to see in stochastic oscillators:
the time to reach $E$ will generally be distributed differently from
the time to reach $S\setminus E$. Under these assumptions, $x_{j}$
and $x_{k}$ are independent when $j\ne k$. Likewise, the pairs $\Delta T_{j}^{S}$
and $x_{k+1}$ as well as $\Delta T_{j}^{S}$ and $\Delta T_{k}^{S}$
are independent when $j\ne k$. In this simple scenario, only one
term survives in each of the components of the TVGR, corresponding
to the first terms in (\ref{eq:FV sum of variances}) and (\ref{eq:funcCV sum of variances}).
Each of the Markov-only, mixed and temporal components of $\mathcal{V}_{E}^{\left(t\right)}$
can be evaluated easily, since $x_{k}$ is a sequence of independent
Bernoulli-distributed random variables governed by the probabilities
$\mathcal{E}$ and $1-\mathcal{E}$. We have via (\ref{eq:FV sum of variances})
and (\ref{eq:funcCV sum of variances}) 
\begin{eqnarray*}
\mathcal{V}_{x\mapsto1_{E}\left(x\right)}^{\left(n\right)} & = & \mathrm{var}\left\{ 1_{E}\left(x_{0}\right)\right\} \\
 & = & \mathcal{E}\left(1-\mathcal{E}\right)\\
\mathcal{CV}_{\left(x\rightarrow x^{\prime}\right)\mapsto1_{E}\left(x\right),\left(x\rightarrow x^{\prime}\right)\mapsto\Delta t\left(x,x^{\prime}\right)}^{\left(n\right)} & = & \mathrm{cov}\left\{ 1_{E}\left(x_{1}\right),\Delta t\left(x_{0},x_{1}\right)\right\} \\
 & = & \mathcal{E}\left(1-\mathcal{E}\right)\left(\overline{\Delta t}_{\left(1\right)}-\overline{\Delta t}_{\left(0\right)}\right)\\
\mathcal{V}_{\left(x\rightarrow x^{\prime}\right)\mapsto\Delta t\left(x,x^{\prime}\right)}^{\left(n\right)} & = & \mathrm{var}\left\{ \Delta t\left(x_{0},x_{1}\right)\right\} \\
 & = & \mathrm{var}\left\{ \Delta T_{0}^{S}\right\} ,
\end{eqnarray*}
where $\overline{\Delta t}_{\left(n\right)}=\mathrm{E}\left\{ \Delta T_{k}^{S}\left|x_{k+1}=n\right.\right\} $
for $n=0,1$.

\begin{figure}
\noindent \begin{centering}
\includegraphics[width=1\textwidth]{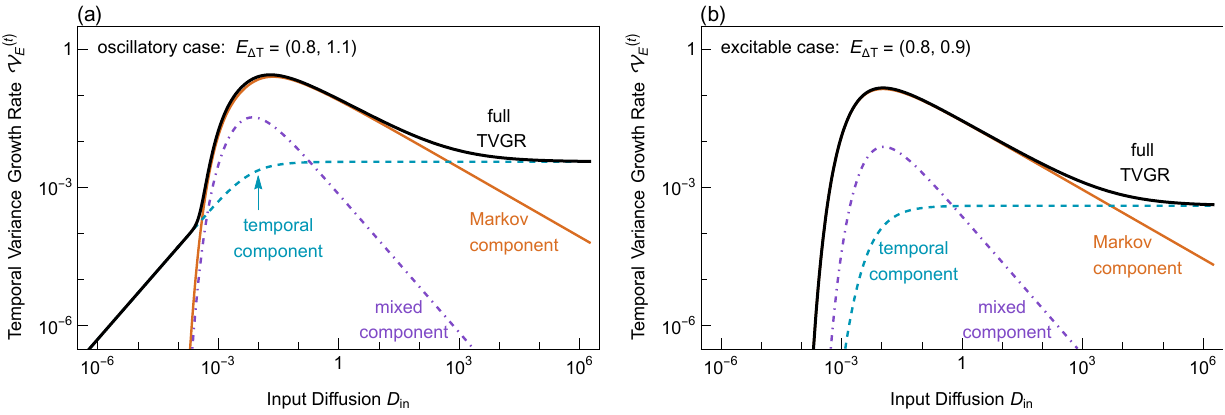}
\par\end{centering}
\caption{\textbf{Temporal variance growth rate $\mathcal{V}_{E}^{\left(t\right)}$
for the 2-state toy model (\ref{eq:toy_model}).} The \textquotedblleft Markov-only\textquotedblright{}
and mixed components of the TVGR are non-monotonic and the temporal
component is monotonic. In (a), the oscillatory case, their sum, the
full TVGR, has the characteristic unruly quality: an initial linear
growth, a strong nonlinear amplification, a maximum, and a subsequent
decrease. The choice of an asymmetric interval $E_{\Delta T}$ about
$\Delta T^{S}=1$ is only made so that $\overline{\Delta t}_{1}\protect\ne\overline{\Delta t}_{0}$
and the mixed term is nonzero. It is not essential to the appearance
of unruliness. In (b), the excitable case, the TVGR is missing the
initial linear growth.\label{fig: toy_model}}
\end{figure}

Recall that we are interested in the TVGR as a function of an input
noise strength $D_{\textit{in}}$. In our toy model, we assert that the effect
of increased input noise will be to increase the variance of the inter-step
times: we take $\Delta T_{k}^{S}$ to be a random variable with mean
$1$ and variance $\sigma^{2}\equiv\frac{D_{\textit{in}}}{2}$. Following (\ref{eq:FanoVar decomposed})
the TVGR can then be written simply as
\begin{equation}
\mathcal{V}_{E}^{\left(t\right)}=\mathcal{E}\left(1-\mathcal{E}\right)\left(1-2\mathcal{E}\left(\overline{\Delta t}_{\left(1\right)}-\overline{\Delta t}_{\left(0\right)}\right)\right)+\mathcal{E}^{2}\,\frac{D_{\textit{in}}}{2},\label{eq:Fano_variance_toy_model}
\end{equation}
where $\mathcal{E}$, $\overline{\Delta t}_{\left(1\right)}$, and
$\overline{\Delta t}_{\left(0\right)}$ each vary with $D_{\textit{in}}$.
When $E_{\Delta T}$ is a contiguous interval and $\Delta T_{n}^{S}$
is unimodally distributed, we expect that each will in fact vary monotonically
with $D_{\textit{in}}$. Then, because of the appearance of the Bernoulli variance
$\mathcal{E}\left(1-\mathcal{E}\right)$, we expect the Markov and
mixed terms to be non-monotonic, but the temporal term to be monotonic.
As a concrete example, we consider the case where $\Delta T_{n}$
is a Gaussian random variable and $E_{\Delta T}$ is an interval $\left(a,\,b\right)$
with $a<1<b$. Note that this choice of $E_{\Delta T}$ implies that
events will be produced regularly when $D_{\textit{in}}=0$, since then $\Delta T_{k}^{S}=1\in E_{\Delta T}$
for all $k$. In this case, the temporal term is indeed monotonic,
producing a roughly linear TVGR in the weak-input-noise regime (Figure
\ref{fig: toy_model}a). And the Markov and mixed terms are non-monotonic
and produce the strong and non-monotonic nonlinearity that is characteristic
of what we call ``unruliness'' (compare Figures \ref{fig: MMO_DEff_vs_DPhase}b
and \ref{fig: toy_model}a).

Although we do not pursue it further in this work, we note that the
toy model can also be placed in an ``excitable'' rather than an
oscillatory regime by choosing $E_{\Delta T}=\left(a,\,b\right)$
with either $a<b<1$ or $1<a<b$. Here event-production requires a
nonzero amount of noise, $D_{\textit{in}}>0$. In contrast with the oscillatory
case, no initial linear growth appears in the TVGR, and the Markov-only
component dominates at low noise (Figure \ref{fig: toy_model}b).

The pattern of behavior of the toy model in the oscillatory regime
is mirrored in the TVGR for planar limit cycle oscillators, as we
show in Section \ref{sec: Main Results - Planar Oscillators}. There,
as is the case here, the strong nonlinear rise in the TVGR roughly
corresponds to $\mathcal{E}$ decreasing from $1$ to $\frac{1}{2}$.
In the oscillator, this reflects an increasing number of phase slips,
i.e. oscillation cycles that miss the event subset $E$ and replace
events with non-events, or vice versa. The TVGR decreases again when
$\mathcal{E}$ decreases past $\frac{1}{2}$, where the phase slips
dominate the oscillations. Thus, the Bernoulli variance $\mathcal{E}\left(1-\mathcal{E}\right)$
underlies the seemingly counterintuitive increased regularity in the
(lack of) events with increased input noise strength.

\section{Temporal Variance Growth Rate for Limit Cycle Oscillators}

\label{sec:Fano-Variance-for-Limit-Cycle}

We now apply the theory developed in Section \ref{sec:Fano-Variance-and-Function-Variance}
to the particular Markov renewal processes arising from Poincare map
dynamics for oscillators. Here we first sketch the case of general
oscillators, which is discussed in detail in SI Sections \ref{sec: SI - Averaging + Phase-Isostable}
and \ref{sec: SI - TVGR Oscillator Derivation}. In Section \ref{sec: Main Results - Planar Oscillators}
we then focus on the implications for unruliness in planar oscillators.

\subsection{Noise-Driven Oscillators in Phase-Isostable Coordinates}

We first write the limit-cycle oscillator in $d+1$ dimensions, (\ref{eq: system}),
in phase-isostable coordinates ($\phi,\psi_{1},\psi_{2},\ldots,\psi_{d}$)
(see \citet{wilson_isostable_2016} and SI Section \ref{subsec: SI - Phase-Isostable Coordinates}
for an extended discussion). For weak noise and assuming small deviations
$\psi_{i}$ from the limit cycle, one has
\begin{eqnarray}
d\phi & = & dt+\sqrt{2D_{\textit{in}}}\vec{Z}\left(\phi\right)^{T}\mathbf{G}_{\textit{LC}}\left(\phi\right)\,d\vec{W}+h.o.t.\label{eq: phase-isostable phase}\\
d\psi_{i} & = & -\kappa_{i}\psi_{i}\,dt+\sqrt{2D_{\textit{in}}}\vec{Y}_{i}\left(\phi\right)^{T}\mathbf{G}_{\textit{LC}}\left(\phi\right)\,d\vec{W}+h.o.t.\,\quad i=1\ldots d\,.\label{eq: phase-isostable isostable}
\end{eqnarray}
Here, $\vec{Z}\left(\phi\right)$ is known as the phase response curve
(PRC) and the $\vec{Y}_{i}\left(\phi\right)$ as isostable response
curves (IRCs). They are the gradients of the phase and isostables
coordinates, respectively, as functions of the state $\vec{y}$ evaluated
at the point on the limit cycle with phase $\phi$. Here, in the limit
of small $D_{\textit{in}}$, they give the responses of the phase and isostable
variables to the perturbation $\sigma\mathbf{G}\left(\vec{y}\right)d\vec{W}\left(t\right)$
that appears in (\ref{eq: system}). At lowest order, $\mathbf{G}\left(\vec{y}\right)$
is replaced by $\mathbf{G}_{\textit{LC}}\left(\Phi\left(\vec{y}\right)\right)$,
where $\mathbf{G}_{\textit{LC}}\left(\phi\right)$ is the value that $\mathbf{G}$
has at the point on the limit cycle with phase $\phi$. The higher
order terms ($h.o.t$.'s) in (\ref{eq: phase-isostable isostable})
and in the following are $\mathcal{O}\left(D_{\textit{in}},\sqrt{D_{\textit{in}}}\left|\vec{\psi}\right|,\left|\vec{\psi}\right|^{2}\right)$,
where $\vec{\psi}\equiv\left(\psi_{1},\psi_{2},\ldots,\psi_{d}\right)$.

We note that, in the context stochastically-forced oscillators, recent
studies have introduced alternative, noise-strength-dependent versions
of phase \citep{schwabedal_phase_2013,thomas_asymptotic_2014} and
isostable coordinates \citep{perez-cervera_isostables_2021}. Here
in (\ref{eq: phase-isostable phase}) and (\ref{eq: phase-isostable isostable}),
we however make use of the phase-isostable coordinate system defined
with respect to the deterministic dynamics (see (\ref{eq: unperturbed_phase})
and \ref{eq: unperturbed_isostable}).

\subsection{Linearized Poincare Map and First Passage Time}

\label{subsec:Linearized-Poincare-Map}Turning now to the Poincare
section and map, we represent the Poincare section $S$ near the limit
cycle by 
\begin{equation}
\phi=\phi_{S}\left(\vec{\psi}\right)=\vec{m}_{S}^{T}\vec{\psi}+\mathcal{O}\left(\left|\vec{\psi}\right|^{2}\right).\label{eq:phi Poincare section}
\end{equation}
Note that this section intersects the limit cycle at $\phi=0$, and
therefore taking $\vec{m}_{S}=0$ chooses the isochron given by $\phi=0$
as the Poincare section. In the general case, $\vec{m}_{S}\ne0$,
the Poincare section does not correspond to an isochron, and so the
time $\Delta T_{n}^{S}$ between consecutive section crossings differs
from the period of the periodic orbit by an amount that reflects the
change in $\vec{\psi}$ between the crossings. One has to leading
order $\mathrm{E}\left\{ \vec{\psi}_{n+1}\left|\vec{\psi}_{n}\right.\right\} \sim\mathbf{\Lambda}\vec{\psi}_{n}$
and $\mathrm{E}\left\{ \Delta T_{n}^{S}\left|\vec{\psi}_{n}\right.\right\} \sim1+\vec{\delta\mu}^{T}\vec{\psi}_{n}$,
where $\Lambda$ is the diagonal matrix with entries $e^{-\kappa_{i}}$
and $\vec{\delta\mu}\equiv m_{S}^{T}\left(\mathbf{I}-\mathbf{\Lambda}\right)$.
Taking into account the contribution of the noisy perturbations at
first order, we expect
\begin{eqnarray}
\Delta T_{n}^{S} & = & 1+\vec{\delta\mu}^{T}\vec{\psi}_{n}+\sqrt{2D_{\textit{in}}}\zeta_{n}+h.o.t.\label{eq: linearized PM}\\
\vec{\psi}_{n+1} & = & \mathbf{\Lambda}\vec{\psi}_{n}+\sqrt{2D_{\textit{in}}}\vec{\eta}_{n}+h.o.t.,\nonumber 
\end{eqnarray}
where $\zeta_{n}$ and $\vec{\eta}_{n}$ are normally distributed
random variables with mean $0$, which, in general, will be correlated.
Because the process is Markovian, $\zeta_{n}$ and $\vec{\eta}_{n}$
are i.i.d. across $n$. The (co)variances of $\zeta_{n}$ and $\vec{\eta}_{n}$
depend on $\vec{m}_{S}$, $\mathbf{\Lambda}$ and the PRC and IRCs.
They are derived along with $\vec{\delta\mu}$ in SI Section \ref{subsec: SI - Linearized Map}.
Note that since we assume stability of the limit cycle, all Floquet
multipliers that appear in the diagonal matrix $\mathbf{\Lambda}$
have magnitude less than $1$. Then in the limit $n\rightarrow\infty$,
where the process becomes stationary, we expect $\vec{\psi}_{n}$
to be distributed as
\begin{equation}
\vec{\psi}_{n}\stackrel{dist.}{\sim}\mathcal{N}\left(0,2D_{\textit{in}}\mathbf{\Gamma}_{\textit{ss}}\right),\label{eq: steady-state distribution}
\end{equation}
where $\mathcal{N}\left(\vec{\mu},\mathbf{\Sigma}\right)$ is the
multivariate normal distribution with mean $\vec{\mu}$ and covariance
matrix $\mathbf{\Sigma}$. The ``steady-state'' matrix $\mathbf{\Gamma}_{\textit{ss}}$
is found via (\ref{eq: linearized PM}) by the condition that the
covariance matrices for $\vec{\psi}_{n}$ and $\vec{\psi}_{n+1}$
are the same, namely $2D_{\textit{in}}\mathbf{\Gamma}_{\textit{ss}}$. Thus,
\[
\mathrm{var}\left\{ \vec{\psi}_{n+1}\right\} =\mathrm{var}\left\{ \mathbf{\Lambda}\vec{\psi}_{n}\right\} +2D_{\textit{in}}\mathbf{\Gamma}
\]
implies 
\[
2D_{\textit{in}}\mathbf{\Gamma}_{\textit{ss}}=2D_{\textit{in}}\left(\mathbf{\Lambda}\mathbf{\Gamma}_{\textit{ss}}\mathbf{\Lambda}+\mathbf{\Gamma}\right)
\]
where $\Gamma$ is the covariance matrix of $\vec{\eta}_{n}$. Equivalently,
\begin{equation}
\mathbf{\Gamma}_{\textit{ss}}=\sum_{k=0}^{\infty}\mathbf{\Lambda}^{k}\mathbf{\Gamma}\mathbf{\Lambda}^{k}.\label{eq: Gamma_ss}
\end{equation}
Further, based on (\ref{eq: linearized PM}) and using elementary
relationships between conditional and joint normal distributions,
we show in SI Section \ref{subsec: SI - Steady-state Densities} that
the joint, stationary distribution for $\vec{\psi}$ across multiple
iterations of the map is given by 
\begin{equation}
\left(\begin{array}{c}
\vec{\psi}_{n+m}\\
\vec{\psi}_{n}
\end{array}\right)\stackrel{dist.}{\sim}\mathcal{N}\left(0,2D_{\textit{in}}\mathbf{C}_{m}\right);\;\mathbf{C}_{m}\equiv\left(\begin{array}{cc}
\mathbf{\Gamma}_{\textit{ss}} & \Lambda^{m}\mathbf{\Gamma}_{\textit{ss}}\\
\mathbf{\Gamma}_{\textit{ss}}\Lambda^{m} & \mathbf{\Gamma}_{\textit{ss}}
\end{array}\right),\label{eq:mth joint distribution}
\end{equation}
again in the limit $n\rightarrow\infty$.

\subsection{Temporal Variance Growth Rate for Linearized Dynamics}

\label{subsec:Fano-Variance_linear}We can compute the temporal variance
growth rate using the distributions and statistic cited above and
derived in SI Section \ref{subsec: SI - Steady-state Densities}.
We use (\ref{eq:FanoVar decomposed}), where the Markov state $x$
is in this case the vector $\vec{\psi}$. The above linearization
of the dynamics is essential for the computation; we make progress
by leveraging the fact that all of the probability distributions are
Gaussian. While we leave the full details of the derivation to SI
Section \ref{subsec: SI - TVGR,  term-by-term}, here we comment on
results for each component of the TVGR, highlighting the dependence
of each on $D_{\textit{in}}$. In the SI and in the following, we take $N\left(\cdot;0,\mathbf{\Sigma}\right)$
to be the probability density function for the multivariate normal
distribution with mean $\vec{0}$ and covariance matrix $\mathbf{\Sigma}$.
Then, for example, (\ref{eq: steady-state distribution}) gives
\begin{equation}
\mathcal{E}=\int_{E}N\left(x;0,2D_{\textit{in}}\mathbf{\Gamma}_{\textit{ss}}\right)dx.\label{eq: calE}
\end{equation}

From (\ref{eq:FV sum of variances}), we see that the Markov-only
component $\mathcal{V}_{x\mapsto1_{E}\left(x\right)}^{\left(n\right)}$
is made up of a series of integrals over $N\left(\cdot;0,\sigma^{2}\mathbf{C}_{m}\right)$,
\begin{equation}
\mathcal{V}_{x\mapsto1_{E}\left(x\right)}^{\left(n\right)}=\left(\mathcal{E}-\mathcal{E}^{2}\right)+2\sum_{m=1}^{\infty}\left(\mathcal{E}_{m}-\mathcal{E}^{2}\right),\label{eq: Markov-only sum}
\end{equation}
where 
\begin{equation}
\mathcal{E}_{m}\equiv\int_{E\times E}N\left(z;0,2D_{\textit{in}}\mathbf{C}_{m}\right)dz.\label{eq: calE_m}
\end{equation}
Note $\mathcal{E}_{0}=\mathcal{E}$ and $\mathcal{E}_{m}\rightarrow\mathcal{E}^{2}$
as $m\rightarrow\infty$ (see the definition of $\mathbf{C}_{m}$
in (\ref{eq:mth joint distribution})), so that in some sense $\mathcal{E}_{m}$
interpolates between $\mathcal{E}$ and $\mathcal{E}^{2}$ as $m$
increases. Notoriously, closed form expressions for integrals like
$\mathcal{E}_{m}$ are not known even if $E$ is simple, e.g. a rectangular
region. One must resort to numerical analysis or, as we do in SI Section
\ref{sec: SI - Bounds} , to bounding procedures to capture their
behavior as a function of $D_{\textit{in}}$.

The mixed component $\mathcal{CV}_{\left(x\rightarrow x^{\prime}\right)\mapsto1_{E}\left(x^{\prime}\right),\left(x\rightarrow x^{\prime}\right)\mapsto\Delta t\left(x,x^{\prime}\right)}^{^{\left(n\right)}}$
involves both $1_{E}$ and $\Delta t\left(x,x^{\prime}\right)$. Under
the linearized dynamics (\ref{eq: linearized PM}), $\Delta t\left(x,x^{\prime}\right)$
is linear in $x$ and $x^{\prime}$ (see (\ref{eq:inter-step interval dist})
in SI Section \ref{subsec: SI - Steady-state Densities}). The mixed
component, therefore, has at its essence the quantity
\begin{equation}
x_{E}\equiv\frac{1}{\mathcal{E}}\int_{E}xN\left(x;0,2D_{\textit{in}}\mathbf{\Gamma}_{\textit{ss}}\right)dx,\label{eq: xE}
\end{equation}
which identifies the center of mass of the event subset $E$ with
respect to the steady-state density $N\left(\cdot;0,2D_{\textit{in}}\Gamma_{\textit{ss}}\right)$.
Indeed, the only dependence $\mathcal{CV}_{\left(x\rightarrow x^{\prime}\right)\mapsto1_{E}\left(x^{\prime}\right),\left(x\rightarrow x^{\prime}\right)\mapsto\Delta t\left(x,x^{\prime}\right)}^{^{\left(n\right)}}$
has on $D_{\textit{in}}$ is via the product $x_{E}\mathcal{E}$ (see SI Section
\ref{subsec: SI - TVGR,  term-by-term}). For many simple choices
of $E$, $x_{E}$ can be written in closed-form in terms of well-known
functions, as we will do in Section \ref{sec: Main Results - Planar Oscillators}.
As a function of $D_{\textit{in}}$, we expect the vector-valued $x_{E}$ to
increase in magnitude from the minimal distance between $x=0$ and
$E$ (when the density $N$ is localized around $x=0$) and converge
to the centroid of $E$ as $D_{\textit{in}}\rightarrow\infty$ (when the density
conditioned on $x\in E$ approaches the uniform density).

The temporal factor, $\mathcal{V}_{\left(x\rightarrow x^{\prime}\right)\mapsto\Delta t\left(x,x^{\prime}\right)}^{^{\left(n\right)}}$,
does not involve the restriction of $x$ to $E$; all integrals found
here will be over the entire domain, and there will be no boundary-induced
effects. Therefore this factor will only depend on $D_{\textit{in}}$ via an
overall scaling by $D_{\textit{in}}$ (see SI Section \ref{subsec: SI - TVGR,  term-by-term}
for details).

In total,
\begin{equation}
\mathcal{V}_{E}^{\left(t\right)}=\left(\mathcal{E}-\mathcal{E}^{2}\right)+2\sum_{m=1}^{\infty}\left(\mathcal{E}_{m}-\mathcal{E}^{2}\right)+\vec{b}^{T}x_{E}\mathcal{E}^{2}+cD_{\textit{in}}\mathcal{E}^{2},\label{eq:FanoVar Poincare}
\end{equation}
$\vec{b}^{T}$ and $c$ and are found in SI Section \ref{subsec: SI - TVGR,  term-by-term}
to be 
\begin{eqnarray}
\vec{b}^{T} & = & 2\mathbf{\Gamma}_{\textit{ss}}^{-1}\left(\mathbf{I}-\mathbf{\Lambda}\right)^{-1}\mathbf{Y}_{G}^{T}\vec{z}_{G}\label{eq:b_general}\\
 & \equiv & 2\mathbf{\Gamma}_{\textit{ss}}^{-1}\left(\mathbf{I}-\mathbf{\Lambda}\right)^{-1}\int_{0}^{1}\mathbf{\Lambda}^{1-s}\mathbf{Y}\left(s\right)^{T}\mathbf{G}_{\textit{LC}}\left(s\right)\mathbf{G}_{\textit{LC}}\left(s\right)^{T}\vec{Z}\left(s\right)ds,\nonumber 
\end{eqnarray}
where the columns of the matrix $\mathbf{Y}\left(\phi\right)$ are
the IRCs $\vec{Y}_{i}\left(\phi\right)$, and

\begin{eqnarray}
c & = & 2\vec{z}_{G}^{T}\vec{z}_{G}\label{eq:c_general}\\
 & \equiv & 2\int_{0}^{1}\vec{Z}\left(s\right)^{T}\mathbf{G}_{\textit{LC}}\left(s\right)\mathbf{G}_{\textit{LC}}\left(s\right)^{T}\vec{Z}\left(s\right)ds,\nonumber 
\end{eqnarray}
respectively. As reflected in the above formulae, we define $\vec{z}_{G}$
to be an averaged version of the PRC $\vec{Z}\left(\phi\right)$ and
the matrix $\mathbf{Y}_{G}$ to be an averaged version of the IRC
matrix $\mathbf{Y}\left(\phi\right)$ (see SI Section \ref{subsec: SI - Averaging}
for an extended discussion of the averaging procedure). $\vec{b}$
and $c$ are therefore quantities that reflect the intrinsic oscillator
dynamics: they do not vary with $D_{\textit{in}}$ or depend on the size and
shape of $E$ within the Poincare section. It will be the focus of
the next section to develop an understanding of the $E$- and $D_{\textit{in}}$-dependent
elements $\mathcal{E}$, $\mathcal{E}_{m}$, and $x_{E}$ and how
they interact to form unruliness given different values of $\vec{b}$
and $c$.

Before moving on, we consider the limiting case where $E=S$. In this
case, since there are no steps in the Markov renewal process that
are not events, no steps $x_{k}$ are distinguished from any other.
We see this in the TVGR (\ref{eq:FanoVar Poincare}), where $x_{E}=0$
and $\mathcal{E}=\mathcal{E}_{m}=1$ when $E=S$. Only the final,
temporal term, $cD_{\textit{in}}=2\vec{z}_{G}^{T}\vec{z}_{G}D_{\textit{in}}$, survives,
and so the effective diffusion coefficient $D_{\textit{eff}}=\frac{1}{2}\mathcal{V}_{E}^{\left(t\right)}$
is precisely the linear phase diffusion $D_{\textit{phase}}$ (\ref{eq: linear phase diffusion})
from the phase reduction theory.

More generally, when $x=0\in E\subset S$, both the Markov-only and
mixed terms are eliminated as $D_{\textit{in}}\rightarrow0$. Specifically,
$\mathcal{V}_{x\mapsto1_{E}\left(x\right)}^{\left(n\right)}$ approaches
$0$ as $D_{\textit{in}}\rightarrow0$, since events are produced with high
probability and little variance. The event center of mass $x_{E}$
also approaches $0$ as $D_{\textit{in}}\rightarrow0$. Once again, only the
temporal term survives. Therefore when the event subsection intersects
the limit cycle ($0\in E$), $D_{\textit{eff}}$ coincides with $D_{\textit{phase}}$
in the weak noise limit. This shows that the effective diffusion coefficient
$D_{\textit{eff}}$ is the result of an augmentation to the phase diffusion
that takes effect above the weak-noise regime. This follows from the
fact that the phase-isostable representation, (\ref{eq: phase-isostable phase})
and (\ref{eq: phase-isostable isostable}), augments the standard
phase reduction. Note that, in particular, oscillator models in $d+1$
dimensions are represented in the full TVGR (\ref{eq:FanoVar Poincare})
by $2d$ additional derived parameters, the $d$ components of each
$\Lambda$ and $\vec{b}$, beyond the scalar $c$, which is associated
with standard phase reduction.

\section{Main Results: Planar Oscillators\label{sec: Main Results - Planar Oscillators}}

We now narrow our focus to one-dimensional Poincare maps that arise
from two-dimensional oscillators. In this case, $b$, $x_{E}$, $\Lambda$,
and $\Gamma_{\textit{ss}}$ become scalar quantities and only one column appears
in $Y_{G}\equiv\vec{y}_{G}$. From (\ref{eq: Gamma_ss}) and (\ref{eq: Gamma in terms of Y})
one obtains 
\begin{equation}
\Gamma_{\textit{ss}}=\frac{1}{1-\Lambda^{2}}\Gamma=\frac{1}{1-\Lambda^{2}}\left\Vert \vec{y}_{G}\right\Vert ^{2}\,,\label{eq: GammaSS 1D}
\end{equation}
and the expression for $b$, (\ref{eq:b_general}), simplifies to
\begin{equation}
b=2\left(1+\Lambda\right)\frac{\vec{z}_{G}^{T}\vec{y}_{G}}{\left\Vert \vec{y}_{G}\right\Vert ^{2}}\,.\label{eq:b in 1D-1}
\end{equation}
The Cauchy-Schwarz inequality applied to $\vec{z}_{G}^{T}\vec{y}_{G}$
shows that $b^{2}$ and $c=2\left\Vert \vec{z}_{G}\right\Vert ^{2}$
cannot be varied completely independently of each other, 
\[
1\ge\frac{\left(\vec{z}_{G}^{T}\vec{y}_{G}\right)^{2}}{\left\Vert \vec{y}_{G}\right\Vert ^{2}\left\Vert \vec{z}_{G}\right\Vert ^{2}}=\frac{b^{2}}{4(1+\Lambda)^{2}}\frac{\left\Vert \vec{y}_{G}\right\Vert ^{2}}{\left\Vert \vec{z}_{G}\right\Vert ^{2}}=\frac{1}{2}\frac{1-\Lambda}{1+\Lambda}\Gamma_{\textit{ss}}\frac{b^{2}}{c}\,.
\]
 Thus, the $b-c$ parameter space is limited to a parabolic region
that depends on $\Lambda$ and $\Gamma_{\textit{ss}}$
\begin{equation}
\frac{c}{\Gamma_{\textit{ss}}}\ge\frac{1}{2}\frac{1-\Lambda}{1+\Lambda}b^{2}.\label{eq: physical region-1}
\end{equation}

We will further limit our investigation to cases where $E$ is a single,
finite interval in the one-dimensional section $S$ and where the
limit cycle intersects $E$, so that events are produced regularly
in the absence of noise. Since $S$ is parameterized by the single
number $x=\psi$, this is equivalent to taking 
\begin{equation}
E=\left(w\left(\delta-\frac{1}{2}\right),w\left(\delta+\frac{1}{2}\right)\right),\label{eq: interval E}
\end{equation}
where $-\frac{1}{2}<\delta<\frac{1}{2}$ and $w$ is the width of
the interval. The TVGR will then depend parametrically on $w$ and
$\delta$ as well as $b$, $c$, and $\Gamma_{\textit{ss}}$:
\begin{eqnarray}
\mathcal{V}_{E}^{\left(t\right)}\left(D_{\textit{in}};w,\delta,b,c,\Gamma_{\textit{ss}}\right) & = & \mathcal{V}_{x\mapsto1_{E}\left(x\right)}^{\left(n\right)}\left(D_{\textit{in}};w,\delta,\Gamma_{\textit{ss}}\right)+\nonumber \\
 &  & +b\,x_{E}\left(D_{\textit{in}};w,\delta,\Gamma_{\textit{ss}}\right)\,\mathcal{E}^{2}\left(D_{\textit{in}};w,\delta,\Gamma_{\textit{ss}}\right)+\nonumber \\
 &  & +c\,D_{\textit{in}}\,\mathcal{E}^{2}\left(D_{\textit{in}};w,\delta,\Gamma_{\textit{ss}}\right).\label{eq:V_fano_b_c}
\end{eqnarray}

The dependence of $\mathcal{V}_{E}^{\left(t\right)}\left(D_{\textit{in}};w,\delta,b,c,\Gamma_{\textit{ss}}\right)$
on its various parameters can be simplified by noting their symmetry
and scaling properties. Since, under linearization of the dynamics,
the relevant joint densities are Gaussian and therefore symmetric,
$\mathcal{E}$ and $\mathcal{E}_{m}$ are invariant and $x_{E}$ is
negated under the reflection of $E$ about $0$, $\delta\rightarrow-\delta$.
And so the TVGR is invariant when $b$ is also negated (see (\ref{eq:V_fano_b_c})),
i.e.
\[
\mathcal{V}_{E}^{\left(t\right)}\left(D_{\textit{in}};w,\delta,b,c,\Gamma_{\textit{ss}}\right)=\mathcal{V}_{E}^{\left(t\right)}\left(D_{\textit{in}};w,-\delta,-b,c,\Gamma_{\textit{ss}}\right).
\]
Therefore we only need to consider intervals $E$, (\ref{eq: interval E}),
with $0\le\delta<\frac{1}{2}$ so that $x_{E}\ge0$. Note also that
$w$ can be absorbed into $D_{\textit{in}}$ in $\mathcal{E}$, $\mathcal{E}_{m}$,
and $x_{E}$ by a rescaling of $b$ and $c$,
\begin{equation}
D_{\textit{in}}\leftarrow\frac{D_{\textit{in}}}{w^{2}},\,b\leftarrow wb,c\leftarrow w^{2}c.\label{eq: w rescaling}
\end{equation}
Thus, a larger event window is compensated for by larger noise. Likewise,
without loss of generality, we could also effectively set $\Gamma_{\textit{ss}}=1$
by absorbing $\Gamma_{\textit{ss}}$ into $D_{\textit{in}}$ and correspondingly rescaling
$c$ (see (\ref{eq:V_fano_b_c})):
\begin{equation}
D_{\textit{in}}\leftarrow\Gamma_{\textit{ss}}D_{\textit{in}},c\leftarrow\frac{c}{\Gamma_{\textit{ss}}}.\label{eq: GammaSS rescaling}
\end{equation}
In total,
\begin{equation}
\mathcal{V}_{E}^{\left(t\right)}\left(D_{\textit{in}};w,\delta,b,c,\Gamma_{\textit{ss}}\right)=\mathcal{V}_{E}^{\left(t\right)}\left(\frac{\Gamma_{\textit{ss}}}{w^{2}}D_{\textit{in}};1,\delta,wb,\frac{w^{2}}{\Gamma_{\textit{ss}}}c,1\right),\label{eq: Gamma_ss and w scaling}
\end{equation}
and so, for any $\Gamma_{\textit{ss}}>0$ and $w>0$, $\mathcal{V}_{E}^{\left(t\right)}$
can be understood by considering the case $\Gamma_{\textit{ss}}=w=1$ and varying
$b$ and $c$. Therefore, we at times only consider the behavior
of $\mathcal{E}\left(D_{\textit{in}};w,\delta,\Gamma_{\textit{ss}}\right)$ and $x_{E}\left(D_{\textit{in}};w,\delta,\Gamma_{\textit{ss}}\right)$
for $\Gamma_{\textit{ss}}=w=1$.

\begin{figure}
\noindent \begin{centering}
\includegraphics[width=1\textwidth]{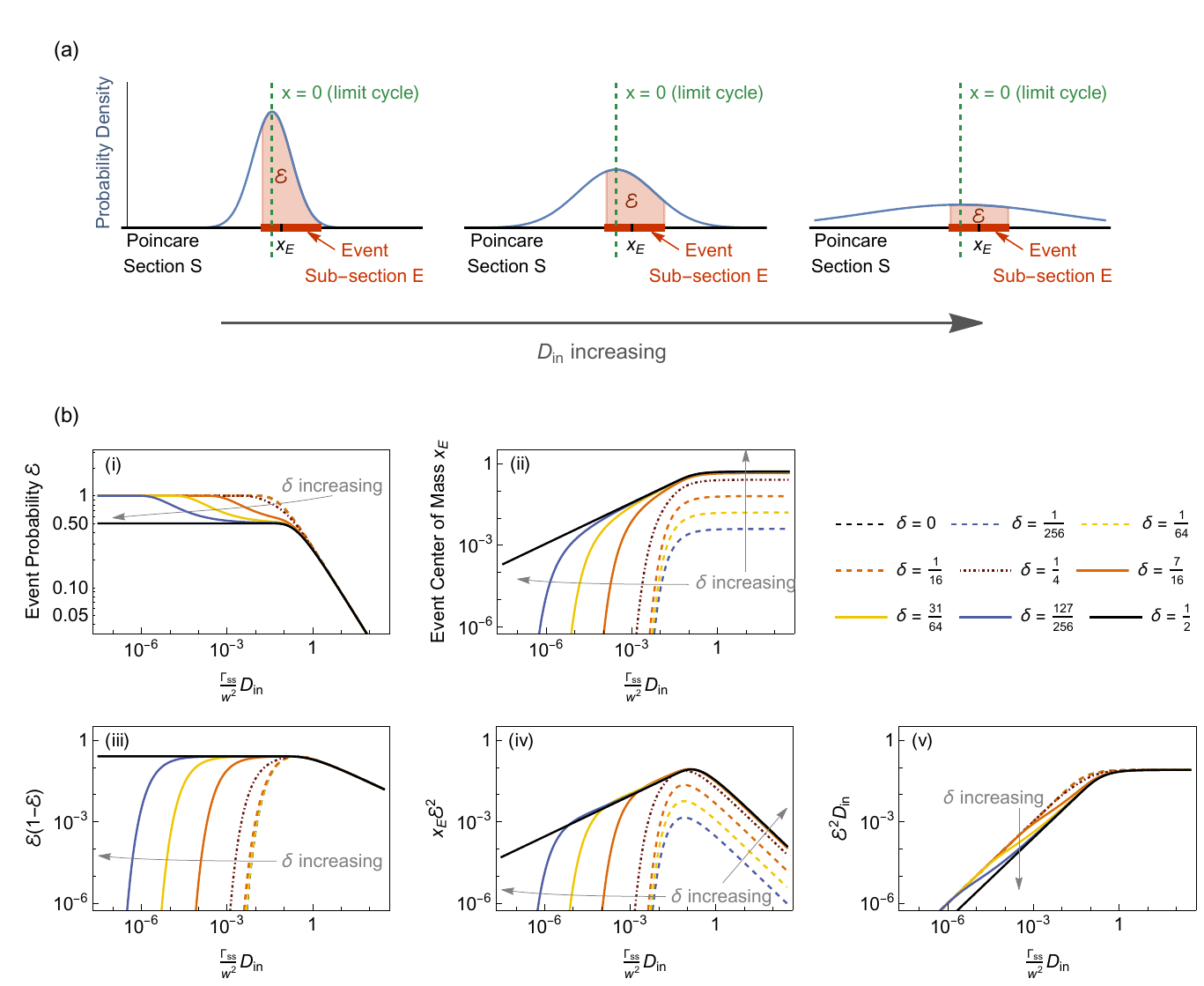}
\par\end{centering}
\caption{\textbf{The $D_{\textit{in}}$-dependent Elements of the TVGR $\mathcal{V}_{E}^{\left(t\right)}$.}
(a): A schematic depiction of the steady-state distribution on the
Poincare section $S$ (black line) for $\delta>0$ and various $D_{\textit{in}}$.
$\mathrm{\mathcal{E}}$ is the probability mass associated with $E$
(red shading above the red interval), and $x_{E}$ is its center of
mass. (b): Plots of the TVGR elements as a function $D_{\textit{in}}$ for
various values of $\delta$. The gray arrows indicate roughly how
the graphs change with increasing $\delta$; all converge to the solid
black curve as $\delta\rightarrow\frac{1}{2}$. The top row shows
the \textquotedblleft raw\textquotedblright{} elements, $\mathcal{E}$
and $x_{E}$. Note that $x_{E}$ is identically $0$ for $\delta=0$.
In the quasi-renewal limit (see Section \ref{subsec:The-Quasi-Renewal-Case}),
the TVGR is a linear combination of the components plotted in (b.iii)
and (b.v), $\mathcal{E}\left(1-\mathcal{E}\right)$ and $D_{\textit{in}}\mathcal{E}^{2}$.\label{fig:sigma-dependent elements}}
\end{figure}

Our main goal is to understand the origin of the unruliness of $\mathcal{V}_{E}^{\left(t\right)}$,
i.e. of its strong rise and subsequent decrease with increasing $D_{\textit{in}}$.
We therefore consider in detail the $D_{\textit{in}}$-dependence of the three
terms in (\ref{eq:V_fano_b_c}). From (\ref{eq: calE}) and (\ref{eq: xE})
we find the exact formulae
\begin{equation}
\mathcal{E}\left(D_{\textit{in}};w,\delta,\Gamma_{\textit{ss}}\right)=\frac{1}{2}\left[\mathrm{erf}\left(\frac{w\left(\delta+\frac{1}{2}\right)}{2\sqrt{\Gamma_{\textit{ss}}D_{\textit{in}}}}\right)-\mathrm{erf}\left(\frac{w\left(\delta-\frac{1}{2}\right)}{2\sqrt{\Gamma_{\textit{ss}}D_{\textit{in}}}}\right)\right]\label{eq: calE 1D}
\end{equation}
and 
\begin{equation}
x_{E}\left(D_{\textit{in}};w,\delta,\Gamma_{\textit{ss}}\right)=\frac{1}{\mathcal{E}}\sqrt{\frac{\Gamma_{\textit{ss}}D_{\textit{in}}}{\pi}}\left(\mathrm{exp}\left(-\frac{w^{2}\left(\delta-\frac{1}{2}\right)^{2}}{4\Gamma_{\textit{ss}}D_{\textit{in}}}\right)-\mathrm{exp}\left(-\frac{w^{2}\left(\delta+\frac{1}{2}\right)^{2}}{4\Gamma_{\textit{ss}}D_{\textit{in}}}\right)\right).\label{eq: xE 1D}
\end{equation}
For any $\delta\in\left[0,\frac{1}{2}\right)$, the event probability
$\mathcal{E}$ is monotonically decreasing as a function of $D_{\textit{in}}$
(Figure \ref{fig:sigma-dependent elements}a,b.i), reflecting the
increased likelihood to miss the event sub-section $E$ as the noise
is increased. In parallel, the event center of mass $x_{E}$, which
is a scalar for planar oscillators, monotonically increases from $0$
to the centroid of $E$ at $x=w\delta$, since the probability density
becomes ever more homogeneous across $E$ (Figure \ref{fig:sigma-dependent elements}a,b.ii).

For the Markov-only term $\mathcal{V}_{x\mapsto1_{E}\left(x\right)}^{\left(n\right)}$
no analytical expression is available for general $\Lambda$ due to
the appearance of the term $\mathcal{E}_{m}$ in (\ref{eq: Markov-only sum}).
As we will see below in Section (\ref{subsec:The-Quasi-Renewal-Case}),
in a limit where $\Lambda\rightarrow0$ and $w\rightarrow0$ simultaneously,
$\mathcal{E}_{m}$ is given by $\mathcal{E}^{2}$ and therefore $\mathcal{V}_{x\mapsto1_{E}\left(x\right)}^{\left(n\right)}=\mathcal{E}\left(1-\mathcal{E}\right)$
is non-monotonic in $D_{\textit{in}}$ (Figure \ref{fig:sigma-dependent elements}b.iii).
As in the toy model, the mixed term $x_{E}\mathcal{E}^{2}$ has a
maximum at intermediate values of $D_{\textit{in}}$ and the temporal term
$D_{\textit{in}}\mathcal{E}^{2}$ increases monotonically, reaching a constant
value asymptotically (Figure \ref{fig:sigma-dependent elements}b.iv
and b.v).

Overall, the $D_{\textit{in}}$-dependence of $\mathcal{V}_{E}^{\left(t\right)}$
depends on the balance between the first two and the third term in
(\ref{eq:V_fano_b_c}). Thus, $b$ and $c$, which represent the specific
dynamics of the oscillator via its averaged PRC and IRC (see (\ref{eq:b_general})
and (\ref{eq:c_general})), are the primary parameters in the TVGR.
We will show in the following sections that for any choice of the
interval $E$ containing $0$, unruliness appears in a set of finite
measure in the $b$-$c$ parameter space. After giving a concrete
criterion for unruliness, we start our investigation in Section \ref{subsec:The-Quasi-Renewal-Case}
with a limiting case of the dynamics where the TVGR can be treated
exactly. In the subsequent sections we then provide analytical estimates
for the Markov-only term that allow us to identify sufficient conditions
on $b$ and $c$ to guarantee unruliness of $\mathcal{V}_{E}^{\left(t\right)}$,
resulting in the phase diagram Figure \ref{fig:phaseDia generic}.

\subsection{The Criterion for Unruliness}

\begin{figure}
\noindent \begin{centering}
\includegraphics[width=1\textwidth]{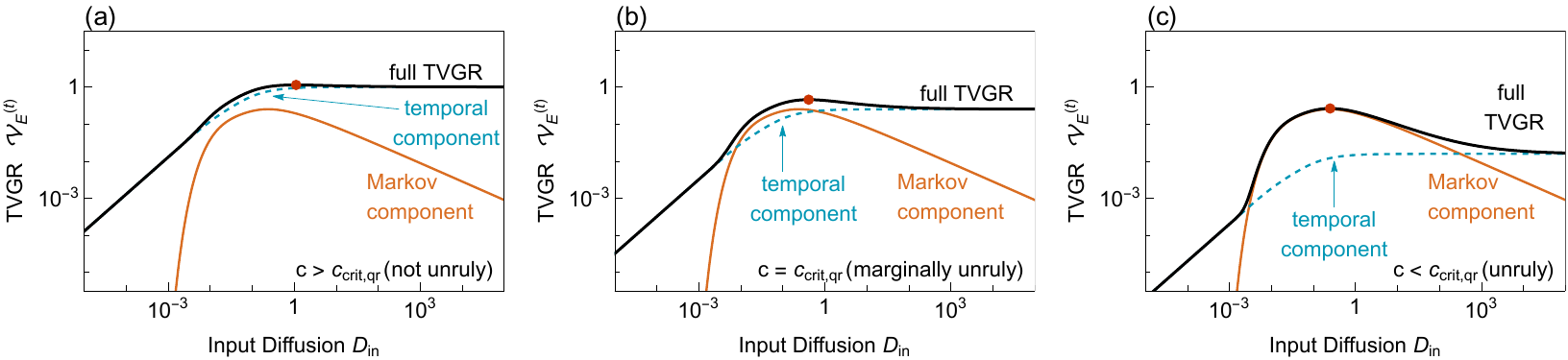}
\par\end{centering}
\caption{\textbf{Unruliness requires $c$ to be sufficiently small. }Sample
TVGRs in the quasi-renewal limit that are (a) not unruly with $c$
above the unruliness threshold $c_{crit,r}$, (b) marginally unruly
with $c$ at the threshold, and (c) unruly with $c$ below the threshold
(\ref{eq: c_crit quasi-renewal}). The presence of a local maximum
(red circle) does not distinguish these three cases. In the plots,
we take $\delta=\frac{1}{4}$, but they are qualitatively representative
for $\delta\in\left[0,\frac{1}{2}\right)$. \label{fig:sampleFanoVariance_quasi-renewal}}
\end{figure}

The main goal of this paper is to demonstrate widespread unruliness
in the event output of noisy oscillators. To make progress, we therefore
need to make the criterion for unruliness concrete and quantitative.
At the same time, unruliness is in some sense a qualitative phenomenon,
where $\mathcal{V}_{E}^{\left(t\right)}$ as a function of $D_{\textit{in}}$
is initially linear, has a \emph{strong} monotonic rise, attains a
maximum, and then eventually decreases. Any quantitative test for
unruliness should, therefore, only be interpreted as a rough ``rule
of thumb''. We briefly consider three criteria, before settling on
one. First, the most obvious feature of unruliness as seen in Figures
\ref{fig: MMO_DEff_vs_DPhase}b and \ref{fig: toy_model}a is the
maximum in the TVGR. A criterion could therefore simply be that $\mathcal{V}_{E}^{\left(t\right)}$
has a local maximum. This is unfortunately too generous, since a local
maximum does not imply a strong rise. For instance, in the simple
quasi-renewal case discussed in Section \ref{subsec:The-Quasi-Renewal-Case}
below, $\mathcal{V}_{E}^{\left(t\right)}$ has a local maximum for
any value of $c\ge0$, but not all cases shown in Figure \ref{fig:sampleFanoVariance_quasi-renewal}
should qualify as unruly. A better criterion is motivated by the visual
appearance of graphs of sums of functions on a log-scale plot. On
the log-scale, addition $f+g$ becomes - very roughly - the maximum
operation $\max\left(\log f,\log g\right)$. This can be seen, e.g.,
in Figure \ref{fig: toy_model}a, where the total TVGR (black) closely
follows the larger of the temporal component (blue, dashed) and the
Markov component (orange, solid). Following that line of reasoning,
the total TVGR would have a prominent maximum if, in the vicinity
of the maximum of the non-monotonic component $\mathcal{V}_{E,non-monotonic}^{\left(t\right)}\equiv\mathcal{V}_{x\mapsto1_{E}\left(x\right)}^{\left(n\right)}+bx_{E}\mathcal{E}^{2}$,
that component dominated the monotonic, temporal component $\mathcal{V}_{E,monotonic}^{\left(t\right)}\equiv cD_{\textit{in}}\mathcal{E}^{2}$.
I.e. 
\begin{equation}
\mathcal{V}_{E,non-monotonic}^{\left(t\right)}(D_{\textit{in}}^{(max)})>\mathcal{V}_{E,monotonic}^{\left(t\right)}(D_{\textit{in}}^{(max)})\qquad\mbox{with \ensuremath{\left.\frac{\partial}{\partial D_{\textit{in}}}\mathcal{V}_{E,non-monotonic}^{\left(t\right)}(D_{\textit{in}})\right|_{D_{\textit{in}}=D_{\textit{in}}^{(max)}}=0\,.}}\label{eq: unruliness - tangency}
\end{equation}
(\ref{eq: unruliness - tangency}) is a reasonable criterion that
effectively separates those cases which we would reasonably classify
as unruly or not unruly. But we found that the analysis of (\ref{eq: unruliness - tangency})
becomes unnecessarily complicated for something meant to be ``just
a rule of thumb''. We therefore opt for a yet stricter criterion
that is more amiable to analysis and interpretation. Given the monotonic
saturation of $\mathcal{V}_{E}^{\left(t\right)}{}^{\textit{monotonic}}$ for
$D_{\textit{in}}\rightarrow\infty$ (see Figure \ref{fig:sigma-dependent elements}b.v),
for $\mathcal{V}_{E}^{\left(t\right)}$ to be classified as unruly
we require that the \emph{maximum} of $\mathcal{V}_{E,non-monotonic}^{\left(t\right)}$
is greater than the \emph{asymptotic} value (and upper bound) of $\mathcal{V}_{E,monotonic}^{\left(t\right)}$,
which is also the asymptotic value of $\mathcal{V}_{E}^{\left(t\right)}$:
\begin{equation}
\max_{D_{\textit{in}}}\left\{ \mathcal{V}_{E,non-monotonic}^{\left(t\right)}\right\} >\lim_{D_{\textit{in}}\rightarrow\infty}\mathcal{V}_{E,monotonic}^{\left(t\right)}.\label{eq: unruliness - max above asymptote}
\end{equation}
This criterion guarantees a prominent maximum in the TVGR; it indeed
picks out those cases that are strongly unruly (Figure \ref{fig:sampleFanoVariance_quasi-renewal}).

\subsection{The Quasi-Renewal Case ($\Lambda\rightarrow0$ with $w\rightarrow0$)\label{subsec:The-Quasi-Renewal-Case}}

We first consider the simple limiting case of strong contraction to
the limit cycle, $\Lambda\rightarrow0$. In this limit, we expect
the steady-state variance on the section, $\Gamma_{\textit{ss}}$, to go to
$0$. Indeed, with (\ref{eq: GammaSS 1D}) and the analysis in SI
Section \ref{subsec: SI - QR Averages}, 
\[
\Gamma_{\textit{ss}}\sim\left\Vert \vec{y}_{G}\right\Vert ^{2}\sim-\frac{1}{\log\Lambda}.
\]
For fixed values of $D_{\textit{in}}$, $w$, and $\delta$, we have then $\mathcal{E}\left(D_{\textit{in}};w,\delta\right)\rightarrow1$,
$\mathcal{E}_{m}\left(D_{\textit{in}};w,\delta\right)\rightarrow1$, and $x_{E}\left(D_{\textit{in}};w,\delta\right)\rightarrow0$;
crossings of the Poincare section are always at $x=0$, i.e. on the
limit cycle. There are no phase slips, since events are produced at
every crossing. This is unsurprising: for strong contraction the standard
phase reduction is an accurate representation of the dynamics, and
the process is effectively a renewal process. Indeed, $\mathcal{V}_{E}^{\left(t\right)}$
reduces in this limit to its phase-reduction value, $\mathcal{V}_{E}^{\left(t\right)}\rightarrow cD_{\textit{in}}=2D_{\textit{phase}}$.

To obtain a non-trivial result in the limit of strong contraction,
we need to take the width of $E$ to 0, as well, in order to keep
the probability of phase slips $1-\mathcal{E}\left(D_{\textit{in}};w,\delta\right)$
unchanged. That requires
\begin{equation}
w\propto\sqrt{\Gamma_{\textit{ss}}}\propto\sqrt{-\frac{1}{\log\Lambda}}\label{eq: w(Lambda) distinguished limit}
\end{equation}
as $\Lambda\rightarrow0$ (see (\ref{eq: calE 1D})). In this case,
it is clear from (\ref{eq: xE 1D}) that $x_{E}\rightarrow0$ still,
but the limiting behavior of $\mathcal{E}_{m}$ requires some care.
Applying the definitions of $E$ from (\ref{eq: interval E}) and
$\mathbf{C}_{m}$ from (\ref{eq:mth joint distribution}), (\ref{eq: calE_m})
becomes
\begin{equation}
\mathcal{E}_{m}=\int_{w\left(\delta-\frac{1}{2}\right)}^{w\left(\delta+\frac{1}{2}\right)}\int_{w\left(\delta-\frac{1}{2}\right)}^{w\left(\delta+\frac{1}{2}\right)}N\left(\left(\begin{array}{c}
x_{1}\\
x_{2}
\end{array}\right);2D_{\textit{in}}\Gamma_{\textit{ss}}\left(\begin{array}{cc}
1 & \Lambda^{m}\\
\Lambda^{m} & 1
\end{array}\right)\right)dx_{1}dx_{2}.\label{eq: calE_m planar integral}
\end{equation}
The dominant contribution to the covariance matrix for $\Lambda\ll1$
is
\[
2D_{\textit{in}}\mathbf{C}_{m}\sim2D_{\textit{in}}\Gamma_{\textit{ss}}\left(\begin{array}{cc}
1 & 0\\
0 & 1
\end{array}\right),
\]
and thus
\begin{equation}
\mathcal{E}_{m}\rightarrow\left[\int_{w\left(\delta-\frac{1}{2}\right)}^{w\left(\delta+\frac{1}{2}\right)}N\left(x;2D_{\textit{in}}\Gamma_{\textit{ss}}\right)dx\right]^{2}=\mathcal{E}^{2}.\label{eq:Em_for_Lambda0}
\end{equation}
In total (\ref{eq: Markov-only sum}) becomes $\mathcal{V}_{x\mapsto1_{E}\left(x\right)}^{\left(n\right)}=\mathcal{E}\left(1-\mathcal{E}\right)$,
reflecting the fact that there are no correlations in events from
one step to the next. We therefore call this limiting case, given
by $\Lambda\rightarrow0$ along with (\ref{eq: w(Lambda) distinguished limit}),
the ``quasi-renewal'' case. Since $x_{E}=0$, the quasi-renewal
TVGR,
\begin{equation}
\mathcal{V}_{E,qr}^{\left(t\right)}\left(D_{\textit{in}};c\right)=\mathcal{E}\left(1-\mathcal{E}\right)+cD_{\textit{in}}\mathcal{E}^{2},\label{eq: quasi-renewal Fano Variance}
\end{equation}
 is independent of $b$. Here the TVGR has essentially the same form
as in the toy model discussed in Section \ref{sec:Fano-Variance-for-Toy-Model}
(cf. (\ref{eq:Fano_variance_toy_model})) and unruliness arises if
$c$ is small enough for the non-monotonic Markov-only term to dominate
over the temporal term. It can be shown that $D_{\textit{in}}\mathcal{E}^{2}\rightarrow\frac{w^{2}}{4\pi\Gamma_{\textit{ss}}}$
independently of $a$ as $D_{\textit{in}}\rightarrow\infty$. Therefore, since
$\mathcal{E}\left(1-\mathcal{E}\right)$ has a maximum value of $\frac{1}{4}$,
$\mathcal{V}_{E,qr}^{\left(t\right)}$ is unruly according to (\ref{eq: unruliness - max above asymptote})
if
\begin{equation}
c<c_{crit,qr}=\pi\frac{\Gamma_{\textit{ss}}}{w^{2}}.\label{eq: c_crit quasi-renewal}
\end{equation}
Sample TVGRs that satisfy (\ref{eq: c_crit quasi-renewal}) are indeed
strongly unruly (Figure \ref{fig:sampleFanoVariance_quasi-renewal}).

\subsection{Symmetric Intervals $E$ ($\delta=0$, $\Lambda>0$)\label{subsec:Symmetric-Intervals-}}

There are two effects as $\Lambda$ is perturbed from $0$: the mixed
term $bx_{E}\mathcal{E}^{2}$ becomes generally non-zero and $\mathcal{E}_{m}$
deviates from $\mathcal{E}^{2}$. As a stepping stone to the most
general case, we first consider only symmetric intervals given by
$\delta=0$. In that case, the event center of mass $x_{E}$ is $0$,
and we can consider $\mathcal{E}_{m}$ alone in the question of how
(\ref{eq: c_crit quasi-renewal}) generalizes.

In order to address unruliness in the TVGR in this case, we will need
to understand the infinite sum in (\ref{eq: Markov-only sum}), $\sum_{m=1}^{\infty}\left(\mathcal{E}_{m}-\mathcal{E}^{2}\right)$.
Bounding $\mathcal{E}_{m}$ in such a way that the sum over $m$ can
be carried out and is well behaved for all $\Lambda\in\left[0,1\right)$
and all $\delta\in\left[0,\frac{1}{2}\right)$ turns out to be quite
involved. We show the full details in SI Section \ref{sec: SI - Bounds}
where we find that 
\[
0\le\mathcal{E}_{m}-\mathcal{E}^{2}\le\Lambda^{m}\mathcal{E}\left(1-\mathcal{E}\right),
\]
and so
\begin{equation}
\mathcal{E}\left(1-\mathcal{E}\right)=\mathcal{V}_{x\mapsto1_{E}\left(x\right)}^{\left(n\right)}\left(\Lambda=0\right)\le\mathcal{V}_{x\mapsto1_{E}\left(x\right)}^{\left(n\right)}<\frac{1+\Lambda}{1-\Lambda}\mathcal{E}\left(1-\mathcal{E}\right),\label{eq:bounds on indicator funcVar}
\end{equation}
via a geometric series. Thus, a lower bound on the non-monotonic part
of the TVGR for $\Lambda>0$ is found by replacing it with its $\Lambda=0$
counterpart, that is by replacing $\mathcal{V}_{x\mapsto1_{E}\left(x\right)}^{\left(n\right)}$
with $\mathcal{E}\left(1-\mathcal{E}\right)$. Therefore, if the maximum
of $\mathcal{E}\left(1-\mathcal{E}\right)$ exceeds the asymptotic
value $\frac{cw^{2}}{4\pi\Gamma_{\textit{ss}}}$ of $cD_{\textit{in}}\mathcal{E}^{2}$,
then so will the maximum value of $\mathcal{V}_{x\mapsto1_{E}\left(x\right)}^{\left(n\right)}$,
and the unruliness criterion (\ref{eq: unruliness - max above asymptote})
will certainly be satisfied. So the quasi-renewal condition for unruliness,
(\ref{eq: c_crit quasi-renewal}), can only underestimate the range
of $c$ for which unruliness appears. At the same time, the upper
bound in (\ref{eq:bounds on indicator funcVar}) guarantees that for
\begin{equation}
c>\frac{1+\Lambda}{1-\Lambda}c_{crit,qr}\label{eq: c crit not unruly}
\end{equation}
$\mathcal{V}_{E}^{\left(t\right)}\left(D_{\textit{in}};w,\delta=0,b,c\right)$
is certainly not unruly.

Thus, despite the uncertainty involved in being able to only bound
$\mathcal{V}_{x\mapsto1_{E}\left(x\right)}^{\left(n\right)}$, we
can make the following statements
\begin{equation}
\mathcal{V}_{E}^{\left(t\right)}\left(D_{\textit{in}};w,\delta=0,b,c\right)\ \mbox{is certainly unruly when}\ c<c_{crit,qr}\label{eq:symmetric_certainly_unruly}
\end{equation}

\begin{equation}
\mathcal{V}_{E}^{\left(t\right)}\left(D_{\textit{in}};w,\delta=0,b,c\right)\ \mbox{is certainly not unruly when}\ c>\frac{1+\Lambda}{1-\Lambda}c_{crit,qr}\label{eq:symmetric_certainly_not_unruly}
\end{equation}
\begin{equation}
\mathcal{V}_{E}^{\left(t\right)}\left(D_{\textit{in}};w,\delta=0,b,c\right)\ \mbox{may be unruly or not unruly for }\ c_{crit,qr}<c<\frac{1+\Lambda}{1-\Lambda}c_{crit,qr}\,.\label{eq:symmetric_uncertain_unruly}
\end{equation}

When we consider $\delta\ne0$ in the following, $b$ and $c$ will
both be relevant parameters. The above statements will generalize
to three regions of the $b$-$c$ parameter space in which $\mathcal{V}_{E}^{\left(t\right)}$
is 1) certainly not unruly, 2) certainly unruly, and 3) may be either.

\subsection{The General Case\label{subsec:General-Case}}

Finally, we consider $\mathcal{V}_{E}^{\left(t\right)}$ in full generality
for $\delta\ne0$ and $\Lambda>0$. Due to the asymmetry $\delta\ne0$,
the mixed term $bx_{E}\mathcal{E}^{2}$ does not drop out and one
has to deal with two non-monotonic terms. Obtaining sharp, general
conditions for the onset of unruliness is therefore quite difficult.
We therefore content ourselves with sufficient conditions.

Before considering more detailed criteria, (\ref{eq: general unruliness criterion 2})-(\ref{eq: general unruliness criterion 1b})
below, we first note that as long as $b\ge0$ and $\delta\ge0$, the
mixed term $bx_{E}\mathcal{E}^{2}$ is non-negative and non-monotonic
(Figure \ref{fig:sigma-dependent elements}b.iv) and can not detract
from unruliness. So the criterion (\ref{eq:symmetric_certainly_unruly}),
which applied in the case that $\delta=0$, will also guarantee unruliness
when $\delta\ge0$ and $b\ge0$:
\[
\mathcal{V}_{E}^{\left(t\right)}\left(D_{\textit{in}};w,\delta,b,c\right)\ \mbox{is certainly unruly when}\ c<c_{crit,qr}\ \mathrm{and}\ b\ge0\,.
\]
Thus, \emph{at least a finite region} of the parameter space is unruly,
bounded by $c=c_{crit,qr}=\pi\frac{\Gamma_{\textit{ss}}}{w^{2}}$ above, $b=0$
to the left and to the right by $b=\sqrt{2\frac{c}{\Gamma_{\textit{ss}}}\frac{1+\Lambda}{1-\Lambda}}$,
which follows from the physicality condition (\ref{eq: physical region-1}).

\begin{figure}
\noindent \begin{centering}
\includegraphics[width=0.9\textwidth]{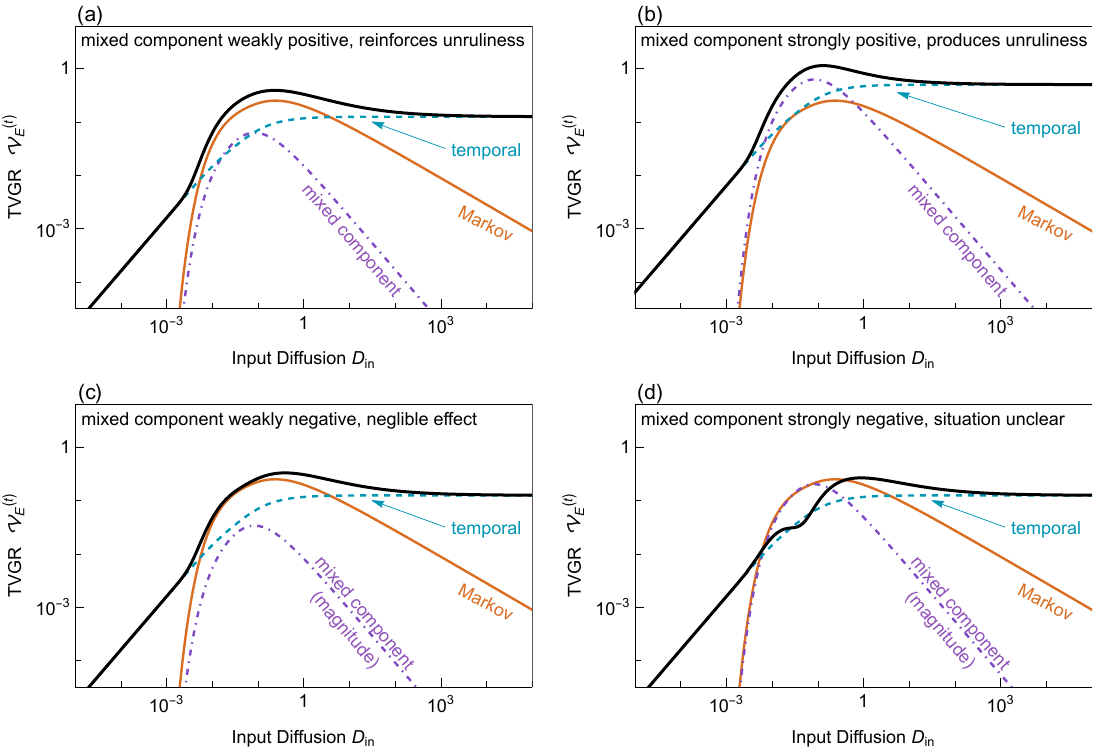}
\par\end{centering}
\caption{\textbf{The mixed component can promote or detract from unruliness.
}Sample lower bounds (from (\ref{eq:bounds on indicator funcVar}))
of TVGRs that are unruly (a) because the Markov component dominates
over the asymptotic value of the TVGR and (b) because $b$ is positive
and large and the mixed component dominates. (c): $b$ is negative,
but the mixed term is not large enough to detract significantly from
the unruliness. (d): a situation where the typical unruly quality
is lost and a local minimum appears because $b$ is sufficiently large
and negative. This situation is outside the scope of our analysis.
\label{fig:sampleFanoVariance_LB_general}}
\end{figure}

But, moreover, for $b>0$ the mixed term $bx_{E}\mathcal{E}^{2}$
and the Markov-only term $\mathcal{V}_{x\mapsto1_{E}\left(x\right)}^{\left(n\right)}$
\emph{both} contribute to the non-monotonicity of $\mathcal{V}_{E}^{\left(t\right)}$
(Figure \ref{fig:sampleFanoVariance_LB_general}a,b). We therefore
choose as a broader, sufficient condition for unruliness that either
of these terms by itself dominate the monotonic temporal term $cD_{\textit{in}}\mathcal{E}^{2}$.
And for $b<0$ unruliness is guaranteed if the mixed term is small
enough to not undermine the dominance of the Markov-only term over
the temporal term (Figure \ref{fig:sampleFanoVariance_LB_general}c).
We therefore require that the mixed term is dominated by the temporal
term, which in turn is dominated by the Markov-only term. These sufficient
conditions can be summarized as either (I) the mixed term dominates,
\begin{equation}
\max_{D_{\textit{in}}}\left\{ bx_{E}\mathcal{E}^{2}\right\} >\lim_{D_{\textit{in}}\rightarrow\infty}\mathcal{V}_{E}^{\left(t\right)}=c\frac{w^{2}}{4\pi\Gamma_{\textit{ss}}},\label{eq: general unruliness criterion 2}
\end{equation}
or (II) the Markov-only term dominates,

\begin{equation}
\max_{D_{\textit{in}}}\left\{ \mathcal{V}_{x\mapsto1_{E}\left(x\right)}^{\left(n\right)}\right\} >\lim_{D_{\textit{in}}\rightarrow\infty}\mathcal{V}_{E}^{\left(t\right)}=c\frac{w^{2}}{4\pi\Gamma_{\textit{ss}}},\label{eq: general unruliness criterion 1a}
\end{equation}
as long as $b$ is not strongly negative,
\begin{equation}
-b<c\min_{D_{\textit{in}}}\left\{ \frac{D_{\textit{in}}}{x_{E}}\right\} .\label{eq: general unruliness criterion 1b}
\end{equation}
If (\ref{eq: general unruliness criterion 1b}) is not satisfied,
a more refined analyses would be required to determine unruliness.
In the following we classify the unruliness in these situations, where
$\mathcal{V}_{E}^{\left(t\right)}$ could even have a local minimum,
as ``unclear'' (Figure \ref{fig:sampleFanoVariance_LB_general}d).

From (\ref{eq: general unruliness criterion 2})-(\ref{eq: general unruliness criterion 1b}),
there will be four boundaries for unruliness in the parameter space.
One boundary, below which unruliness certainly occurs, follows from
(\ref{eq: general unruliness criterion 2}). Because of the range
of values that $\mathcal{V}_{x\mapsto1_{E}\left(x\right)}^{\left(n\right)}$
can take on (see (\ref{eq:bounds on indicator funcVar})), two more
follow from (\ref{eq: general unruliness criterion 1a}): the boundary
$c=\pi\frac{\Gamma_{\textit{ss}}}{w^{2}}$ below which unruliness certainly
occurs and the boundary $c=\frac{1+\Lambda}{1-\Lambda}\pi\frac{\Gamma_{\textit{ss}}}{w^{2}}$
below which unruliness may possibly occur. These are generalizations
of (\ref{eq:symmetric_certainly_unruly}) and (\ref{eq:symmetric_uncertain_unruly})
for $\delta\ge0$, and they apply as long as the condition on $b$,
(\ref{eq: general unruliness criterion 1b}), is satisfied. Thus,
the two regions corresponding to certain and possible unruliness will
be cut short by the fourth boundary $-b=c\min_{D_{\textit{in}}}\left\{ \frac{D_{\textit{in}}}{x_{E}}\right\} $
beyond which the negative mixed term could diminish the unruliness.

\begin{figure}
\noindent \begin{centering}
\includegraphics[width=0.9\textwidth]{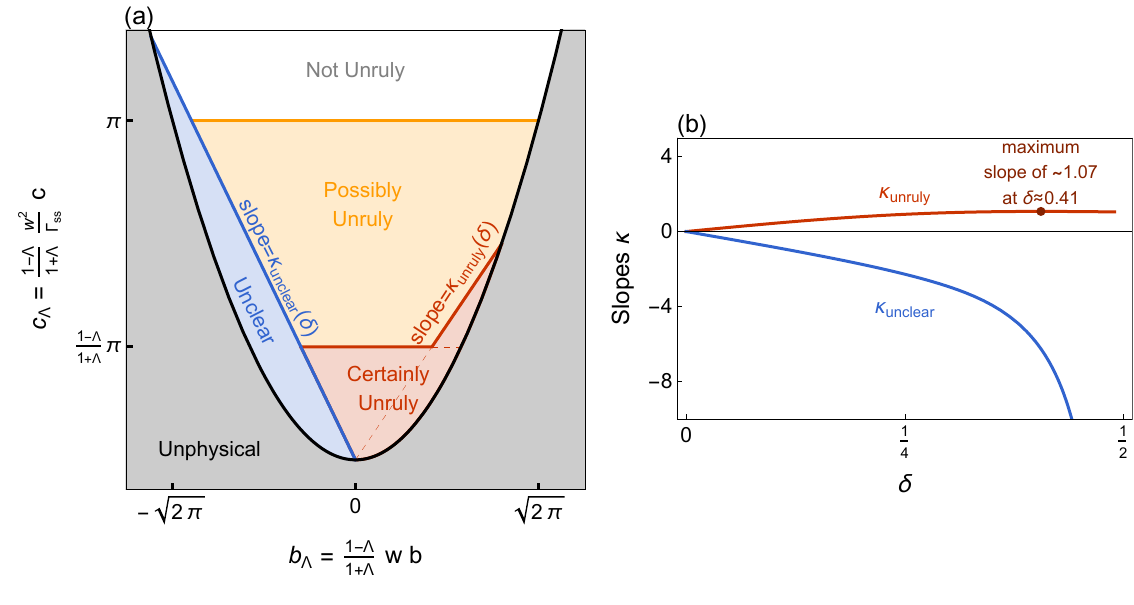}
\par\end{centering}
\caption{\textbf{Dependence of unruliness on $b_{\Lambda}$ and $c_{\Lambda}$
(cf. (\ref{eq:bLambda_cLambda})).} (a): The $b_{\Lambda}$-\textbf{$c_{\Lambda}$
}phase diagram, representative for all $\Lambda>0$ and $0<\delta<\frac{1}{2}$.
The physical portion of the diagram is divided into four regions,
reflecting TVGRs that are not unruly (white), certainly unruly (red),
and may be either (orange). The fourth, blue region corresponds to
situations that may be different altogether, with a local minimum
possibly appearing in the TVGR. The boundaries between physical regions
are all piece-wise linear, and the pieces are either fixed ($c_{\Lambda}=\pi$),
vary with $\Lambda$ ($c_{\Lambda}=\frac{1-\Lambda}{1+\Lambda}\pi$),
or have slopes that vary with $\delta$ ($c_{\Lambda}=\kappa\left(\delta\right)b_{\Lambda}$).
(b): The $\delta$-dependent slopes $\kappa_{\textit{unruly}}$ and $\kappa_{\textit{uncertain}}$.
\label{fig:phaseDia generic}}
\end{figure}

In order to represent all of the possibilities graphically in a single
phase diagram (Figure \ref{fig:phaseDia generic}), we minimize the
number of parameters that need to be considered. We have already shown
in (\ref{eq: Gamma_ss and w scaling}) that the event interval width
$w$ and the steady-state variance coefficient $\Gamma_{\textit{ss}}$ can
be absorbed by rescaling $D_{\textit{in}}$, $b$, and $c$. In the following
we make this rescaling explicit by defining 
\begin{equation}
b_{\Lambda}\equiv\frac{1-\Lambda}{1+\Lambda}wb\qquad\mbox{and}\qquad c_{\Lambda}\equiv\frac{1-\Lambda}{1+\Lambda}w^{2}\frac{c}{\Gamma_{\textit{ss}}}.\label{eq:bLambda_cLambda}
\end{equation}
To be precise, the noise strength should also be explicitly scaled,
$D_{\textit{in}}\leftarrow\frac{\Gamma_{\textit{ss}}}{w^{2}}D_{\textit{in}}$. But that just
shifts the noise level at which features in the TVGR, like unruliness,
occur, and does not change whether they appear or not. We choose to
include also a $\Lambda$-dependent factor in this rescaling, since
it largely frees us from considering $\Lambda$ explicitly. The physicality
condition (\ref{eq: physical region-1}) is simply $c_{\Lambda}\ge\frac{1}{2}b_{\Lambda}^{2}$.
And the threshold beyond which unruliness cannot occur, (\ref{eq: c crit not unruly}),
becomes $c_{\Lambda}=\pi$, for \emph{any} value of $\Lambda>0$.
The tradeoff is that the threshold in $c_{\Lambda}$ below which unruliness
certainly does occur, $c_{\Lambda}=\frac{1-\Lambda}{1+\Lambda}\pi$,
depends on $\Lambda$. The other criteria, (\ref{eq: general unruliness criterion 2})
and (\ref{eq: general unruliness criterion 1b}), remain independent
of $\Lambda$. They correspond to linear boundaries that pass through
the origin in the $b_{\Lambda}$-$c_{\Lambda}$ plane (Figure \ref{fig:phaseDia generic}a),
with $\delta$-dependent slopes given by 
\begin{equation}
\kappa_{\textit{unclear}}\left(\delta\right)=-\max_{D_{\textit{in}}}\left\{ \frac{x_{E}\left(D_{\textit{in}};\delta\right)}{D_{\textit{in}}}\right\} \label{eq:slope_unclear}
\end{equation}
 and 
\begin{equation}
\kappa_{\textit{unruly}}\left(\delta\right)=4\pi\max_{D_{\textit{in}}}\left\{ x_{E}\left(D_{\textit{in}};\delta\right)\mathcal{E}^{2}\left(D_{\textit{in}};\delta\right)\right\} \,,\label{eq:slope_unruly}
\end{equation}
respectively (Figure \ref{fig:phaseDia generic}b). Note that in the
case $\delta=0$, which was discussed in Section \ref{subsec:Symmetric-Intervals-},
$\kappa_{\textit{unclear}}=\kappa_{\textit{unruly}}=0$: the phase diagram is divided
into horizontal bands corresponding to (\ref{eq:symmetric_certainly_unruly})-(\ref{eq:symmetric_uncertain_unruly}).
And, as $\delta\rightarrow\frac{1}{2}$, $\kappa_{\textit{unclear}}$ diverges:
in this limit, the entire left hand of the phase diagram becomes ``unclear''. 

\begin{figure}
\noindent \begin{centering}
\includegraphics[width=0.9\textwidth]{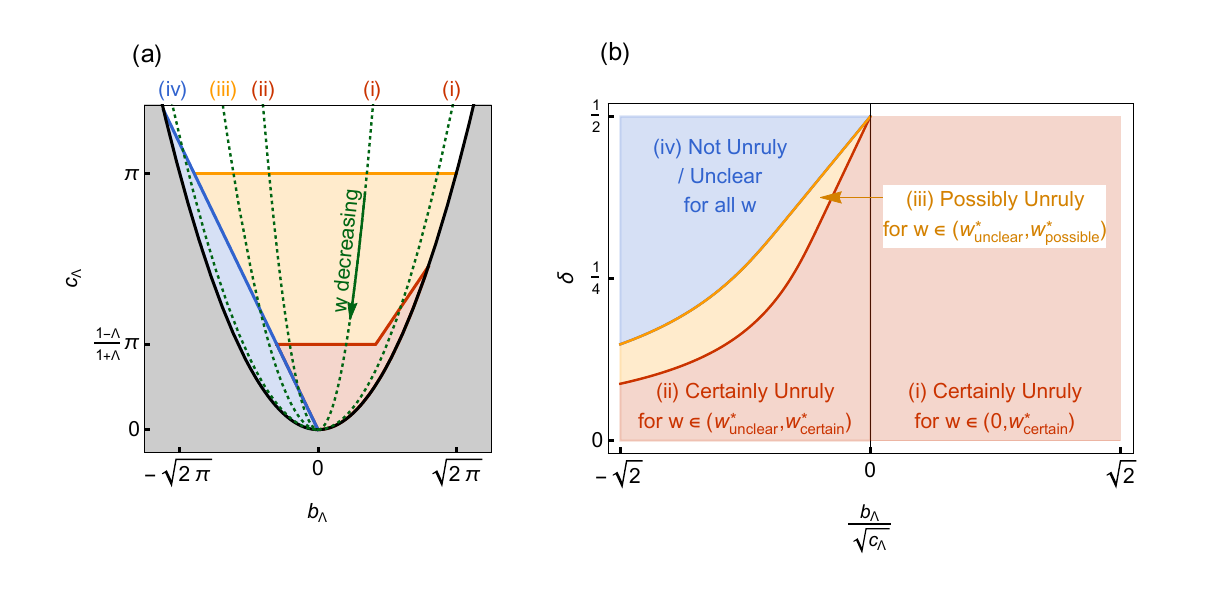}
\par\end{centering}
\caption{\textbf{Dependence of unruliness on the event interval $E$.} (a):
The effect of varying the interval width, $w$, in the $b_{\Lambda}$-\textbf{$c_{\Lambda}$
}phase diagram is to trace out parabolic curves of constant $\frac{b_{\Lambda}}{\sqrt{c_{\Lambda}}}$.
Four qualitatively different behaviors appear, depending on $\frac{b_{\Lambda}}{\sqrt{c_{\Lambda}}}$:
(i) unruliness will certainly appear for some interval $w\in\left(0,w_{\textit{certain}}^{*}\right)$,
(ii) unruliness will certainly appear for $w\in\left(w_{\textit{unclear}}^{*},w_{\textit{certain}}^{*}\right),$
(iii) unruliness could possibly appear for $w\in\left(w_{\textit{unclear}}^{*},w_{\textit{possible}}^{*}\right)$,
and (iv) the oscillator is either not unruly or the situation is unclear
for all $w$. (b): A phase diagram over $\frac{b_{\Lambda}}{\sqrt{c_{\Lambda}}}$
and $\delta$ summarizes the appearance of the four different unruliness
behaviors (i)-(iv).\label{fig:phaseDia w-delta}}
\end{figure}

It is worthwhile considering what happens when the width of the event
interval $w$ is changed. Varying $w$ while keeping the remaining
parameters fixed amounts to tracing out parabolic curves in the $\left(b_{\Lambda},c_{\Lambda}\right)$-plane
that are defined by $b_{\Lambda}=\mathcal{B}\,\sqrt{c_{\Lambda}}$
(cf. (\ref{eq:bLambda_cLambda})) with $\mathcal{B}$ a constant (dashed
lines in Figure \ref{fig:phaseDia w-delta}a). Depending on $\mathcal{B}$
different scenarios arise (Figure \ref{fig:phaseDia w-delta}b). Examples
of the traces corresponding to the different regimes in Figure \ref{fig:phaseDia w-delta}b
are shown as dashed lines in Figure \ref{fig:phaseDia w-delta}a.
Considering the limit $w\rightarrow0$, $\mathcal{V}_{E}^{\left(t\right)}$
``certainly'' becomes unruly if $b>0$ (scenario (i)), while for
$b<0$ the outcome is ``unclear'' (scenarios (ii)-(iv)). As $w$
is increased from 0, $\mathcal{V}_{E}^{\left(t\right)}$ can become
``certainly unruly'' (scenario (ii)) or ``possibly unruly'' (scenarios
(ii) and (iii)) even when $b<0$. The various transitions define the
values $w_{\textit{unclear}}^{*}$, $w_{\textit{certain}}^{*}$, and $w_{\textit{possible}}^{*}$,
respectively, which are determined in SI Section \ref{sec: SI - w-dependence}.

In summary, since the physically accessible parameter range for planar
oscillators is given by
\begin{equation}
-1\le\frac{1}{\sqrt{2}}\frac{b_{\Lambda}}{\sqrt{c_{\Lambda}}}=\frac{\vec{z}_{G}^{T}\vec{y}_{G}}{\left\Vert \vec{z}_{G}\right\Vert \left\Vert \vec{y}_{G}\right\Vert }\le1\label{eq: reduced internal parameter}
\end{equation}
(see (\ref{eq: physical region-1}) and (\ref{eq:bLambda_cLambda})),
and all finite event intervals $E$ that intersect the limit cycle
satisfy $0\le\delta<\frac{1}{2}$ up to reflection, Figure \ref{fig:phaseDia w-delta}b
captures all planar oscillators and their relevant event intervals.
It shows that for any value of $\mathcal{B}=b_{\Lambda}/\sqrt{c_{\Lambda}}$,
which characterizes the internal dynamics of the oscillator, there
is a range of event intervals, characterized by $\delta$ and $w$,
for which the TVGR $\mathcal{V}_{E}^{\left(t\right)}$ is certainly
unruly.

\subsection{Example: Hopf Normal Form}

One might rightly complain that while the $b_{\Lambda}$-$c_{\Lambda}$
plane (Figure \ref{fig:phaseDia generic}a) shows a substantial region
of unruliness, $b$ and $c$ are derived parameters, whose relationship
to the parameters that originally appear in a given oscillator model
are not clear. Likewise, while \emph{some} choice of $E$ produces
unruliness in any oscillator, that choice may not be of interest in
a given model or application. In short, it is not clear that the unruly
region is reachable by reasonable choices of the model's parameters
and a reasonable choice of interval $E$. Here we address this concern
by investigating the prototypical limit-cycle oscillator. We compute
the reduced parameters $b$ and $c$ for the unfolded Hopf normal
form, also known as the Stuart-Landau oscillator, with added noise,
\[
d\vec{y}=2\pi\left[\left(\begin{array}{cc}
\epsilon & -\alpha\\
\alpha & \epsilon
\end{array}\right)\vec{y}-\left(\begin{array}{cc}
\beta & -\gamma\\
\gamma & \beta
\end{array}\right)\left|\vec{y}\right|^{2}\vec{y}\right]\,dt+\sigma\mathbf{G}\,d\vec{W}.
\]
In this coordinate system, the limit cycle is a circle centered at
the origin with radius $\sqrt{\frac{\epsilon}{\beta}}$ and has a
period of $\alpha-\frac{\epsilon\gamma}{\beta}$. Without loss of
generality, we rescale $\vec{y}$ and $t$ so that the radius and
period are both fixed at $1$. We can equivalently set $\beta=\epsilon$
and $\alpha=\gamma+1$.

We also make three simplifying choices of parameters in our analysis.
Despite these choices, the oscillator will show unruliness over a
wide range of its remaining natural parameters. First, we set $\gamma=0$,
eliminating the amplitude-dependence of the deterministic angular
velocity. This yields what we call the ``Hopf oscillator'',
\begin{equation}
d\vec{y}=2\pi\left[\left(\begin{array}{cc}
\epsilon & -1\\
1 & \epsilon
\end{array}\right)\vec{y}-\left(\begin{array}{cc}
\epsilon & 0\\
0 & \epsilon
\end{array}\right)\left|\vec{y}\right|^{2}\vec{y}\right]\,dt+\sigma\mathbf{G}\,d\vec{W},\label{eq:Hopf oscillator}
\end{equation}
for $\epsilon>0.$ We also choose the Poincare section to be along
the positive $y_{1}$ axis, and leave the event interval $E$ unspecified
for now, investigating below what is required of it to produce unruliness.
Lastly, we allow the additive noise to be correlated in the $y_{1}$
and $y_{2}$ components but have the same marginal variance in each;
we parameterize the covariance matrix as
\[
\sigma^{2}\mathbf{G}\mathbf{G}^{T}=\sigma^{2}\left(\begin{array}{cc}
1 & \rho\\
\rho & 1
\end{array}\right),
\]
where $\rho\in\left[-1,1\right]$ is the correlation coefficient between
the components.

In order to rewrite the system in phase-isostable coordinates, (\ref{eq: phase-isostable isostable}),
we first note that, with the choice $\gamma=0$, the geometric angle
$\theta=\arctan\frac{y_{2}}{y_{1}}$ evolves with a constant rate
when the oscillator is unforced ($\sigma=0$). So the isochrons are
radial lines and we will take $\phi=\frac{\theta}{2\pi}$ as the phase
coordinate. The radial coordinate $r=\sqrt{y_{1}^{2}+y_{2}^{2}}$
evolves by
\[
\dot{r}=2\pi\epsilon r\left(1-r^{2}\right).
\]
$\psi\sim r-1$ is the isostable coordinate to first order approximation,
since it has linear dynamics near the limit cycle in the absence of
noise,
\[
\dot{\psi}=-4\pi\epsilon\psi+\mathcal{O}\left(\psi^{2}\right)
\]
(cf. (\ref{eq: phase-isostable isostable})). We therefore have $\kappa=4\pi\epsilon$
and $\Lambda=\mathrm{e}^{-4\pi\epsilon}$. The PRC $\vec{Z}\left(\phi\right)$
and the IRC $\vec{Y}\left(\phi\right)$ follow by evaluating the appropriate
gradients on the limit cycle:
\begin{eqnarray*}
\vec{Z}\left(\phi\right) & = & \left.\frac{d\Phi}{d\vec{y}}\right|_{r=1}=\frac{1}{2\pi}\left(\begin{array}{c}
-\sin\left(2\pi\phi\right)\\
\cos\left(2\pi\phi\right)
\end{array}\right)\\
\vec{Y}\left(\phi\right) & = & \left.\frac{d\Psi}{d\vec{y}}\right|_{r=1}=\left(\begin{array}{c}
\cos\left(2\pi\phi\right)\\
\sin\left(2\pi\phi\right)
\end{array}\right),
\end{eqnarray*}
which, once averaged, yield via (\ref{eq: zG-zG inner product})-(\ref{eq: yG-yG inner product}),
\begin{eqnarray*}
\vec{z}_{G}^{T}\vec{z}_{G} & = & \frac{1}{4\pi^{2}}\\
\vec{z}_{G}^{T}\vec{y}_{G} & = & \frac{1-\mathrm{e}^{-4\pi\epsilon}}{8\pi^{2}}\frac{\epsilon\rho}{1+\epsilon^{2}}\\
\vec{y}_{G}^{T}\vec{y}_{G} & = & \frac{1-\mathrm{e}^{-8\pi\epsilon}}{8\pi\epsilon}\frac{1-2\epsilon\rho+4\epsilon^{2}}{1+4\epsilon^{2}}.
\end{eqnarray*}
Then from (\ref{eq:b in 1D-1}) and (\ref{eq:c_general}),
\begin{eqnarray}
b & = & \frac{2}{\pi}\frac{\epsilon^{2}}{1+\epsilon^{2}}\frac{1+4\epsilon^{2}}{1-2\epsilon\rho+4\epsilon^{2}}\rho\label{eq: b Hopf}\\
\frac{c}{\Gamma_{\textit{ss}}} & = & \frac{4}{\pi}\epsilon\frac{1+4\epsilon^{2}}{1-2\epsilon\rho+4\epsilon^{2}}\label{eq:c Hopf}
\end{eqnarray}
and from (\ref{eq:bLambda_cLambda}) and (\ref{eq: reduced internal parameter}),
\[
\frac{c_{\Lambda}}{b_{\Lambda}}=2w\frac{1+\epsilon^{2}}{\epsilon\rho}
\]
and
\[
\frac{b_{\Lambda}}{\sqrt{c_{\Lambda}}}=\frac{1}{\sqrt{\pi}}\sqrt{\frac{1-\mathrm{e}^{-4\pi\epsilon}}{1+\mathrm{e}^{-4\pi\epsilon}}}\frac{\epsilon^{\nicefrac{3}{2}}}{1+\epsilon^{2}}\sqrt{\frac{1+4\epsilon^{2}}{1-2\epsilon\rho+4\epsilon^{2}}}\,\rho.
\]
Note that $\frac{b_{\Lambda}}{\sqrt{c_{\Lambda}}}$ has the same sign
as $\rho$ and is monotonically increasing in $\rho$.

Recall that there are two thresholds beyond which unruliness will
certainly occur: (\ref{eq: general unruliness criterion 2}) and (\ref{eq: general unruliness criterion 1a}),
which correspond to the line with slope $\kappa_{\textit{unruly}}\left(\delta\right)$
and the lower horizontal line segment, respectively, in Figure \ref{fig:phaseDia generic}a.
For the Hopf oscillator, it turns out that the second criterion, $c<c_{\textit{crit}}=\pi\frac{\Gamma_{\textit{ss}}}{w^{2}}$
with $\frac{c_{\Lambda}}{b_{\Lambda}}>\kappa_{\textit{unclear}}\left(\delta\right)$,
is sufficient to describe the onset of certain unruliness. That is
because $\frac{b_{\Lambda}}{\sqrt{c_{\Lambda}}}$ and $\kappa_{\textit{unruly}}\left(\delta\right)$
are both bounded above in such a way that the first criterion is met
only if the second is already satisfied. In particular, as $w$ is
decreased, the parameterized curve $\left(b_{\Lambda}\left(w\right),c_{\Lambda}\left(w\right)\right)$
always intersects the line $c=c_{\textit{crit}}$ before the line $\frac{c_{\Lambda}}{b_{\Lambda}}=\kappa_{\textit{unruly}}\left(\delta\right)$.
(This is similar to the left parabolic curve labelled by ``(i)''
in Figure \ref{fig:phaseDia w-delta}a and in contrast with the rightmost
curve.) We therefore need only consider how the Hopf oscillator's
parameters $\epsilon$, $\rho$, $\delta$, and $w$ interact with
the boundary $c=c_{\textit{crit}}$ for certain unruliness and the boundary
$\frac{c_{\Lambda}}{b_{\Lambda}}=\kappa_{\textit{unclear}}\left(\delta\right)$
for the ``unclear'' region.

\begin{figure}
\noindent \begin{centering}
\includegraphics[width=1\textwidth]{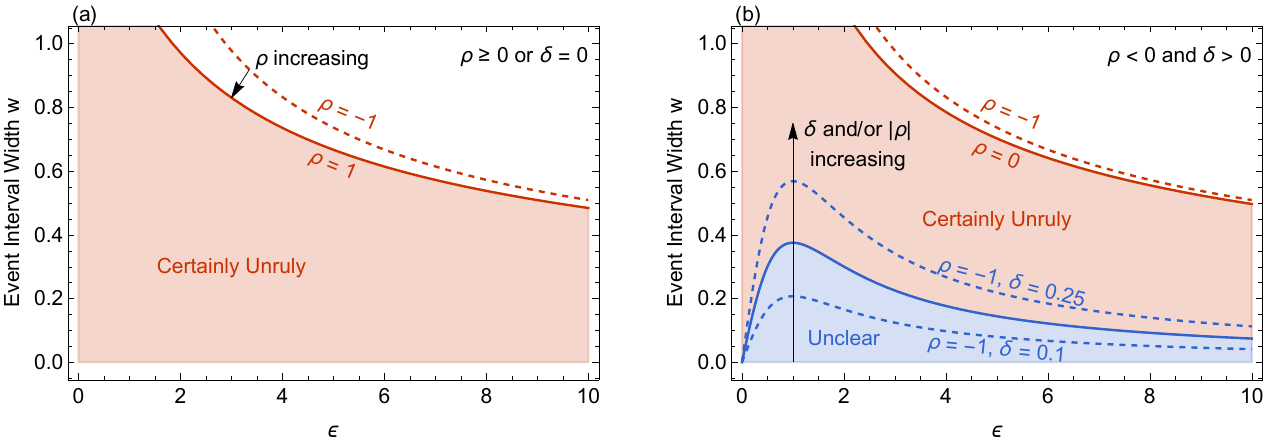}
\par\end{centering}
\caption{\textbf{Dependence of unruliness on $\epsilon$ and $w$ in the Hopf
Oscillator.} Regions of certain unruliness appear prominently in the
Hopf Oscillator. (a): For\textbf{ }$\rho\ge0$ or $\delta=0$,\textbf{
}there is no unclear region and certain unruliness appears when either
$w$ or $\epsilon$ take on moderate or small values. (b): When both
$\rho<0$ and $\delta>0$, the unclear region appears for small $w$.
Regions of possible unruliness are not shown.\textbf{\label{fig:phaseDia Hopf}}}
\end{figure}

For the Hopf oscillator, the second unruliness criterion (\ref{eq: general unruliness criterion 1a})
becomes
\[
\frac{4}{\pi}\epsilon\frac{1+4\epsilon^{2}}{1-2\epsilon\rho+4\epsilon^{2}}=\frac{c}{\Gamma_{\textit{ss}}}<\frac{c_{\textit{crit}}}{\Gamma_{\textit{ss}}}=\frac{\pi}{w^{2}}
\]
or
\[
w<\frac{\pi}{2}\sqrt{\frac{1-2\epsilon\rho+4\epsilon^{2}}{\epsilon\left(1+4\epsilon^{2}\right)}},
\]
which is satisfied for moderate and small values of each $w$ and
$\epsilon$ (red region, Figure \ref{fig:phaseDia Hopf}a and b).
There is no ``unclear'' region when $\rho\ge0$, since the mixed
component of the TVGR is positive and can only enhance the unruliness.
In this case, there is a prominent portion of the parameter space
which shows certain unruliness, and it is realized with very reasonable
parameter choices, e.g. $w\lesssim\frac{1}{2}$ with $\epsilon<10$.
Recall that in Hopf oscillator has a limit cycle of radius $1$; only
event interval widths $w$ that are less than $1$ are sensible. The
parameter $\epsilon$ measures the strength of contraction to the
limit cycle, and corresponds to a damping time constant of $\frac{1}{2\epsilon}$.
Since the limit cycle period is fixed to be $1$, without further,
specific information about the system, $\epsilon$ should be ``expected''
to be $\mathcal{O}\left(1\right)$. Figure \ref{fig:phaseDia Hopf}a
shows that for such choices of $w$ and $\epsilon$, the TVGR is largely
unruly.

When $\rho<0$ and $\delta\ne0$, $\frac{b_{\Lambda}}{\sqrt{c_{\Lambda}}}$
and, therefore, the mixed component of the TVGR is negative, a region
where the situation is unclear appears in the $\epsilon$-$w$ plane.
The nature of the TVGR is unclear when (\ref{eq: general unruliness criterion 1b})
is not satisfied. That happens when
\[
2w\frac{1+\epsilon^{2}}{\epsilon\rho}=\frac{c_{\Lambda}}{b_{\Lambda}}<\kappa_{\textit{unclear}}\left(\delta\right)
\]
or
\[
w<\frac{1}{2}\frac{\epsilon}{1+\epsilon^{2}}\rho\kappa_{\textit{unclear}}\left(\delta\right).
\]
Since $\kappa_{\textit{unclear}}\left(\delta\right)$ is negative and increasing
in magnitude with $a,$ the corresponding region grows with $\delta$
and $\left|\rho\right|$ (Figure \ref{fig:phaseDia Hopf}b). For moderate
asymmetry $\delta$ in the event interval, there is still a prominent
region of certain unruliness even in this case.

\subsection{Comparison of Theoretical Bounds, Numerics, and Simulation\label{subsec:comparisonTheoryNumericsSimulation}}

We also make use the Hopf oscillator example to empirically validate
our theory by comparing with numerics and simulations. We consider
two parameter sets: $\left(\epsilon=1,\rho=0,\delta=0\right)$ and
$\left(\epsilon=0.01,\rho=1,\delta=0.25\right)$. The first corresponds
to ``typical'' parameter choices for the Hopf oscillator: $\epsilon=1$
corresponds to an $\mathcal{O}\left(1\right)$ time constant for the
contraction to the limit cycle that is comparable to the limit cycle
period, $\rho=0$ to isotropic additive noise, and $\delta=0$ to
the event subset $E$ placed symmetrically around the limit cycle.
Comparing the theory and simulation using (in addition) the second,
non-trivial parameter set validates our derivation of the TVGR (\ref{eq:FanoVar Poincare})
and the bounds (\ref{eq:bounds on indicator funcVar}) we place on
it in the case of planar oscillators.

We consider the following computations of the TVGR:
\begin{enumerate}
\item The theoretical bounds (see (\ref{eq:FanoVar Poincare}) and (\ref{eq:bounds on indicator funcVar})),
\[
\left(\mathcal{E}-\mathcal{E}^{2}\right)+bx_{E}\mathcal{E}^{2}+\frac{c}{2}\sigma^{2}\mathcal{E}^{2}\leq\mathcal{V}_{E}^{\left(t\right)}\leq\frac{1+\Lambda}{1-\Lambda}\left(\mathcal{E}-\mathcal{E}^{2}\right)+bx_{E}\mathcal{E}^{2}+\frac{c}{2}\sigma^{2}\mathcal{E}^{2}.
\]
\item Numerical approximation of the theoretical results via a truncation
of the series in $\mathcal{V}_{x\mapsto1_{E}\left(x\right)}^{\left(n\right)}$
to $M=50$ terms (cf. (\ref{eq: Markov-only sum})),
\[
\mathcal{V}_{E}^{\left(t\right)}\approx\left(\mathcal{E}-\mathcal{E}^{2}\right)+2\sum_{m=1}^{M}\left(\mathcal{E}_{m}-\mathcal{E}^{2}\right)+bx_{E}\mathcal{E}^{2}+\frac{c}{2}\sigma^{2}\mathcal{E}^{2},
\]
where the double integral (\ref{eq: calE_m planar integral}) for
$\mathcal{E}_{m}$ is computed exactly along one dimension and via
numerical integration in the other. Note that, like the theoretical
bounds, this numerical estimate is subject to the first order asymptotic
approximations in the noise strength and isostable coordinate developed
in Section \ref{sec:Fano-Variance-for-Limit-Cycle} and the corresponding
SI Sections \ref{sec: SI - Averaging + Phase-Isostable} and \ref{sec: SI - TVGR Oscillator Derivation}.
\item Simulation of the Hopf oscillator model with 8192 independent noise
realizations. We use the Euler-Maruyama time-stepping scheme with
a step size of $\Delta t=10^{-6}$ for a total time of $t_{\textit{run}}=900$
(after a burn-in period of $100$) and determine the times $T_{k}^{E}$
at which each realization crosses the line segment $E$ via linear
interpolation between the Euler-Maruyama steps. We then empirically
estimate the asymptotic TVGR as (see the definitions of ``Event Rate''
and ``Event Dispersion Rate'' in Table \ref{tab:Point-process-statistics})
\[
\mathcal{V}_{E}^{\left(t\right)}\approx n^{2}\frac{\mathrm{\hat{var}}\left\{ T_{n}^{E}\right\} }{\hat{\mathrm{E}}\left\{ T_{n}^{E}\right\} ^{3}},
\]
where the hatted, sample statistics are computed across the 8192 realizations,
and $n$ is the largest value such that all of the realizations produce
at least $n$ events within a time of $t_{\textit{run}}$. For weak noise,
the above empirical quantity seems to converge much faster than the
more obvious empirical estimate of $\frac{1}{t_{\textit{run}}}\hat{\mathrm{var}}\left\{ N_{t_{\textit{run}}}^{E}\right\} $.
\end{enumerate}
\begin{figure}
\noindent \begin{centering}
\includegraphics[width=1\textwidth]{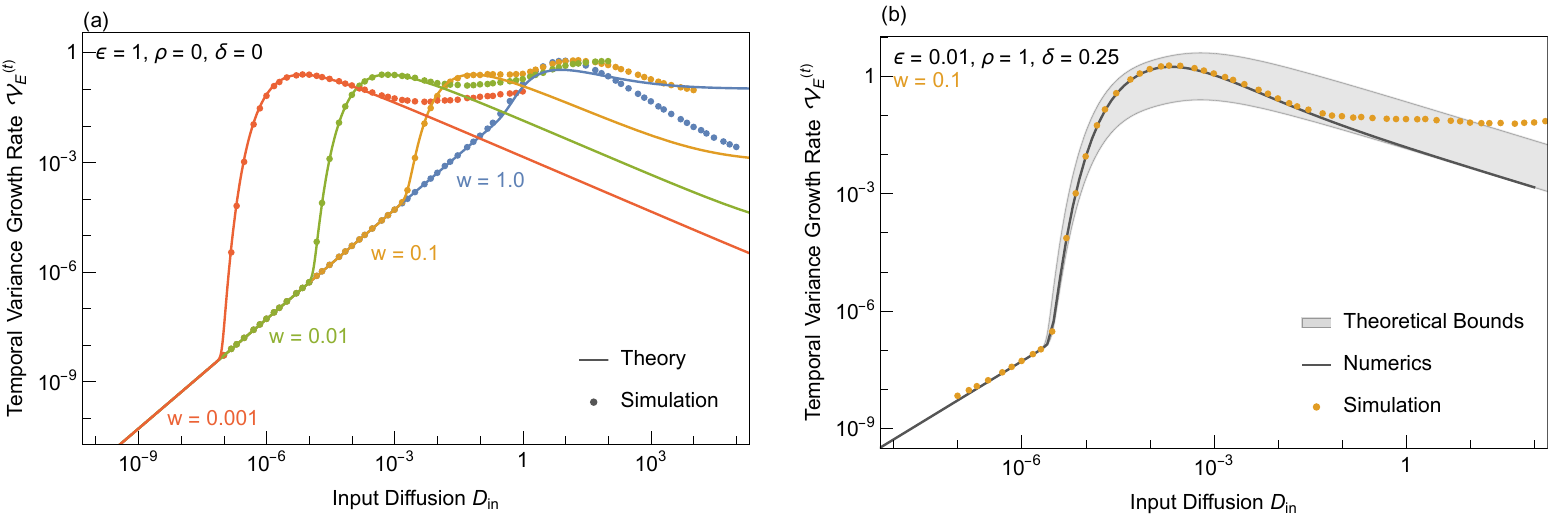}
\par\end{centering}
\caption{\textbf{Comparison of theory, numerics, and simulation for the Hopf
Oscillator.}\label{fig: FanoVar Hopf} (a): The TVGR for the Hopf
oscillator as a function of the input noise strength $D_{\textit{in}}$ for
a variety of event subset sizes $w$ and $\epsilon=1$, $\rho=0$,
and $\delta=0$. Since $\Lambda=\mathrm{e}^{-4\pi\epsilon}\approx3.4\times10^{-6}$
is so small in this case, the theoretical upper bound, the lower bound,
and the numerical approximation are indistinguishable at this scale.
Only the theoretical lower bounds (solid curves) are plotted in comparison
with the simulation results (points). (b): The TVGR for $w=0.01$
and $\epsilon=0.01$, $\rho=1$, and $\delta=0.25$. The shaded region
shows the range of the theoretical estimates, the solid line the numerical
approximation and the points the simulation results.}
\end{figure}

For the first parameter set $\left(\epsilon=1,\rho=0,\delta=0\right)$,
note that $\Lambda=\mathrm{e}^{-4\pi\epsilon}\approx3.4\times10^{-6}$
is quite small and so the theoretical upper and lower bounds and the
numerical approximation offer nearly identical results. We focus on
the accuracy of those predictions in comparison with simulations as
a function of the noise strength $D_{\textit{in}}$ and the event subset size
$\left|E\right|=w$. For smaller event intervals, the unruliness phenomenon
arises at weaker noise strengths (see (\ref{eq: Gamma_ss and w scaling})
for the theoretical effective scaling of $D_{\textit{in}}$ by $w^{2}$). Since
the theory is accurate in the limit of weak noise, we expect that
the theory will better capture the unruliness in the graph of $\mathcal{V}_{E}^{\left(t\right)}\left(D_{\textit{in}};w\right)$
for smaller $w$. This is indeed the case (Figure \ref{fig: FanoVar Hopf}a);
for $w=0.001$ and $w=0.01$, the theory quantitatively recovers the
nonlinear rise and local maximum in $\mathcal{V}_{E}^{\left(t\right)}\left(D_{\textit{in}};w\right)$,
which the standard phase reduction can not capture even qualitatively.
In all cases, the Markov renewal theory offers accuracy at noise levels
orders of magnitude greater than where the linear estimate $\mathcal{V}_{E,linear}^{\left(t\right)}=cD_{\textit{in}}$
from the standard phase reduction theory applies. Note that the Euler-Maruyama
scheme captures the stochastic forcing with a term with magnitude
of order $\sqrt{\Delta tD_{\textit{in}}}$. For the scheme to be accurate it
must be able to resolve the passage through $E$, and the simulation
should not be trusted when $\sqrt{\Delta tD_{\textit{in}}}\gtrsim w$ or, equivalently,
$D_{\textit{in}}\gtrsim10^{6}w^{2}$. In Figure \ref{fig: FanoVar Hopf}, we
have restricted the simulation results for each $w$ accordingly.

The second parameter set $\left(\epsilon=0.01,\rho=1,\delta=0.25\right)$
tests the accuracy of the theory when the mixed term in the TVGR is
non-zero (since $\delta\ne0$ and $\rho\ne0$) and when the lower
bound, upper bound, and numerical approximation are distinguishable
(since $\Lambda=\mathrm{e}^{-4\pi\cdot0.01}\approx0.88$ is significant).
Naturally, the numerical estimate will lie between the theoretical
upper and lower bound. But we should not necessarily expect the numerical
approximation and the simulation to coincide, since the numerical
estimate is subject to the first order approximation in the noise
strength and isostable coordinate (Section \ref{sec:Fano-Variance-for-Limit-Cycle}).
Nevertheless, for an event interval width of $w=0.1$, both the theoretical
bound and numerics recover the unruly quality of the TVGR found via
simulation (Figure \ref{fig: FanoVar Hopf}b).

\section{Extensions}

\label{sec:Extensions}

In this section we briefly investigate situations in which alternative
types of unruliness appear: a semi-infinite event interval $E$ and
oscillators with higher dimensional state spaces. We proceed heuristically,
only considering the limiting, quasi-renewal case, where the TVGR
is given by (\ref{eq: quasi-renewal Fano Variance}). As we saw for
planar oscillators with a finite interval $E$, the $\mathbf{\Lambda}\ne0$
case is significantly more complicated but often produces qualitatively
similar results.

\subsection{Semi-infinite $E$ \label{subsec: Semi-infinite E}}

In Section \ref{sec: Main Results - Planar Oscillators}, we have
only considered the cases where $E$ is a finite subset of $S$. There
is another natural case to consider: $E$ semi-infinite. For a planar
oscillator, we might for example take $E=\left(x_{\textit{min}},\infty\right)$,
so that events occur when the crossing position is above a threshold
$x_{\textit{min}}$. This is perhaps the case most relevant to neural oscillators,
where only a sufficiently large voltage peak may be classified as
an ``output event''.

\begin{figure}
\noindent \begin{centering}
\includegraphics[width=0.5\textwidth]{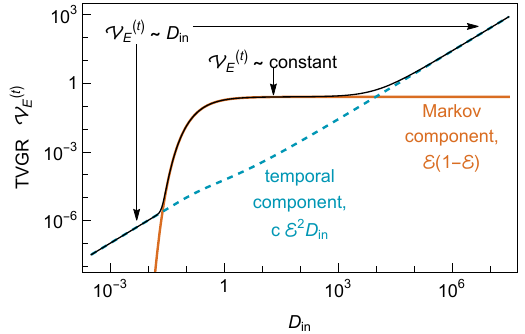}
\par\end{centering}
\caption{\textbf{An alternative, monotonic, form of \textquotedblleft unruliness\textquotedblright{}
for a semi-infinite interval $E$.} Shown are the two components of
the quasi-renewal TVGR $\mathcal{V}_{E,qr}^{\left(t\right)}$ and
their total for $c=10^{-4}$. We make this extreme choice and $c$
for illustrative purposes: each component is dominant for some range
of $D_{\textit{in}}$, and thus this plot shows all of the possible trends
that can appear in $\mathcal{V}_{E,qr}^{\left(t\right)}$. $c\mathcal{E}^{2}D_{\textit{in}}$
will always be dominant for small and very large noise, producing
a linear trend. But if $c$ is relatively small, $\mathcal{E}\left(1-\mathcal{E}\right)$
will then produce the characteristically strong nonlinear rise for
moderate noise strengths. It is possible that $\mathcal{E}\left(1-\mathcal{E}\right)$
then contributes to a range over moderately large input noise strengths
where $\mathcal{V}_{E}^{\left(t\right)}$ is roughly constant.\label{fig:semi-infinite E}}
\end{figure}

In the quasi-renewal limit, $\Lambda$ and $x_{\textit{min}}$ are jointly
taken to $0$ such that $x_{\textit{min}}\propto\sqrt{\Gamma_{\textit{ss}}}\propto\sqrt{-\frac{1}{\log\Lambda}}$.
Then, as was the case with the finite interval $E$ (cf. Section \ref{subsec:The-Quasi-Renewal-Case}),
\[
\mathcal{E}=\frac{1}{2}\left[1-\mathrm{erf}\left(\frac{x_{\textit{min}}}{2\sqrt{\Gamma_{\textit{ss}}D_{\textit{in}}}}\right)\right]
\]
is left invariant, $x_{E}\rightarrow0$, and $\mathcal{E}_{m}\rightarrow\mathcal{E}^{2}$
so that $\mathcal{V}_{E,qr}^{\left(t\right)}\left(D_{\textit{in}};c\right)=\mathcal{E}\left(1-\mathcal{E}\right)+cD_{\textit{in}}\mathcal{E}^{2}$.
The proportionality constant between $x_{\min}$ and $\sqrt{\Gamma_{\textit{ss}}}$
can be absorbed into $D_{\textit{in}}$, so that $\mathcal{E}=\frac{1}{2}\left[1+\mathrm{erf}\left(\frac{1}{2\sqrt{D_{\textit{in}}}}\right)\right]$,
and we find that $\mathcal{E}$ is monotonically decreasing with $D_{\textit{in}}$.
But, unlike the situation with a finite interval $E$, $\mathcal{E}\left(1-\mathcal{E}\right)$
is monotonic as well (Figure \ref{fig:semi-infinite E}, orange curve).
Thus, overall $\mathcal{V}_{E}^{\left(t\right)}$ will not be non-monotonic.
Instead, we find an alternative ``unruly'' quality for semi-infinite
intervals $E$: a linear trend for low $D_{\textit{in}}$, a substantial nonlinear
rise for moderate $D_{\textit{in}}$, and a monotonically increasing return
to a linear trend for large $D_{\textit{in}}$ (Figure \ref{fig:semi-infinite E},
black curve). Recall that, in contrast, for finite $E$ the TVGR $\mathcal{V}_{E}^{\left(t\right)}$
is asymptotically constant for large $D_{\textit{in}}$.

\subsection{Oscillators of Dimension $d+1>2$}

Considering the general analysis in Section \ref{sec:Fano-Variance-for-Limit-Cycle},
perhaps the most obvious generalization is an extension to higher-dimensional
oscillators. This becomes a challenge immediately: the probability
distributions of interest, though they might be Gaussian (Section
\ref{subsec:Linearized-Poincare-Map}), are all multi-dimensional.
For an arbitrary choice of the event subset $E$, the integrals $\mathcal{E}_{m}$
(\ref{eq: calE_m}) and $x_{E}$ (\ref{eq: xE}) will not in general
be found exactly or easily bounded. In this section, we briefly consider
two choices for $E$ for which the calculation in the $\Lambda\rightarrow0$
limit is tractable. They serve as prototypical examples and may give
insights into the general unruliness phenomena in higher dimensions.

\subsubsection{Semi-infinite $E$}

The discussion of Section \ref{subsec: Semi-infinite E} applies directly
to higher-dimensional oscillators when one takes $E$ to be a semi-infinite
region bounded by a hyperplane: $E=\left\{ \vec{\psi}\left|\vec{a}^{T}\vec{\psi}>x_{\textit{min}}\right.\right\} $.
In that case, the dimensions orthogonal to the vector $\vec{a}$ can
be ``integrated out'' reducing the analysis to the one dimensional
case. And, since all probability densities are Gaussian, the marginal
distributions along the coordinate $y$ defined by $y=\vec{a}^{T}\vec{\psi}$
are Gaussian as well. We must note, however, that this analysis only
covers the case where the boundary of $E$ is a hyperplane in the
phase-isostable coordinates. This may not be a natural choice for
every oscillator model. Even so, the high-dimensional neural mixed-mode
oscillator (Figure \ref{fig: MMO_DEff_vs_DPhase}b) has a linear trend
for large $D_{\textit{in}}$ (as is predicted here for a semi-infinite $E$,
cf. Figure \ref{fig:semi-infinite E}). In that model, any voltage
peak large enough qualifies as a ``spike'' event, i.e. $E$ is bounded
by a hyperplane orthogonal to the voltage axis, which, in general,
is not linear in phase-isostable coordinates. We speculate that the
semi-infinite quality, rather than the specific choice of the boundary
of $E$ is responsible for the large-noise linear trend.

\subsubsection{Finite $E$ with $0\in E$}

\begin{figure}
\noindent \begin{centering}
\includegraphics[width=1\textwidth]{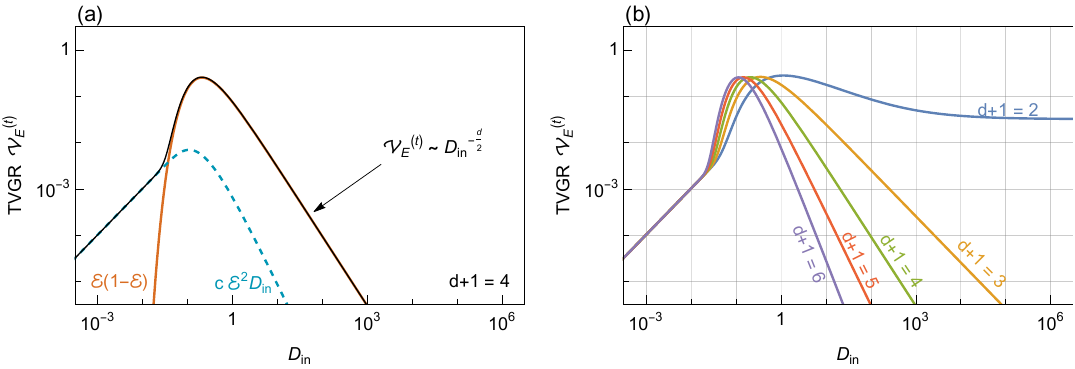}
\par\end{centering}
\caption{\textbf{Alternative forms of unruliness in higher dimensions with
symmetric $E$.} Shown is the quasi-renewal TVGR $\mathcal{V}_{E,qr}^{\left(t\right)}$
with $c=0.1$. (a): The two non-zero components of the TVGR for oscillator
dimension $d+1=4$. For $d+1>3$, the TVGR is dominated by $\mathcal{E}\left(1-\mathcal{E}\right)$
for large $D_{\textit{in}}$. (b): The TVGR for various $d$. For $d+1>2$,
it is asymptotically decreasing.\label{fig: higherD finite E}}
\end{figure}

As the prototypical example for finite $E$, take $E$ to be an ellipsoid
centered at $\vec{\psi}=0$, shaped like the level-sets of the steady-state
probability density $\mathcal{N}\left(0,2D_{\textit{in}}\mathbf{\Gamma}_{\textit{ss}}\right)$
on the $d$-dimensional section $S$. In this case, due to the symmetry
in $E$, the event center of mass $x_{E}=0$. In the quasi-renewal
limit, we take $\Lambda$ and the size of $E$ to $0$ jointly so
that $\mathcal{E}$ is fixed. With the appropriate constants absorbed
into $D_{\textit{in}}$, we find
\[
\mathcal{E}=1-\frac{\Gamma\left(\frac{d}{2},\frac{1}{4D_{\textit{in}}}\right)}{\Gamma\left(\frac{d}{2},0\right)},
\]
where $\Gamma$ is the incomplete gamma function. Since $0\in E$,
the event probability $\mathcal{E}$ approaches $1$ for small $D_{\textit{in}}$,
as before. For large $D_{\textit{in}}$, $\mathcal{E}$ goes like $D_{\textit{in}}^{-\frac{d}{2}}$.
And so, in contrast with the planar case, the temporal term $c\mathcal{E}^{2}D_{\textit{in}}\sim cD_{\textit{in}}^{1-d}$
is actually asymptotically decreasing for state-space dimension $d+1>2$
. The result is a yet different form of unruliness, where the TVGR
is dominated by $\mathcal{E}\left(1-\mathcal{E}\right)$ at moderate
and large noise (Figure \ref{fig: higherD finite E}). The prominent
maximum at moderate noise will appear so long as $c$ is sufficiently
small (we set $c=0.1$ in Figure \ref{fig: higherD finite E}). Note
that in the large noise limit $\mathcal{E}\left(1-\mathcal{E}\right)\sim\mathcal{E}$
goes like $D_{\textit{in}}^{-\frac{d}{2}}$, which dominates $D_{\textit{in}}^{1-d}$
when $d+1>3$. For all $d+1>2$, the TVGR is asymptotically decreasing.

Consider for the moment a similar ellipsoid $E$, which is, however,
asymmetric, such that the center of mass of $E$ is non-zero, $x_{E}\ne0$.
Since the mixed term in the TVGR is $\vec{b}\cdot x_{E}\mathcal{E}^{2}$,
the contribution of the asymmetry to $\mathcal{V}_{E}^{\left(t\right)}$
in the large-noise limit is subdominant, $\mathcal{O}\left(\mathcal{E}^{2}\right)$.
Therefore we expect that the small- and large- noise asymptotic trends
and the alternative form of unruliness found for the unit ball $E$
will be reproduced for mild asymmetry in $E$. We speculate that it
is also reproduced for other reasonable choices of finite $E$, e.g.
those that are convex.

\section{Discussion}

\label{sec:Discussion}

In this work, we have demonstrated that beyond the linear regime predicted
by the standard phase-reduction analysis, the event-based, diffusive
response of noise-driven limit-cycle oscillators can be ``unruly'':
with increasing noise strength it can exhibit an enormous nonlinear
amplification and a subsequent decrease for yet stronger noise. Such
behavior had been observed previously in numerical simulations of
mixed-mode oscillations \citep{karamchandani_pulse-coupled_2018}.
Here we extracted the origin of such unruly behavior by considering
the point process that arises from crossings of a Poincare section
$S$ on which a subset $E$ is distinguished as event-producing. Since
we make use of the phase-isostable representation, the work presented
here covers a large class of systems with stable limit cycles. Indeed,
for generic planar oscillators and $E$ a finite interval, we have
argued that unruliness appears in a finite region of the natural parameter
space, and, via a linearization of the dynamics, we have shown this
explicitly. We found that unruliness can appear even in a simple,
prototypical oscillator driven by additive noise. And, through that
example, we confirmed quantitatively the predictions of the point-process
approach by direct numerical simulations.

The specification of the event surface $E$ as a proper subset of
the Poincare section $S$ is essential to our results; it partitions
the points on the section $S$ into events and non-events. Using the
linearization of the Poincare map (Section \ref{subsec:Linearized-Poincare-Map})
but excluding the partitioning, there would be no unruliness (see
the discussion at the end of Section \ref{subsec:Fano-Variance_linear}).
What justifies introducing such a partitioning? Recall that our inspiration
follows from the mixture of large-amplitude voltage spikes and small-amplitude
STOs that appear in neural mixed-mode oscillators (Figure \ref{fig: MMO_DEff_vs_DPhase}a).
There, only the spikes correspond to the action potentials through
which the neurons interact with each other; therefore the spikes are
the events of interest. Indeed, in our previous work, \citet{karamchandani_pulse-coupled_2018},
we found that replacing the standard phase diffusion coefficient $D_{\textit{phase}}$
in the Fokker-Planck equation (\ref{eq:Fokker-Planck}) with the event-based
effective diffusion coefficient $D_{\textit{eff}}=\frac{1}{2}\mathcal{V}_{E}^{\left(t\right)}$,
which is unruly and substantially larger, was essential and sufficient
to capture the dynamics of a globally-coupled coupled population of
noisy neural mixed-mode oscillators. That replacement enables (\ref{eq:Fokker-Planck})
to reproduce even the bistability and non-trivial population dynamics
that appear in the full simulations \citep{karamchandani_pulse-coupled_2018}.
Notably, because the effective diffusion coefficient $D_{\textit{eff}}$ is
a property of noise-response of individual oscillators, we could measure
it efficiently via simulations of a single, uncoupled oscillator and
then apply the result to the population dynamics.

The designation of events by $E$ can be interpreted more generally
as a relevant read-out of an oscillator's dynamics. Examples can also
be found outside of neuroscience. For instance, in synthetic biology
the repressilator is captured experimentally via large amplitude oscillations
in the read-out fluorescence \citep{elowitz_synthetic_2000}. And,
in fact, a recent, mathematical study of a repressilator model makes
use of a finite Poincare section (equivalently, a finite subset $E$
of a section $S$) to capture the distribution of times between oscillation
peaks \citep{potapov_multi-stable_2015}. In contrast with the original
repressilator model, Potapov et. al.'s version supports bistability
between a fixed point and a limit cycle via a subcritical Hopf bifurcation.
That leads to noise-induced small amplitude oscillations near the
fixed point and noisy large amplitude oscillations near the limit
cycle, with stochastic transitions between them. While those authors
did not consider it, the observed variability in the oscillator output
resulting from the transitions could be quantified using the effective
diffusion coefficient $D_{\textit{eff}}$.

Regular transitions between oscillatory and quiescent phases in the
absence of noise are also found in a wide range of natural systems.
These ``bursters'' often arise from the coupling of a bistable oscillator
to an additional slow mode \citep{rinzel_formal_1987}. When the large
oscillations are identified as events and the system is forced with
noise, our results suggest that the event variability of the bursters
often could be unruly. Indeed, in a numerical study \citep{karamchandani_reduced_2022},
we found such unruliness in a burster arising in an unfolding of the
Hopf-zero bifurcation normal form (see e.g. \citet{guckenheimer_codimension_1981}
for a discussion of the unfolding of this codimension-2 bifurcation).
The effect of noise in both bursting and mixed-mode oscillator systems
is of considerable theoretical interest \citep{longtin_autonomous_1997,su_effects_2004,berglund_random_2015}.
It would be an important extension of our work to go beyond the computational
results of \citet{karamchandani_pulse-coupled_2018,karamchandani_reduced_2022}
and capture theoretically the effective diffusion coefficient in such
systems, where, in particular, oscillations of variable amplitude
appear already without noise. The analysis is likely to be more complicated
than what we have covered here: when the noiseless limit cycle shows
both small and large oscillation amplitudes, typical Poincare sections
would be pierced by it multiple times. The technical elements of our
derivation in Section \ref{sec:Fano-Variance-for-Limit-Cycle} and
SI Section \ref{sec: SI - TVGR Oscillator Derivation} will then be
more involved, with a key challenge being a reduced representation
of the more complicated Poincare map and first passage time.

In this work, we have focused on the variance in the number of events
appearing in a given time interval normalized by that time interval
in order to quantify the variability in events produced by oscillators.
This ``temporal variance growth rate'' (TVGR), which is proportional
to the effective diffusion coefficient, is just one of a many statistical
attributes of a point process. The distributions of the individual
time intervals between events, e.g. inter-spike-intervals (ISIs) for
neuronal oscillators, are also of broad interest. And, interestingly,
for certain oscillators the coefficient of variation of the ISIs can
be non-monotonic as a function of input noise strength \citep{nesse_oscillation_2008,craft_effects_2022}.
Our event-based framework may prove useful in that context as well,
offering efficient models for complex (e.g. multi-modal) ISIs distributions
and complementing existing, theoretical \citep{nesse_oscillation_2008}
and phenomenological \citep{craft_effects_2022} approaches. Besides
the TVGR and the ISI coefficient of variation, other natural, variance-like,
summary statistics have been previously investigated in the literature.
Some, like the dispersion of event times and the Fano factor, are
related to the TVGR by factors of the mean event interval in the long-time
limit (Table \ref{tab:Point-process-statistics}). Thus, the unruly
behavior in the TVGR we characterized here may well be related to
the so-called coherence resonance and incoherence maximization other
authors have found in the Fano factor \citep{pikovsky_coherence_1997,lindner_maximizing_2002}.
It would be an interesting extension of our work to characterize the
mean event interval and thus connect all of the statistics that appear
in Table \ref{tab:Point-process-statistics} in application to limit-cycle
oscillators. We note that in much of the literature (e.g. in \citet{pyke_limit_1964,jones_markov_2004}),
the TVGR and the related quantities $\mathcal{V}_{x\mapsto f\left(x\right)}^{\left(n\right)}$
and $\mathcal{V}_{\left(x,x^{\prime},\Delta T\right)\mapsto w\left(x,x^{\prime},\Delta T\right)}^{\left(t\right)}$,
which we considered in Sections \ref{sec:Fano-Variance-and-Function-Variance}
and \ref{sec: SI - Event TVGR Formula}, appear in the context of
central limit theorems. Informally, those theorems indicate that those
mean and variance statistics are largely sufficient in describing
the event point process over long time scales.

We provided a general decomposition of the TVGR, (\ref{eq:FanoVar decomposed}),
and applied it to a simple toy model (Section \ref{sec:Fano-Variance-for-Toy-Model}).
Notably, the toy model does not include any memory between states
at one step and the next, but nevertheless shows unruliness. Thus,
the Markov-renewal process (and the memory across steps it offers)
is not strictly necessary; a renewal process would be sufficient to
generate unruliness in the TVGR. Indeed, the analyses of coherence
resonance and incoherence maximization found in the literature often
assume renewal dynamics, e.g. by considering only the marginal distribution
of single inter-event time intervals. However, the Poincare map dynamics
for limit cycle oscillators are generally Markovian, reflecting the
deterministic dynamics of the underlying limit cycle. Moreover, those
non-renewal dynamics are not only widespread but may also play a functional
role, for instance, in the context of neuroscience, in information
processing in the brain \citep{farkhooi_serial_2009,avila-akerberg_nonrenewal_2011,ramlow_interspike_2021}.
One of our contributions in this work is an examination of the TVGR
in its entirety within the Markov renewal framework, including the
effect of correlations between successive states of the process. In
our analysis of planar oscillators, we carefully bound the size of
those effects (see (\ref{eq:bounds on indicator funcVar}) and SI
Section \ref{sec: SI - Bounds}) and find that the correlations contribute
positively towards the non-monotonic, Markov-only component of the
TVGR (see (\ref{eq:bounds on indicator funcVar})), and therefore
may enhance its unruly quality.

The formula we derive for the TVGR (\ref{eq:FanoVar decomposed})
is exact, but we apply it to limit cycle oscillators in Section \ref{sec:Fano-Variance-for-Limit-Cycle}
and the corresponding SI Section \ref{sec: SI - TVGR Oscillator Derivation}
via a linearization of the dynamics. Even though the linearization
is an approximation, we anticipate that generically there will be
a degree of structural stability as nonlinear effects are ``turned
on''. In particular, we expect that the event probability $\mathcal{E}$
continues to be a monotonically decreasing function of the noise strength
$D_{\textit{in}}$ (as in Figure \ref{fig:sigma-dependent elements}a,b.i),
leading to the non-monotonicity in the Markov-only component and the
TVGR overall. And, so, we anticipate that the unruliness we have discussed
here will appear in many (weakly) nonlinear oscillator models.

In real, nonlinear systems the events of interest and the non-events
may well be produced in different regions of phase space, which may
have very different dynamics. The partitioning of $S$ by $E$ is
then a first step in taking this qualitative difference into account.
The boundary $\partial E$ constitutes a dynamical separatrix of some
type (see \citet{rowat_interspike_2007,hitczenko_bursting_2009,berglund_mixed-mode_2012}
for specific examples and analyses). In general, we would not expect
the Poincare map to vary smoothly across such a boundary, and the
linearization we have considered in this work fails to be an accurate
representation. It would be an interesting extension of this work
to consider a piece-wise linear approximation to the Poincare map
that is segmented at $\partial E$. That would be analogous to piece-wise
linear approximations to the differential equations of nonlinear systems
that have been used to study, e.g., coherence and stochastic resonance
\citep{lindner_coherence_2000}.

Additionally, one may consider a relaxation of the hard threshold
in the partitioning of events and non-events. In the broader context
of Markov renewal reward processes that we discuss in Section \ref{sec:Fano-Variance-and-Function-Variance},
there is no particular reason to limit an asymptotic variance growth
rate calculation to rewards given by the indicator function $1_{E}$.
In this view, $1_{E}$ is just one of many observation functions on
the Poincare section that can be used to probe the nonlinear, stochastic
dynamics of the oscillator. We would expect the unruly behavior that
we have investigated in this work to be reflected in some way also
in the asymptotic variance growth rates for other observation functions.

Finally, we return to our original motivation: a reduced model whose
event statistics (Table \ref{tab:Point-process-statistics}) agree
with those of the limit-cycle oscillator in the long-time limit. In
considering the events, we eschewed the phase variable in favor of
a point process. Notably, extending the phase reduction by considering
isostable variables, (\ref{eq: phase-isostable phase})-(\ref{eq: phase-isostable isostable}),
does not offer corrections to the phase dynamics at first order in
the noise strength (compare (\ref{eq: phase}) and (\ref{eq: phase_averaged})).
In contrast, since the event subset $E$ distinguishes different amplitudes,
the phase-isostable representation along with $E$ \emph{does} offer
an enhanced approximation of the event point process statistics even
at linear order. Our work offers one example where first identifying
within the dynamics a non-renewal point process and then capturing
the point process as a reduced Markov renewal-reward process proved
beneficial. There may be many other circumstances in which complex
noise-driven nonlinear dynamics can be captured effectively by Markov-renewal
point processes.

\section*{Acknowledgements}

The author gratefully acknowledges support from NSF via CMMI-1435358,
DMS-1547394, and DMS-1937229. The author thanks Hermann Riecke for
very helpful discussions and feedback on this manuscript and Kevin
Lin for interesting discussion.

\bibliography{eventsNoiseDrivenOscillators}

\section*{\newpage}

\setcounter{figure}{0} \renewcommand{\thefigure}{S\arabic{figure}} 

\setcounter{equation}{0}\renewcommand\theequation{S\arabic{equation}}

\setcounter{section}{0}\renewcommand\thesection{S\arabic{section}}

\setcounter{page}{1}\renewcommand\thepage{S\arabic{page}}

\part*{Supplementary Information}

\section{Derivation of the Event TVGR Formula\label{sec: SI - Event TVGR Formula}}

In this Supplementary Information (SI) section, we offer two derivations
of a generalization of (\ref{eq:FanoVar from funcVar}), (\ref{eq: Markov renewal variance formula})
below, wherein the temporal variance growth rate $\mathcal{V}_{\left(x,x^{\prime},\Delta T\right)\mapsto w\left(x,x^{\prime},\Delta T\right)}^{\left(t\right)}$
(TVGR) is related to a sequential variance growth rate $\mathcal{V}_{\left(x\rightarrow x^{\prime}\right)\mapsto f_{w}\left(x,x^{\prime}\right)}^{\left(n\right)}$
(SVGR), and a derivation of (\ref{eq:FanoVar decomposed}), where
the TVGR is decomposed into three components. In Section \ref{subsec: SI - TVGR Formula by Analogy},
we offer a non-rigorous, plausibility argument that motivates the
form of (\ref{eq:FanoVar from funcVar simple}). In Section \ref{subsec: SI - TVGRFormula Discrete}
we give a more careful derivation that is only valid for Markov chains
with discrete state spaces and yet yields the same formula. Between
the plausibility argument given in \ref{subsec: SI - TVGR Formula by Analogy},
the derivation of limited scope given in \ref{subsec: SI - TVGRFormula Discrete},
and the empirical evidence we provide in Section \ref{subsec:comparisonTheoryNumericsSimulation},
we are content to take (\ref{eq:FanoVar from funcVar simple}) as
fact for the further analyses presented in this work.

\subsection{The TVGR Formula by Rough Analogy\label{subsec: SI - TVGR Formula by Analogy}}

(\ref{eq:Fano variance}) shows that the event TVGR is associated
with the function $1_{E}(x)$ on the Markov renewal process. We will
consider more generally any ``reward'' $w$ on the Markov renewal
process, which is typically taken to be a (possibly random) function
of the current state $x$, the next state $x^{\prime}$, and the time
interval $\Delta T$ between them. The temporal variance growth rate
associated with $w$ is therefore
\begin{equation}
\mathcal{V}_{\left(x\rightarrow x^{\prime},\Delta T\right)\mapsto w\left(x,x^{\prime},\Delta T\right)}^{\left(t\right)}\equiv\lim_{t\rightarrow\infty}\frac{1}{t}\mathrm{var}\left\{ \sum_{k=1}^{N_{t}}w\left(x_{k},x_{k+1},\Delta T_{k}\right)\right\} .\label{eq: MRR variance}
\end{equation}
We recall that the random variable $\Delta T_{k}$ will in general
depend on $x_{k}$ and $x_{k+1}$, and, once conditioned on $x_{k}$
and $x_{k+1}$, $\Delta T_{k}$ is independent of $x_{j}$, $j\ne k,k+1$.
The same is true of the reward $w\left(x_{k},x_{k+1},\Delta T_{k}\right)$
conditioned on $x_{k}$ and $x_{k+1}$. We can imagine, therefore,
constructing the Markov process $x_{k}$ first, independently of the
time intervals and reward, and then add to the process the random
time intervals $\Delta T_{k}$ and the random reward $w$. To that
end, we take $\Delta t\left(x_{k},x_{k+1}\right)$ to be a random
function that has the same distribution as $\Delta T_{k}$ when conditioned
on $x_{k}$ and $x_{k+1}$ and rewrite $w\left(x_{k},x_{k+1},\Delta T_{k}\right)$
as $w\left(x_{k},x_{k+1},\Delta t\left(x_{k},x_{k+1}\right)\right)$:
\begin{equation}
\mathcal{V}_{\left(x\rightarrow x^{\prime},\Delta T\right)\mapsto w\left(x,x^{\prime},\Delta T\right)}^{\left(t\right)}=\lim_{t\rightarrow\infty}\frac{1}{t}\mathrm{var}\left\{ \sum_{k=1}^{N_{t}}w\left(x_{k},x_{k+1},\Delta t\left(x_{k},x_{k+1}\right)\right)\right\} .\label{eq: MRR variance DeltaT function}
\end{equation}
As compared with (\ref{eq: MRR variance}), (\ref{eq: MRR variance DeltaT function})
is a step closer to the generalized SVGR (\ref{eq: genearlized function variance}),
since the quantity in the sum is a random function of the states $x_{j}$
alone. This will be helpful in the following, where we find a formula
for $\mathcal{V}_{\left(x,x^{\prime},\Delta T\right)\mapsto w\left(x,x^{\prime},\Delta T\right)}^{\left(t\right)}$
in terms of a SVGR.

In Section \ref{subsec: SI - TVGRFormula Discrete}, we offer a derivation
of such a formula for the special case of a Markov renewal process
on a discrete state space. But here we find the same formula ((\ref{eq: Markov renewal variance formula}),
below) in a more general setting by analogy. To set up the analogy,
first compare the SVGR of a Markov process $x_{k}$ with that of an
i.i.d. process $y_{k}$. For independent samples $y_{k}$,
\begin{equation}
\frac{1}{n}\mathrm{var}\left\{ \sum_{k=1}^{n}f\left(y_{k}\right)\right\} =\mathrm{var}\left\{ f\left(y\right)\right\} ,\label{eq:var_sum_iid}
\end{equation}
while in general for large $n$, 
\begin{equation}
\frac{1}{n}\mathrm{var}\left\{ \sum_{k=1}^{n}f\left(x_{k}\right)\right\} \sim\mathcal{V}_{x\mapsto f\left(x\right)}^{\left(n\right)}\ne\mathrm{var}\left\{ f\left(x\right)\right\} .\label{eq: var_sum_Markovian}
\end{equation}
So the SVGR $\mathcal{V}_{x\mapsto f\left(x\right)}^{\left(n\right)}$
can be thought of as the asymptotic, effective variance of $f\left(x\right)$
that takes into account the fact that $x_{k}$ and $x_{l}$ are interdependent.
Now, in the analogy, the SVGR (\ref{eq: var_sum_Markovian}) is to
the i.i.d variance (\ref{eq:var_sum_iid}) as the TVGR $\mathcal{V}_{\left(x,x^{\prime},\Delta T\right)\mapsto w\left(x,x^{\prime},\Delta T\right)}^{\left(t\right)}$
for $w$ is to a TVGR $\mathcal{V}_{\Delta T\mapsto w\left(\Delta T\right)}^{\left(t\right)}$
for a \emph{renewal-reward} process:
\begin{equation}
\mathcal{V}_{x\mapsto f\left(x\right)}^{\left(n\right)}\;:\;\mathrm{var}\left\{ f\left(y\right)\right\} \;::\;\mathcal{V}_{\left(x,x^{\prime},\Delta T\right)\mapsto w\left(x,x^{\prime},\Delta T\right)}^{\left(t\right)}\;:\;\mathcal{V}_{\Delta T\mapsto w\left(\Delta T\right)}^{\left(t\right)}.\label{eq: function to Fano-like analogy}
\end{equation}
As compared with a Markov renewal-reward process, a renewal-reward
process has no state space and the time intervals are i.i.d. The reward
can only be a random function of the time interval, and therefore
the corresponding temporal variance growth rate is
\begin{equation}
\mathcal{V}_{\Delta T\mapsto w\left(\Delta T\right)}^{\left(t\right)}\equiv\lim_{t\rightarrow\infty}\frac{1}{t}\mathrm{var}\left\{ \sum_{k=1}^{N_{t}}w\left(\Delta T_{k}\right)\right\} .\label{eq: RR variance}
\end{equation}
Renewal-reward theory \citep{smith_regenerative_1955} offers the
required formula for (\ref{eq: RR variance}), 
\begin{equation}
\mathcal{V}_{\Delta T\mapsto w\left(\Delta T\right)}^{\left(t\right)}=\frac{1}{\mathrm{E}\left\{ \Delta T\right\} }\mathrm{var}\left\{ w\left(\Delta T\right)-\frac{\mathrm{E}\left\{ w\left(\Delta T\right)\right\} }{\mathrm{E}\left\{ \Delta T\right\} }\Delta T\right\} .\label{eq: renewal variance formula}
\end{equation}
Note that like the i.i.d variance $\mathrm{var}\left\{ f\left(y\right)\right\} $
and in contrast with the function variance $\mathcal{V}_{x\mapsto f\left(x\right)}^{\left(n\right)}$
(cf. (\ref{eq:FV sum of variances})), (\ref{eq: renewal variance formula})
only includes the variance evaluated over one step of the process,
reflecting the fact that the steps in the renewal-reward process are
independent.

To complete the analogy (\ref{eq: function to Fano-like analogy})
and find $\mathcal{V}_{\left(x,x^{\prime},\Delta T\right)\mapsto w\left(x,x^{\prime},\Delta T\right)}^{\left(t\right)}$,
we must replace the variance $\mathrm{var}$ in (\ref{eq: renewal variance formula})
with its asymptotic, effective counterpart $\mathcal{V}^{\left(n\right)}$
that takes into account the non-renewal quality of Markov renewal
processes (compare (\ref{eq:var_sum_iid}) and (\ref{eq: var_sum_Markovian})).
We also generalize the function $w\left(\Delta T\right)$ as $w\left(x,x^{\prime},\Delta t\left(x,x^{\prime}\right)\right)$,
concluding that 
\begin{equation}
\mathcal{V}_{\left(x,x^{\prime},\Delta T\right)\mapsto w\left(x,x^{\prime},\Delta T\right)}^{\left(t\right)}=\frac{1}{\mathrm{E}\left\{ \Delta T\right\} }\mathcal{V}_{\left(x\rightarrow x^{\prime}\right)\mapsto f_{w}\left(x,x^{\prime}\right)}^{\left(n\right)},\label{eq: Markov renewal variance formula}
\end{equation}
where 
\[
f_{w}\left(x,x^{\prime}\right)\equiv w\left(x,x^{\prime},\Delta t\left(x,x^{\prime}\right)\right)-\frac{\mathrm{E}\left\{ w\left(x,x^{\prime},\Delta T\right)\right\} }{\mathrm{E}\left\{ \Delta T\right\} }\Delta t\left(x,x^{\prime}\right).
\]
Note that the left hand side of (\ref{eq: Markov renewal variance formula})
is the TVGR-like quantity, defined in (\ref{eq: MRR variance}) with
respect to the time $t$, while the right hand side includes a function
variance, defined via (\ref{eq:function variance}) and (\ref{eq: genearlized function variance})
with respect to the discrete steps $n$. We make use of (\ref{eq: Markov renewal variance formula})
and the formula for function variances, (\ref{eq:FV sum of variances}),
to evaluate the reward TVGR $\mathcal{V}_{\left(x,x^{\prime},\Delta T\right)\mapsto w\left(x,x^{\prime},\Delta T\right)}^{\left(t\right)}$.

For the case of interest in this work, the reward is the indicator
function for events, $w\left(x,x^{\prime},\Delta T\right)=1_{E}\left(x\right)$,
and (\ref{eq: Markov renewal variance formula}) yields (\ref{eq:FanoVar from funcVar}).

\subsection{The TVGR Formula for Discrete State Spaces\label{subsec: SI - TVGRFormula Discrete}}

\begin{figure}
\begin{centering}
\includegraphics{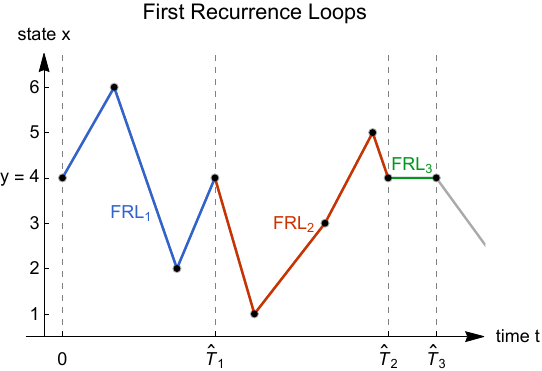}
\par\end{centering}
\caption{A schematic depiction of Cinlar's analysis for discrete-state Markov-renewal
processes \citep{cinlar_markov_1975}. The \textquotedblleft first
recurrence loops\textquotedblright{} (FRLs) each with initial and final
state $y$ are distinguished by color. The time durations $\Delta\hat{T}_{l}=\hat{T}_{l+1}-\hat{T}_{l}$
of the FRLs are independent and identically distributed; the visits
to $y$ form the renewal process $\hat{T}_{l}$. \label{fig:Cinlar_schematic}}
\end{figure}

Here we derive (\ref{eq: Markov renewal variance formula}) for Markov
renewal processes for a discrete rather than continuous state space.
We begin by referring to \citet{cinlar_markov_1975}, who approaches
the Markov-renewal-reward process by connecting it to a renewal-reward
process. He considers a partitioning of the sequence $x_{k}$ for
$k\in\left\{ 0,1,\ldots\right\} $ into what we call ``first recurrence
loops'' (FRLs), which are all the subsequences that start at the
initial state $y=x_{0}$ and end at the same state $y$ without visiting
it in between (Figure \ref{fig:Cinlar_schematic}). Note that this
requires the state space to be discrete, since otherwise the return
to state $y$ happens with probability $0$. Once the first FRL back
to $y$ is complete, the next one is a completely independent realization
of a FRL starting at $y$, reflecting the Markov character of the
underlying process. Thus, the time $\Delta\hat{T}_{l}$ elapsed between
subsequent visits to $y$ (during the $l^{\textit{th}}$ FRL, $\mathrm{FRL}_{l}$)
along with the reward $\hat{W}_{l}$ accumulated during those intervals
is a renewal-reward process. Renewal-reward theory then gives \citep{smith_regenerative_1955}\footnote{In \citet{cinlar_markov_1975}, the factor of $\mathrm{E}\left\{ \Delta\hat{T}_{1}\right\} ^{-1}$
is missing. It appears correctly in Equation 5.2.5 of \citet{smith_regenerative_1955}.}
\begin{equation}
\mathcal{V}_{\left(x,x^{\prime},\Delta T\right)\mapsto w\left(x,x^{\prime},\Delta T\right)}^{\left(t\right)}=\frac{1}{\mathrm{E}\left\{ \Delta\hat{T}_{1}\right\} }\mathrm{var}\left\{ \hat{W}_{1}-\frac{\mathrm{E}\left\{ \hat{W}_{1}\right\} }{\mathrm{E}\left\{ \Delta\hat{T}_{1}\right\} }\Delta\hat{T}_{1}\right\} ,\label{eq: MRR variance from renewal}
\end{equation}
where $\Delta\hat{T}_{1}=\sum_{k=1}^{\hat{n}_{1}}\Delta T_{k}=T_{\hat{n}_{1}}$
and $\hat{W}_{1}=\sum_{k=1}^{\hat{n}_{1}}w\left(x_{k},x_{k+1},\Delta T_{k}\right)$,
defining $\hat{n}_{l}$ to be the (random) number of steps in the
$l^{\textit{th}}$ FRL. Since the FRLs parameterized by $l$ are independent,
the choice to set $l=1$ in (\ref{eq: MRR variance from renewal})
is as good as any. $\mathrm{E}\left\{ \Delta\hat{T}_{1}\right\} $
is the mean first recurrence time for state $y$. It is proportional
to the expected transition time $\mathrm{E}\left\{ \Delta T\right\} $
across all pairs of states,
\[
\mathrm{E}\left\{ \Delta\hat{T}_{1}\right\} =p\,\mathrm{E}\left\{ \Delta T\right\} ,
\]
where the constant of proportionality $p\equiv\frac{1}{\pi\left(y\right)}=\mathrm{E}\left\{ \hat{n}_{1}\right\} $
is the mean first recurrence time as measured in steps \citep{cinlar_markov_1975}.
Similarly, $\mathrm{E}\left\{ \hat{W}_{1}\right\} =p\mathrm{E}\left\{ w\right\} $,
so that $\nicefrac{\mathrm{E}\left\{ \hat{W}_{1}\right\} }{\mathrm{E}\left\{ \Delta\hat{T}_{1}\right\} }=\nicefrac{\mathrm{E}\left\{ w\right\} }{\mathrm{E}\left\{ \Delta T\right\} }$.

We are now equipped to meet our goal - to rewrite $\mathcal{V}_{\left(x,x^{\prime},\Delta T\right)\mapsto w\left(x,x^{\prime},\Delta T\right)}^{\left(t\right)}$
as a SVGR that can be computed via (\ref{eq:FV sum of variances}).
We first note that the definition of the TVGR for a general reward,
(\ref{eq: MRR variance}), covers the generalized SVGR case, $\mathcal{V}_{\left(x\rightarrow x^{\prime}\right)\mapsto f\left(x,x^{\prime}\right)}^{\left(n\right)}$,
considered in Section \ref{subsec: SVGR}. Namely, when\noun{ }the
Markov renewal process is defined so that the intervals $\Delta T_{k}$
are always $1$, 
\begin{eqnarray}
\mathcal{V}_{\left(x\rightarrow x^{\prime}\right)\mapsto f\left(x,x^{\prime}\right)}^{\left(n\right)} & = & \mathcal{V}_{\left(x,x^{\prime},\Delta T\right)\mapsto f\left(x,x^{\prime}\right)}^{\left(t\right)}\nonumber \\
 & = & \frac{1}{\mathrm{E}\left\{ \hat{n}_{1}\right\} }\mathrm{var}\left\{ \hat{F}_{1}-\mathrm{E}\left\{ f\right\} \hat{n}_{1}\right\} ,\label{eq: funcVar from renewal}
\end{eqnarray}
where $\Delta\hat{T}_{1}$ is replaced by $\hat{n}_{1}$ and $\hat{F}_{1}=\sum_{k=1}^{\hat{n}_{1}}f\left(x_{k},x_{k+1}\right)$
is determined from $f$ like $\hat{W}_{1}$ is from $w$. Now consider
the randomly-valued function 
\[
f_{w}\left(x,x^{\prime}\right)\equiv w\left(x,x^{\prime},\Delta t\left(x,x^{\prime}\right)\right)-\frac{\mathrm{E}\left\{ w\right\} }{\mathrm{E}\left\{ \Delta T\right\} }\Delta t\left(x,x^{\prime}\right),
\]
where $\Delta t\left(x,x^{\prime}\right)$ is such that $\Delta t\left(x_{k-1},x_{k}\right)$
has the same distribution as $\Delta T_{k}$ conditioned on $\text{x}_{k-1}$
and $x_{k}$. Note that 
\begin{eqnarray*}
\hat{W}_{1}-\frac{\mathrm{E}\left\{ w\right\} }{\mathrm{E}\left\{ \Delta T\right\} }\Delta\hat{T}_{1} & = & \sum_{k=1}^{\hat{n}_{1}}\left[w\left(x_{k-1},x_{k},\Delta T_{k}\right)-\frac{\mathrm{E}\left\{ w\right\} }{\mathrm{E}\left\{ \Delta T\right\} }\Delta T_{k}\right]\\
\hat{F}_{w1}-\mathrm{E}\left\{ f_{w}\right\} \hat{n}_{1} & = & \sum_{k=1}^{\hat{n}_{1}}\left[f_{w}\left(x_{k-1},x_{k}\right)-\mathrm{E}\left\{ f_{w}\right\} \right].
\end{eqnarray*}
And, since $\mathrm{E}\left\{ f_{w}\right\} =0$, $f_{w}\left(x_{k-1},x_{k}\right)-\mathrm{E}\left\{ f_{w}\right\} $
has the same distribution as $w\left(x_{k-1},x_{k},\Delta T_{k}\right)-\frac{\mathrm{E}\left\{ w\right\} }{\mathrm{E}\left\{ \Delta T\right\} }\Delta T_{k}$
when each is conditioned on $x_{k-1}$ and $x_{k}$. In turn, since
$\hat{n}_{1}$ is completely determined by the sequence of states,
$x_{0},x_{1},\ldots$, the sums $\hat{F}_{w1}-\mathrm{E}\left\{ f_{w}\right\} \hat{n}_{1}$
and $\hat{W}_{1}-\frac{\mathrm{E}\left\{ w\right\} }{\mathrm{E}\left\{ \Delta T\right\} }\Delta\hat{T}_{1}$
conditioned on $x_{0},x_{1},\ldots$ have the same mean and variance.
It then follows from the law of total variance that $\mathrm{var}\left\{ \hat{F}_{w1}-\mathrm{E}\left\{ f_{w}\right\} \hat{n}_{1}\right\} =\mathrm{var}\left\{ \hat{W}_{1}-\frac{\mathrm{E}\left\{ w\right\} }{\mathrm{E}\left\{ \Delta T\right\} }\Delta\hat{T}_{1}\right\} $.
Thus, comparing (\ref{eq: MRR variance from renewal}) and (\ref{eq: funcVar from renewal})
yields (\ref{eq: Markov renewal variance formula}).

As mentioned earlier, the reasoning Cinlar and we present regarding
the FRLs - in particular the hatted quantities that appear in (\ref{eq: MRR variance from renewal})
and subsequent equations - are sensical only for Markov chains with
finitely-many and perhaps countably-many states. However, the rough
argument by analogy presented in Section \ref{subsec: SI - TVGR Formula by Analogy}
does not require a restriction of the state-space and yet yields the
same result, suggesting a more general validity.

\subsection{Decomposition of the TVGR\label{subsec: SI - TVGR Decomposition}}

In this section we expand the temporal variance growth rate, $\mathcal{V}_{E}^{\left(t\right)}=\mathcal{V}_{\left(x\rightarrow x^{\prime}\right)\mapsto\left(1_{E}\left(x\right)-\mathcal{E}\Delta t\left(x,x^{\prime}\right)\right)}$,
to produce (\ref{eq:FanoVar decomposed}). From (\ref{eq:FV sum of variances}),
we have
\[
\mathcal{V}_{E}^{\left(t\right)}=\mathrm{var}\left\{ 1_{E,0}-\mathcal{E}\Delta t_{0}\right\} +2\sum_{k=1}^{\infty}\mathrm{cov}\left\{ 1_{E,0}-\mathcal{E}\Delta t_{0},1_{E,k}-\mathcal{E}\Delta t_{k}\right\} ,
\]
where $1_{E,k}\equiv1_{E}\left(x_{k}\right)$ and $\Delta t_{k}=\Delta t\left(x_{k},x_{k+1}\right)$.
Note that since $1_{E}\left(x^{\prime}\right)-\mathcal{E}\Delta t\left(x,x^{\prime}\right)$
is a scalar function, we can and have combined the two sums that appear
in (\ref{eq:FV sum of variances}). Then by the standard sum properties
of variance and covariance,
\[
\mathrm{var}\left\{ 1_{E,0}-\mathcal{E}\Delta t_{0}\right\} =\mathrm{var}\left\{ 1_{E,0}\right\} -2\mathcal{E}\mathrm{cov}\left\{ 1_{E,0},\Delta t_{0}\right\} +\mathcal{E}^{2}\mathrm{var}\left\{ \Delta t_{0}\right\} 
\]
and
\[
\mathrm{cov}\left\{ 1_{E,0}-\mathcal{E}\Delta t_{0},1_{E,k}-\mathcal{E}\Delta t_{k}\right\} =\mathrm{cov}\left\{ 1_{E,0},1_{E,k}\right\} -\mathcal{E}\mathrm{cov}\left\{ 1_{E,0},\Delta t_{k}\right\} -\mathcal{E}\mathrm{cov}\left\{ \Delta t_{0},1_{E,k}\right\} +\mathcal{E}^{2}\mathrm{cov}\left\{ \Delta t_{0},\Delta t_{k}\right\} 
\]
The first terms in each of the above two lines give
\[
\mathrm{var}\left\{ 1_{E,0}\right\} +2\sum_{k=1}^{\infty}\mathrm{cov}\left\{ 1_{E,0},1_{E,k}\right\} =\mathcal{V}_{x\mapsto1_{E}\left(x\right)}^{\left(n\right)},
\]
and similarly the last terms give $\mathcal{E}^{2}\mathcal{V}_{\left(x\rightarrow x^{\prime}\right)\mapsto\Delta t\left(x,x^{\prime}\right)}^{\left(n\right)}$.
The remaining terms - those with pre-factors of $\mathcal{E}$ - are
\[
-2\mathcal{E}\left[\mathrm{cov}\left\{ 1_{E,0},\Delta t_{0}\right\} +\sum_{k=1}^{\infty}\mathrm{cov}\left\{ 1_{E,0},\Delta t_{k}\right\} +\sum_{k=1}^{\infty}\mathrm{cov}\left\{ \Delta t_{0},1_{E,k}\right\} \right]=-2\mathcal{E}\,\mathcal{CV}_{\left(x\rightarrow x^{\prime}\right)\mapsto1_{E}\left(x^{\prime}\right),\left(x\rightarrow x^{\prime}\right)\mapsto\Delta t\left(x,x^{\prime}\right)}^{\left(n\right)},
\]
using the definition of $\mathcal{CV}$ given in (\ref{eq:funcCV sum of variances}).
Together, these give the decomposition cited in (\ref{eq:FanoVar decomposed}).

\section{Averaged System in Phase-Isostable Coordinates\label{sec: SI - Averaging + Phase-Isostable}}

In order to capture the linearized Poincare map dynamics, we make
use of the phase and isostable coordinates \citep{wilson_isostable_2016},
which we find reflect the oscillator dynamics simply and facilitate
a comparison to the standard phase reduction. Here in Section \ref{subsec: SI - Phase-Isostable Coordinates}
we discuss the phase-isostable coordinate system, and in Section \ref{subsec: SI - Averaging}
we introduce a form of averaging to further simplify the equations.

\subsection{Phase-Isostable Coordinates\label{subsec: SI - Phase-Isostable Coordinates}}

The phase-isostable coordinates are based on isochrons, equi-phase
surfaces that foliate the basin of attraction of the limit cycle.
The isochrons can be thought of as level sets of the so-called asymptotic
phase function $\Phi\left(\vec{y}\right)$, which extends the definition
of a phase coordinate $\phi$ on the limit cycle to the basin of attraction.
In turn, the isostable coordinates $\psi_{i}$, defined via level
sets of functions $\Psi_{i}\left(\vec{y}\right)$, parameterize the
isochrons in such a way that $\Psi_{i}\left(\vec{y}\right)=0$ identifies
the limit cycle. Thus, the $\psi_{i}=\Psi_{i}\left(\vec{y}\right)$
can be thought of as amplitude coordinates. $\Phi$ and $\Psi_{i}$
are defined so that that flows $\vec{y}_{0}\left(t\right)$ of the
unperturbed system (i.e. (\ref{eq: system}) with $D_{\textit{in}}=0$) satisfy
\begin{eqnarray}
\frac{d}{dt}\Phi\left(\vec{y}_{0}\left(t\right)\right) & = & 1\label{eq: unperturbed_phase}\\
\frac{d}{dt}\Psi_{i}\left(\vec{y}_{0}\left(t\right)\right) & = & -\kappa_{i}\Psi_{i}\left(\vec{y}_{0}\left(t\right)\right).\label{eq: unperturbed_isostable}
\end{eqnarray}
Note we have implicitly defined time such that the limit cycle has
period $1$ in the absence of perturbations; the phase $\phi$ goes
through a unit interval in a complete cycle. Note also that when the
constants $\kappa_{i}$ have positive real parts, the isostable functions
relax towards the limit cycle given by $\psi_{i}=0$ as $t\rightarrow\infty$.
This reflects the fact that the limit cycle is stable. In this work,
we will limit our consideration to the cases where $\kappa_{i}$ are
real and positive, in which case the functions $\Psi_{i}$ are also
real-valued. We refer the reader to \citet{wilson_isostable_2016}
for a detailed introduction to isostable coordinates and to \citet{wilson_isostable_2019}
for a treatment of the situations where some $\kappa_{i}$ are complex.
We also note that, since they satisfy the linear (\ref{eq: unperturbed_isostable}),
the isostable functions $\Psi_{i}$ are eigenfunctions of the Koopman
operator. But not all Koopman eigenfunctions produce isostable coordinates.
Isostables correspond in particular to the so-called ``principal''
eigenfunctions, which are non-degenerate: they have non-zero gradients
on the limit cycle. See \citet{kvalheim_existence_2021} for a rigorous
investigation of when the principal eigenfunctions exist and are unique.

The non-degeneracy of the isostables comes into play when we consider
the noisy perturbative forcing of the oscillator (\ref{eq: system}).
In the change of coordinates from $\vec{y}$ to phase and isostables,
the gradients of $\Phi$ and $\Psi_{i}$ gives the responses of the
phase and isostables to the forcing:
\begin{eqnarray*}
d\phi & = & dt+\sigma\left(\vec{\nabla}\Phi\left(\vec{y}\right)\right)^{T}\mathbf{G}\left(\vec{y}\right)\,d\vec{W}+\mathcal{O}\left(\sigma^{2}\right)\\
d\psi_{i} & = & -\kappa_{i}\psi_{i}\,dt+\sigma\left(\vec{\nabla}\Psi_{i}\left(\vec{y}\right)\right)^{T}\mathbf{G}\left(\vec{y}\right)\,d\vec{W}+\mathcal{O}\left(\sigma^{2}\right)\,\quad i=1\ldots d\,,
\end{eqnarray*}
where $\sigma\equiv\sqrt{2D_{\textit{in}}}.$ (\ref{eq: phase-isostable phase})
and (\ref{eq: phase-isostable isostable}) then follow from assumption
that the deviation $\vec{\psi}\equiv\left(\psi_{1},\psi_{2},\ldots,\psi_{d}\right)$
from the limit cycle is $\mathcal{O}\left(\sigma\right)$, and, therefore,
at first order approximation the gradients can be replaced with their
values on the limit cycle. Thus, in (\ref{eq: phase-isostable phase})
and (\ref{eq: phase-isostable isostable}),
\begin{eqnarray*}
\vec{Z}\left(\phi\right) & = & \left.\vec{\nabla}\Phi\left(\vec{y}\right)\right|_{\vec{y}_{\textit{LC}}\left(\phi\right)}\\
\vec{Y}_{i}\left(\phi\right) & = & \left.\vec{\nabla}\Psi_{i}\left(\vec{y}\right)\right|_{\vec{y}_{\textit{LC}}\left(\phi\right)},
\end{eqnarray*}
where $\vec{y}_{\textit{LC}}\left(\phi\right)$ is the point on the limit cycle
with phase $\phi$. Note that, since we work to first order in $\sigma$,
the result is the same under either the Ito or Stratonovich interpretation
of (\ref{eq: system}).

\subsection{Averaging\label{subsec: SI - Averaging}}

For small $\sigma$, we expect that the accumulation of the noisy
perturbation is slow in comparison with the fast oscillation of $\phi$.
Therefore, within the first order of approximation, we use a form
of averaging to simplify the $\phi$-dependent terms $\vec{Z}\left(\phi\right)^{T}\mathbf{G}\left(\phi\right)$
and $\vec{Y}_{i}^{T}\left(\phi\right)\mathbf{G}_{\textit{LC}}\left(\phi\right)$
that appear in (\ref{eq: phase-isostable phase}) and (\ref{eq: phase-isostable isostable}).
The end result of our averaging are the equations
\begin{eqnarray}
d\phi & = & dt+\sigma\vec{z}_{G}^{T}\,d\vec{W}\left(t\right)+h.o.t.\label{eq: phase_averaged}\\
d\psi_{i} & = & -\kappa_{i}\psi_{i}\,dt+\sigma\mathrm{e}^{\kappa_{i}\left(1-\mathrm{mod}_{1}\phi\right)}\vec{y}_{G,i}^{T}\,d\vec{W}\left(t\right)+h.o.t.,\label{eq: isostable_with_y_redef}
\end{eqnarray}
where $\mathrm{mod}_{1}\left(\phi\right)=\phi-\left\lfloor \phi\right\rfloor $,
a $1$-periodic function, is the fractional part of $\phi$. In Section
\ref{subsec: SI - Averaged PRC and IRCs} we derive expressions for
the constant vectors $\vec{z}_{G}$ and $\vec{y}_{G,i}$, and in Section
\ref{subsec: SI - Constant IRC Limitations} we explain the (admittedly
odd) appearance of $\mathrm{e}^{-\kappa_{i}\mathrm{mod}_{1}\left(\phi\right)}$
in (\ref{eq: isostable_with_y_redef}). We consider a limiting case
in Section \ref{subsec: SI - QR Averages} and offer a brief discussion
of this unusual form of averaging in Section \ref{subsec: SI - Averaging Discussion}.

\subsubsection{Averaged PRC and IRCs\label{subsec: SI - Averaged PRC and IRCs}}

In order to apply averaging, we first rewrite the equations for $\psi_{i}$,
(\ref{eq: phase-isostable isostable}), using an integrating factor,
\begin{equation}
d\left(\mathrm{e}^{\kappa_{i}t}\psi_{i}\right)=\sigma\mathrm{e}^{\kappa_{i}t}\vec{Y}_{i}\left(\phi\right)^{T}\mathbf{G}_{\textit{LC}}\left(\phi\right)\,d\vec{W}\left(t\right)+h.o.t.,\label{eq: psi_integrating_factor}
\end{equation}
where, as in Section \ref{sec:Fano-Variance-for-Limit-Cycle}, the
higher order terms ($h.o.t.$'s) are $\mathcal{O}\left(\sigma^{2},\sigma\left|\vec{\psi}\right|,\left|\vec{\psi}\right|^{2}\right)$.
Averaging (\ref{eq: isostable with time}) is non-trivial, since $\mathrm{e}^{\kappa_{i}t}$
is non-periodic, and we will in fact not average in the usual, long-time-scale
sense. We will be interested in the solutions $\phi\left(t\right)$
and $\psi_{i}\left(t\right)$ for times $t\in\mathbb{Z}$ that are
roughly the times of crossings of the Poincare section. So, instead,
by ``averaging'' we here mean replacing $\vec{Z}\left(\phi\right)^{T}\mathbf{G}_{\textit{LC}}\left(\phi\right)$
and $\vec{Y}_{i}\left(\phi\right)^{T}\mathbf{G}_{\textit{LC}}\left(\phi\right)$
with the constant real vector $\vec{z}_{G}^{T}$ and a simplified
expression $\mathrm{e}^{\kappa_{i}\left(1-\mathrm{mod}_{1}\phi\right)}\vec{y}_{G,i}^{T}$
(their respective ``averages'') in such a way that the distribution
of the solutions $\phi$ and $\psi_{i}$ remains unchanged at integer
values of $t$.

Consider initial conditions $\phi\left(0\right)=0+\mathcal{O}\left(\sigma,\left|\vec{\psi}\right|\right)$
and $\psi_{i}\left(0\right)=\psi_{i,0}$, which will be relevant to
the Poincare map dynamics. For both the unaveraged and averaged equations,
$\phi=t+\mathcal{O}\left(\sigma,\left|\vec{\psi}\right|\right)$.
From the unaveraged equations (\ref{eq: phase-isostable phase}) and
(\ref{eq: psi_integrating_factor}), we obtain linear, time-varying
equations,
\begin{eqnarray}
d\phi & = & dt+\sigma\vec{Z}\left(t\right)^{T}\mathbf{G}_{\textit{LC}}\left(t\right)\,d\vec{W}\left(t\right)+h.o.t.\label{eq: phase with time}\\
d\left(\mathrm{e}^{\kappa_{i}t}\psi_{i}\right) & = & \sigma\mathrm{e}^{\kappa_{i}t}\vec{Y}_{i}\left(t\right)^{T}\mathbf{G}_{\textit{LC}}\left(t\right)\,d\vec{W}\left(t\right)+h.o.t.,\label{eq: isostable with time}
\end{eqnarray}
at first order approximation. Similarly, from the averaged equations
(\ref{eq: phase_averaged}) and (\ref{eq: isostable_with_y_redef}),
\begin{eqnarray}
d\phi & = & dt+\sigma\,\vec{z}_{G}^{T}d\vec{W}\left(t\right)+h.o.t.\label{eq: phase averaged with time}\\
d\left(\mathrm{e}^{\kappa_{i}t}\psi_{i}\right) & = & \sigma\mathrm{e}^{\kappa_{i}\left(t+1-\mathrm{mod}_{1}\left(t\right)\right)}\vec{y}_{G,i}^{T}\,d\vec{W}\left(t\right)+h.o.t.\nonumber \\
 & = & \sigma\mathrm{e}^{\kappa_{i}\left(\left\lfloor t\right\rfloor +1\right)}\vec{y}_{G,i}^{T}\,d\vec{W}\left(t\right)+h.o.t.\label{eq: isostable averaged with time}
\end{eqnarray}
The solutions of both the unaveraged system (\ref{eq: phase with time})
and (\ref{eq: isostable with time}),
\begin{eqnarray}
\phi\left(t\right) & = & \phi_{0}+t+\sigma\int_{0}^{t}\vec{Z}\left(s\right)^{T}\mathbf{G}_{\textit{LC}}\left(s\right)d\vec{W}\left(s\right)+h.o.t.\label{eq: phi solution with time}\\
\mathrm{e}^{\kappa_{i}t}\psi_{i}\left(t\right) & = & \psi_{i,0}+\sigma\int_{0}^{t}\mathrm{e}^{\kappa_{i}s}\vec{Y}_{i}\left(s\right)^{T}\mathbf{G}_{\textit{LC}}\left(s\right)d\vec{W}\left(s\right)+h.o.t.,\label{eq: psi solution with time}
\end{eqnarray}
and the averaged system (\ref{eq: phase averaged with time}) and
(\ref{eq: isostable averaged with time}), 
\begin{eqnarray}
\phi\left(t\right) & = & \phi_{0}+t+\sigma\vec{z}_{G}^{T}\vec{W}\left(t\right)+h.o.t.\label{eq: phi solution averaged}\\
\mathrm{e}^{\kappa_{i}t}\psi_{i}\left(t\right) & = & \psi_{i,0}+\sigma\int_{0}^{t}\mathrm{e}^{\kappa_{i}\left(\left\lfloor s\right\rfloor +1\right)}\vec{y}_{G,i}^{T}d\vec{W}\left(s\right)+h.o.t.,\label{eq: psi solution averaged}
\end{eqnarray}
are Gaussian processes at lowest order. Both sets of equations are
in agreement already about the means, $\mathrm{E}\left\{ \phi\left(t\right)\right\} =t$
and $\mathrm{E}\left\{ \psi_{i}\left(t\right)\right\} =\mathrm{e}^{-\kappa_{i}t}\psi_{i,0}$.
What remains is to match the variances and covariances, $\mathrm{var}\left\{ \phi\left(t\right)\right\} $,
$\mathrm{cov}\left\{ \phi\left(t\right),\psi_{i}\left(t\right)\right\} $,
and $\mathrm{cov}\left\{ \psi_{i}\left(t\right),\psi_{j}\left(t\right)\right\} $,
between the two sets of solutions.

It is sufficient to match after one period, i.e. at $t=1$: since
$\mathbf{G}_{\textit{LC}}\left(s\right)$, $\vec{Y}_{i}\left(s\right)$ and
$\vec{Z}\left(s\right)$ are $1$-periodic, their influence on the
(co)variances from $t=0$ to $t=1$ is similar to that between $t=l$
and $t=l+1$ for $l\in\mathbb{Z}$. Indeed, the (co)variances at integer
values of $t$ only depend on their values at $t=1$ and a time-dependent
prefactor that depends parametrically at most on $\kappa_{i}$. For
$\mathrm{var}\left\{ \phi\left(t\right)\right\} $, this is easy to
see, since the amount of variance that accumulated in each period
is the same:
\begin{eqnarray*}
\mathrm{var}\left\{ \phi\left(t\right)\right\}  & = & \sigma^{2}\int_{0}^{t}\vec{Z}\left(s\right)^{T}\mathbf{G}_{\textit{LC}}\left(s\right)\mathbf{G}_{\textit{LC}}\left(s\right)^{T}\vec{Z}\left(s\right)ds\\
 & = & \sigma^{2}t\int_{0}^{1}\vec{Z}\left(s\right)^{T}\mathbf{G}_{\textit{LC}}\left(s\right)\mathbf{G}_{\textit{LC}}\left(s\right)^{T}\vec{Z}\left(s\right)ds\\
 & = & t\mathrm{var}\left\{ \phi\left(1\right)\right\} ,\;\left(t\in\mathbb{Z}\right),
\end{eqnarray*}
and so the aforementioned prefactor for $\mathrm{var}\left\{ \phi\left(t\right)\right\} $
is $t$. For $\mathrm{cov}\left\{ \phi\left(t\right),\psi_{i}\left(t\right)\right\} $,
the prefactor turns out to be $\frac{1-\mathrm{\mathrm{e}^{-\kappa_{i}t}}}{1-\mathrm{e}^{-\kappa_{i}}}$:
\begin{eqnarray*}
\mathrm{cov}\left\{ \phi\left(t\right),\psi_{i}\left(t\right)\right\}  & = & \sigma^{2}\mathrm{e}^{-\kappa_{i}t}\int_{0}^{t}\mathrm{e}^{\kappa_{i}s}\vec{Z}\left(s\right)^{T}\mathbf{G}_{\textit{LC}}\left(s\right)\mathbf{G}_{\textit{LC}}\left(s\right)^{T}\vec{Y}_{i}\left(s\right)ds\\
 & = & \sigma^{2}\mathrm{e}^{-\kappa_{i}t}\sum_{l=0}^{t-1}\mathrm{e}^{\kappa_{i}l}\int_{0}^{1}\mathrm{e}^{\kappa_{i}s}\vec{Z}\left(s\right)^{T}\mathbf{G}_{\textit{LC}}\left(s\right)\mathbf{G}_{\textit{LC}}\left(s\right)^{T}\vec{Y}_{i}\left(s\right)ds\\
 & = & \frac{1-\mathrm{\mathrm{e}^{-\kappa_{i}t}}}{1-\mathrm{e}^{-\kappa_{i}}}\mathrm{cov}\left\{ \phi\left(1\right),\psi_{i}\left(1\right)\right\} \;\left(t\in\mathbb{Z}\right).
\end{eqnarray*}
Similarly, $\mathrm{cov}\left\{ \psi_{i}\left(t\right),\psi_{j}\left(t\right)\right\} =\frac{1-\mathrm{e}^{-\left(\kappa_{i}+\kappa_{j}\right)t}}{1-\mathrm{e}^{-\left(\kappa_{i}+\kappa_{j}\right)}}\mathrm{cov}\left\{ \psi_{i}\left(1\right),\psi_{j}\left(1\right)\right\} $
for $t\in\mathbb{Z}$. Since the time-dependent prefactors $t$, $\frac{1-\mathrm{\mathrm{e}^{-\kappa_{i}t}}}{1-\mathrm{e}^{-\kappa_{i}}}$
and $\frac{1-\mathrm{e}^{-\left(\kappa_{i}+\kappa_{j}\right)t}}{1-\mathrm{e}^{-\left(\kappa_{i}+\kappa_{j}\right)}}$
are independent of $\mathbf{G}_{\textit{LC}}\left(t\right)$, $\vec{Y}_{i}\left(t\right)$
and $\vec{Z}\left(t\right)$, they are unchanged when $\vec{Y}_{i}\left(t\right)^{T}\mathbf{G}_{\textit{LC}}\left(t\right)$
and $\vec{Z}\left(t\right)^{T}\mathbf{G}_{\textit{LC}}\left(t\right)$ are
averaged. Therefore, we need only to match the (co)variances of the
unaveraged and averaged solutions at a single integer value of $t$.
We choose $t=1$.

At $t=1$, the expression for the averaged $\psi_{i}$, (\ref{eq: psi solution averaged}),
simplifies to 
\begin{equation}
\mathrm{e}^{\kappa_{i}}\psi_{i}\left(1\right)\sim\psi_{i,0}+\frac{\sigma}{\sqrt{\kappa_{i}}}\mathrm{e}^{\kappa_{i}}\vec{y}_{G,i}^{T}\vec{W}\left(s\right),\label{eq: psi solution averaged t=00003D1}
\end{equation}
and the (co)variances of interest for averaged solutions, (\ref{eq: phi solution averaged})
and (\ref{eq: psi solution averaged t=00003D1}) are 
\begin{eqnarray*}
\mathrm{var}\left\{ \phi\left(1\right)\right\}  & = & \sigma^{2}\vec{z}_{G}^{T}\vec{z}_{G}\\
\mathrm{cov}\left\{ \phi\left(1\right),\psi_{i}\left(1\right)\right\}  & = & \sigma^{2}\vec{z}_{G}^{T}\vec{y}_{G,i}\\
\mathrm{cov}\left\{ \psi_{i}\left(1\right),\psi_{j}\left(1\right)\right\}  & = & \sigma^{2}\vec{y}_{G,i}^{T}\vec{y}_{G,j}.
\end{eqnarray*}
Define $\sigma^{2}\mathbf{\Xi}$ as the covariance matrix for $\left(\phi\left(1\right);\psi_{1}\left(1\right);\psi_{2}\left(1\right);\ldots;\psi_{d}\left(1\right)\right)$
with entries $\sigma^{2}\xi_{i,j}$ as they follow from the unaveraged
solutions, (\ref{eq: phi solution with time}) and (\ref{eq: psi solution with time}).
Then for the averaged and unaveraged version of the (co)variances
to match, we require that
\begin{eqnarray}
\vec{z}_{G}^{T}\vec{z}_{G} & = & \xi_{1,1}\equiv\int_{0}^{1}\vec{Z}\left(s\right)^{T}\mathbf{G}_{\textit{LC}}\left(s\right)\mathbf{G}_{\textit{LC}}\left(s\right)^{T}\vec{Z}\left(s\right)ds\label{eq: zG-zG inner product}\\
\vec{z}_{G}^{T}\vec{y}_{G,i} & = & \xi_{1,i+1}\equiv\mathrm{e}^{-\kappa_{i}}\int_{0}^{1}\mathrm{e}^{\kappa_{i}s}\vec{Z}\left(s\right)^{T}\mathbf{G}_{\textit{LC}}\left(s\right)\mathbf{G}_{\textit{LC}}\left(s\right)^{T}\vec{Y}_{i}\left(s\right)ds\label{eq: zG-yG inner product}\\
\vec{y}_{G,i}^{T}\vec{y}_{G,j} & = & \xi_{i+1,j+1}\equiv\mathrm{e}^{-\left(\kappa_{i}+\kappa_{j}\right)}\int_{0}^{1}\mathrm{e}^{\left(\kappa_{i}+\kappa_{j}\right)s}\vec{Y}_{i}\left(s\right)^{T}\mathbf{G}_{\textit{LC}}\left(s\right)\mathbf{G}_{\textit{LC}}\left(s\right)^{T}\vec{Y}_{j}\left(s\right)ds,\label{eq: yG-yG inner product}
\end{eqnarray}
for $1\le i,j\leq d$. Equivalently, we require that $\mathbf{\Xi}$
can be decomposed as $\mathbf{\Xi}=\mathbf{R}_{G}^{T}\mathbf{R}_{G}$,
where
\begin{equation}
\mathbf{R}_{G}=\left(\begin{array}{ccccc}
\vec{z}_{G}, & \vec{y}_{G,1}, & \vec{y}_{G,2}, & \cdots, & \vec{y}_{G,d}\end{array}\right).\label{eq: Xi factor}
\end{equation}
Since $\mathbf{\Xi}$ is proportional to a covariance matrix, it is
positive semi-definite. The decomposition is therefore possible, but
it is not unique. $\mathbf{\Xi}$ is invariant under orthogonal transformations
$\mathbf{O}$ of $\mathbf{R}_{G}$, $\mathbf{R}_{G}\mapsto\mathbf{O}\mathbf{R}_{G}$.
But $\mathbf{O}$ only has the effect of rotating or reflecting the
independent components of $d\vec{W}\left(t\right)$. Indeed, as can
be seen in Section (\ref{sec: SI - TVGR Oscillator Derivation}) below,
the temporal variance growth rate only depends on $\mathbf{\Xi}$,
not on the specific choice of vectors $\vec{z}_{G}$ and $\vec{y}_{G,i}$.

\subsubsection{Limitations of a Constant IRC Average\label{subsec: SI - Constant IRC Limitations}}

We now justify the odd inclusion of $\mathrm{e}^{\kappa_{i}\left(1-\mathrm{mod}_{1}\phi\right)}$
in the average of $\vec{Y}_{i}\left(\phi\right)^{T}\mathbf{G}_{\textit{LC}}\left(\phi\right)$.
Consider an equivalence class of all $1$-periodic, vector-valued
functions $\vec{Y}_{i}\left(\phi\right)^{T}\mathbf{G}_{\textit{LC}}\left(\phi\right)$
that produce the same (co)variance statistics for $\phi$ and $\psi_{i}$.
In the averaging discussed above, we have chosen the $1$-periodic
vector $\mathrm{e}^{\kappa_{i}\left(1-\mathrm{mod}_{1}\phi\right)}\vec{y}_{G,i}$
as a simple representative of that equivalence class. Would a constant
vector $\vec{y}_{G}^{\textit{constant}}$ not be a simpler choice? In other
words, why can one not replace $\vec{Y}_{i}\left(\phi\right)^{T}\mathbf{G}_{\textit{LC}}\left(\phi\right)$
with a constant vector $\left(\vec{y}_{G}^{\textit{constant}}\right)^{T}$,
like we do for $\vec{Z}\left(\phi\right)^{T}\mathbf{G}_{\textit{LC}}\left(\phi\right)$?

As it turns out, constant averages can not capture the full range
of possible correlations between $\phi$ and the $\psi_{i}$. Consider
a planar oscillator where $\vec{Z}\left(\phi\right)^{T}\mathbf{G}_{\textit{LC}}\left(\phi\right)=\vec{z}_{G}$
and $\vec{Y}_{1}\left(\phi\right)^{T}\mathbf{G}_{\textit{LC}}\left(\phi\right)=\vec{y}_{G,1}^{\textit{constant}}$
are constant. The correlation coefficient between $\phi\left(1\right)$
and $\psi_{1}\left(1\right)$ is (following (\ref{eq: zG-zG inner product})-(\ref{eq: yG-yG inner product}))
\begin{eqnarray*}
\rho_{\phi\left(1\right),\psi_{1}\left(1\right)} & = & \frac{\mathrm{\mathrm{cov}\left\{ \phi\left(1\right),\psi_{1}\left(1\right)\right\} }}{\sqrt{\mathrm{var}\left\{ \phi\left(1\right)\right\} \mathrm{var}\left\{ \psi_{1}\left(1\right)\right\} }}\\
 & = & \frac{\mathrm{e}^{-\kappa_{1}}\int_{0}^{1}\mathrm{e}^{\kappa_{1}s}ds}{\sqrt{\left(\int_{0}^{1}ds\right)\left(\mathrm{e}^{-2\kappa_{1}}\int_{0}^{1}\mathrm{e}^{2\kappa_{1}s}ds\right)}}\frac{\vec{z}_{G}^{T}\vec{y}_{G,1}^{\textit{constant}}}{\left\Vert \vec{z}_{G}\right\Vert \left\Vert \vec{y}_{G,1}^{\textit{constant}}\right\Vert }\\
 & = & \sqrt{\frac{2}{\kappa_{1}}\frac{\mathrm{e}^{\kappa_{1}}-1}{\mathrm{e}^{\kappa_{1}}+1}}\frac{\vec{z}_{G}^{T}\vec{y}_{G,1}^{\textit{constant}}}{\left\Vert \vec{z}_{G}\right\Vert \left\Vert \vec{y}_{G,1}^{\textit{constant}}\right\Vert }.
\end{eqnarray*}
The second factor is the cosine of the angle between $\vec{z}_{G}$
and $\vec{y}_{G,1}^{\textit{constant}}$ and can take on values between $-1$
and $1$, but the first factor is strictly less than $1$ for $\kappa_{1}>0$.
The choice of constant averages, $\vec{z}_{G}$ and $\vec{y}_{G,1}^{\textit{constant}}$,
only accommodates correlation coefficients between $-\sqrt{\frac{2}{\kappa_{1}}\frac{\mathrm{e}^{\kappa_{1}}-1}{\mathrm{e}^{\kappa_{1}}+1}}$
and $\sqrt{\frac{2}{\kappa_{1}}\frac{\mathrm{e}^{\kappa_{1}}-1}{\mathrm{e}^{\kappa_{1}}+1}}$.
It would not be a problem if no choice of $\vec{Z}\left(\phi\right)^{T}\mathbf{G}_{\textit{LC}}\left(\phi\right)$
and $\vec{Y}_{1}\left(\phi\right)^{T}\mathbf{G}_{\textit{LC}}\left(\phi\right)$
yields a correlation coefficient outside of that range: $\vec{z}_{G}$
and $\vec{y}_{G,1}^{\textit{constant}}$ could be made to produce the same
statistics. But counterexamples are easy to find. Consider the simple
scenario where
\[
\vec{Z}\left(\phi\right)^{T}\mathbf{G}_{\textit{LC}}\left(\phi\right)=\vec{Y}_{1}\left(\phi\right)^{T}\mathbf{G}_{\textit{LC}}\left(\phi\right)=\left(\begin{array}{cc}
\sin\left(2\pi\phi\right) & 0\end{array}\right).
\]
The correlation coefficient is
\begin{eqnarray*}
\rho_{\phi\left(1\right),\psi_{1}\left(1\right)} & = & \frac{\mathrm{e}^{-\kappa_{1}}\int_{0}^{1}\mathrm{e}^{\kappa_{1}s}\sin^{2}\left(2\pi s\right)ds}{\sqrt{\left(\int_{0}^{1}\sin^{2}\left(2\pi\phi\right)ds\right)\left(\mathrm{e}^{-2\kappa_{1}}\int_{0}^{1}\mathrm{e}^{2\kappa_{1}s}\sin^{2}\left(2\pi s\right)ds\right)}}\\
 & = & \frac{8\pi\sqrt{4\pi^{2}+\kappa_{1}^{2}}}{16\pi^{2}+\kappa_{1}^{2}}\sqrt{\frac{2}{\kappa_{1}}\frac{\mathrm{e}^{\kappa_{1}}-1}{\mathrm{e}^{\kappa_{1}}+1}}
\end{eqnarray*}
which exceeds $\sqrt{\frac{2}{\kappa_{1}}\frac{\mathrm{e}^{\kappa_{1}}-1}{\mathrm{e}^{\kappa_{1}}+1}}$
when $\kappa_{1}$ is between $0$ and about $17.8$.

Using the averages $\vec{z}_{G}$ and $\mathrm{e}^{\kappa_{i}\left(1-\mathrm{mod}_{1}\phi\right)}\vec{y}_{G,i}$,
the correlation coefficient $\rho_{\phi\left(1\right),\psi_{i}\left(1\right)}$
is - by design - just the cosine of the angle between $\vec{z}_{G}$
and $\vec{y}_{G,i}$. Similarly, $\rho_{\psi_{i}\left(1\right),\psi_{j}\left(1\right)}$
is the cosine of the angle between $\vec{y}_{G,i}$ and $\vec{y}_{G,j}$.
The averages $\vec{z}_{G}$ and $\mathrm{e}^{\kappa_{i}\left(1-\mathrm{mod}_{1}\phi\right)}\vec{y}_{G,i}$
thus cover all possible correlation values and can be made to represent
any $\vec{Z}\left(\phi\right)^{T}\mathbf{G}_{\textit{LC}}\left(\phi\right)$
and $\vec{Y}_{i}\left(\phi\right)^{T}\mathbf{G}_{\textit{LC}}\left(\phi\right)$.

\subsubsection{Averages in the Quasi-Renewal Limit\label{subsec: SI - QR Averages}}

The analysis of the temporal variance growth rate is greatly simplified
in the $\Lambda\rightarrow0$ ($\kappa_{i}\rightarrow\infty$) limit.
Here we discuss the behavior of $\vec{z}_{G}^{T}\vec{z}_{G}$, $\vec{z}_{G}^{T}\vec{y}_{G,i}$,
and $\vec{y}_{G,i}^{T}\vec{y}_{G,j}$ (given in (\ref{eq: zG-zG inner product})-(\ref{eq: yG-yG inner product}))
in that ``quasi-renewal'' limit. $\vec{z}_{G}^{T}\vec{z}_{G}$ does
not depend on $\kappa_{i}$ and is unchanged. $\vec{z}_{G}^{T}\vec{y}_{G,i}$
and $\vec{y}_{G,i}^{T}\vec{y}_{G,j}$ on the other hand both go to
$0$ as $\kappa_{i}\rightarrow\infty$. The asymptotic form of the
integrals in (\ref{eq: zG-yG inner product}) and (\ref{eq: yG-yG inner product})
can be found via Laplace's method (see e.g. Section 6.4 of \citet{bender_advanced_1999}
for a pedagogical treatment). Both $\vec{z}_{G}^{T}\vec{y}_{G,i}$
and $\vec{y}_{G,i}^{T}\vec{y}_{G,j}$ are of the form
\[
I\left(\kappa\right)=\int_{0}^{1}\mathrm{e}^{\kappa\left(s-1\right)}g\left(s\right)ds
\]
where $g\left(s\right)$ is $1$-periodic and $\kappa$ play the role
of $\kappa_{i}$ or $\kappa_{i}+\kappa_{j}$. Since $\mathrm{e}^{\kappa\left(s-1\right)}$
is maximized at $s=1$, we replace $g\left(s\right)$ with $g\left(1\right)=g\left(0\right)$
at first order approximation. To the same order of approximation,
since the integrand is exponentially small for $s<1$, the integral
can be taken over $s$ from $-\infty$ to $1$, yielding
\[
I\left(\kappa\right)\sim\left(\int_{-\infty}^{1}\mathrm{e}^{\kappa\left(s-1\right)}\right)g\left(0\right)=\frac{1}{\kappa}g\left(0\right).
\]
For $\vec{z}_{G}^{T}\vec{y}_{G,i}$ and $\vec{y}_{G,i}^{T}\vec{y}_{G,j}$,
this becomes (cf. (\ref{eq: zG-yG inner product}) and (\ref{eq: yG-yG inner product}))
\begin{eqnarray*}
\vec{z}_{G}^{T}\vec{y}_{G,i} & \sim & \frac{1}{\kappa_{i}}\vec{Z}\left(0\right)^{T}\mathbf{G}_{\textit{LC}}\left(0\right)\mathbf{G}_{\textit{LC}}\left(0\right)^{T}\vec{Y}_{i}\left(0\right)\\
\vec{y}_{G,i}^{T}\vec{y}_{G,j} & \sim & \frac{1}{\kappa_{i}+\kappa_{j}}\vec{Y}_{i}\left(0\right)^{T}\mathbf{G}_{\textit{LC}}\left(0\right)\mathbf{G}_{\textit{LC}}\left(0\right)^{T}\vec{Y}_{j}\left(0\right).
\end{eqnarray*}
In particular, in a planar oscillator, where there is a single vector
$\vec{y}_{G}$ and $\Lambda=\mathrm{e}^{-\kappa}$ is a scalar, $\left\Vert \vec{y}_{G}\right\Vert ^{2}$
goes like $-\frac{1}{\log\Lambda}$, a fact the we make use of in
Section \ref{subsec:The-Quasi-Renewal-Case}.

\subsubsection{Some Comments About Averaging\label{subsec: SI - Averaging Discussion}}

As we discussed above, averaging in the usual long-time-scale sense
is not desirable, since - in deriving the Poincare map dynamics -
we integrate (\ref{eq: phase_averaged}) and (\ref{eq: isostable_with_y_redef})
over an $\mathcal{O}\left(1\right)$ time interval. The $\phi$-dependent
factor $\mathrm{e}^{\kappa_{i}\left(1-\mathrm{mod}_{1}\phi\right)}$
is required to capture the joint statistics of $\phi\left(t\right)$
and $\psi_{i}\left(t\right)$ at $t\sim1$, and, in some sense, $\vec{z}_{G}^{T}$
and $\mathrm{e}^{\kappa_{i}\left(1-\mathrm{mod}_{1}\phi\right)}\vec{y}_{G,i}^{T}$
are the simplest PRC and IRCs that can reproduce the statistics from
any given oscillator model. This reflects the fact that additional
time scales introduced by the appearance of $\kappa_{i}$ in the isostable
equations, (\ref{eq: phase-isostable isostable}), remain relevant
over the $\mathcal{O}\left(1\right)$ time interval. In fact, the
initial condition $\phi\left(0\right)\sim0$ also remains relevant.
$\vec{y}_{G,i}$ not only varies with $\kappa_{i}$ but also depends
on our choice of where $\phi=0$: a redefinition of $\phi$ by a shift
would not leave $\vec{y}_{G,i}$ invariant (see (\ref{eq: zG-yG inner product})
and (\ref{eq: yG-yG inner product})). Since we choose $\phi=0$ to
identify the point where the Poincare section intersects the limit
cycle, this ultimately means that the averaged IRCs $\vec{y}_{G,i}$
depend the location of that intersection.

The standard phase reduction with the standard averaging, (\ref{eq: phase})
and (\ref{eq: linear phase diffusion}), follows under the assumption
of strong contraction to the limit cycle ($\kappa_{i}\gg1$). In that
case $\psi_{i}\sim0$ and can be ignored. Note that the phase diffusion
$D_{\textit{phase}}=\vec{z}_{G}^{T}\vec{z}_{G}D_{\textit{in}}$ is invariant under shifts
of the phase variable (see (\ref{eq: zG-zG inner product})), and
it is thus independent of where the Poincare section intersects the
limit cycle.

We must also note that the averaging step we have presented here is
not strictly necessary in our derivation of the event temporal variance
growth rate, which could follow directly from the linearized, but
unaveraged system, (\ref{eq: phase-isostable phase}) and (\ref{eq: phase-isostable isostable}).
But, by averaging, we reveal what minimal, extra information about
the oscillator beyond the standard phase reduction is required to
capture events statistically in the long-time limit. The averaged
quantities give an interpretation to the oscillator-dependent parameters
that appear in the formula for the TVGR (\ref{eq:FanoVar Poincare}):
$\vec{b}$ roughly measures the overlap between the averaged PRC $\vec{z}_{G}$
and each of the averaged IRCs $\vec{y}_{G,i}$ (see (\ref{eq:b_general})),
and $c$ measures the magnitude of the PRC (see (\ref{eq:c_general})).

\section{Temporal Variance Growth Rate for Limit-cycle Oscillators\label{sec: SI - TVGR Oscillator Derivation}}

Here we detail the derivation of the TVGR for limit cycle oscillators,
starting from the averaged equations, (\ref{eq: phase_averaged})
and (\ref{eq: isostable_with_y_redef}). In Section \ref{subsec: SI - Linearized Map},
we adapt the derivation given by \citet{hitczenko_poincare_2013}
of the linearized Poincare map and first passage time, (\ref{eq: linearized PM}),
for the averaged system. We derive the associated joint distributions
and statistics for the positions on the Poincare section and the first
passage times in Section \ref{subsec: SI - Steady-state Densities}.
In Section \ref{subsec: SI - TVGR,  term-by-term}, we use those distributions
and statistics to produce an expression for the temporal variance
growth rate (\ref{eq:FanoVar Poincare}).

\subsection{Linearized Stochastic Poincare Map\label{subsec: SI - Linearized Map}}

As in \citet{hitczenko_poincare_2013} (hereafter referred to as ``HM''),
our derivation of the first passage time and linearized Poincare map,
(\ref{eq: linearized PM}), follows from the approximate solution
in a moving coordinates representation of the noise-driven limit cycle
oscillator. While HM use amplitude coordinates $\vec{\rho}$ that
are orthogonal to the limit cycle, we use isostable coordinates that
parameterize isochrons, which in general are not orthogonal to the
limit cycle. We first contrast the two choices, noting the simplicity
of the system in isostable coordinates. In the moving orthogonal coordinates,
the system to lowest order is (Equations 2.8-2.9 of HM)
\begin{eqnarray}
d\theta & \sim & \left[1+\vec{a}\left(\theta\right)^{T}\vec{\rho}\right]\,dt+\sigma\vec{h}\left(\theta\right)^{T}\,d\vec{W}\left(t\right)\label{eq: moving orthogonal}\\
d\vec{\rho} & \sim & \mathbf{R}\left(\theta\right)\vec{\rho}+\sigma\mathbf{H}\left(\theta\right)^{T}d\vec{W}\left(t\right),\label{eq: moving orthogonal 2}
\end{eqnarray}
while in isostable coordinates with averaging, it is (\ref{eq: phase_averaged})
and (\ref{eq: isostable_with_y_redef}), or
\begin{eqnarray}
d\phi & \sim & dt+\sigma\vec{z}_{G}^{T}\,d\vec{W}\left(t\right)\label{eq: phase averaged vec}\\
d\vec{\psi} & \sim & -\mathbf{K}\vec{\psi}\,dt+\sigma\mathrm{e}^{\mathbf{K}\left(1-\mathrm{mod}_{1}\phi\right)}\mathbf{Y}_{G}^{T}\,d\vec{W}\left(t\right),\label{eq: isostable averaged vec}
\end{eqnarray}
where $\sigma\equiv\sqrt{2D_{\textit{in}}}$, $\mathbf{K}$ is the diagonal
matrix with entries $\kappa_{i}$, and $\mathbf{Y}_{G}$ is the matrix
with columns $\vec{y}_{G,i}$. Note the inclusion of the vector $\vec{a}$
in (\ref{eq: moving orthogonal}) that captures shear in the flow
in the vicinity of the limit cycle. Since the shape of the isochrons
takes the shear into account, no such term appears in phase equation
(\ref{eq: phase averaged vec}) under phase-isostable coordinates.
The dynamics of the amplitude coordinates are also simpler. In the
absence of perturbative inputs, the isostable coordinates are decoupled,
evolving by a constant diagonal matrix $\mathbf{K}$ (cf. the generally
non-diagonal $\mathbf{R}\left(\theta_{t}\right)$ in (\ref{eq: moving orthogonal})
that varies with the phase variable $\theta$). Since we apply averaging
beforehand, the noise terms in (\ref{eq: phase averaged vec}) and
(\ref{eq: isostable averaged vec}) are simplified. We note that over
the course of HM's derivation, the averaging calculation presented
in SI Section \ref{subsec: SI - Averaging}, as it applies to the
time interval $t\in\left[0,1\right]$, is done implicitly. In the
end, the results we obtain by averaging first are equivalent to those
found by applying HM's work to the unaveraged system (\ref{eq: phase-isostable isostable}).
Formally, the averaged system, (\ref{eq: phase averaged vec}) and
(\ref{eq: isostable averaged vec}), which are the starting point
of our analysis, are a special case of (\ref{eq: moving orthogonal})
and (\ref{eq: moving orthogonal 2}). HM study (\ref{eq: moving orthogonal})
and (\ref{eq: moving orthogonal 2}) in their full generality, and
so their results apply to isostable coordinates - up to one limitation,
discussed below - simply by making the following substitutions:
\begin{equation}
\theta\rightarrow\phi,\;\vec{\rho}\rightarrow\vec{\psi}\label{eq: HM substituion 1}
\end{equation}
\begin{equation}
\vec{a}\left(\theta\right)\rightarrow0,\;\mathbf{R}\left(\theta\right)\rightarrow-\mathbf{K},\;h\left(\theta\right)\rightarrow z_{G},\mathbf{H}\left(\theta\right)\rightarrow\mathbf{Y}_{G}\mathrm{e}^{\mathbf{K}\left(1-\mathrm{mod}_{1}\phi\right)}.\label{eq: HM substituion 2}
\end{equation}
Recall, however, that we define a general Poincare section by $\phi=\phi_{S}\left(\vec{\psi}\right)\sim\vec{m}_{S}^{T}\vec{\psi}$
(see the discussion around (\ref{eq:phi Poincare section})). In contrast,
HM's analysis is limited to the case where the Poincare section is
given by $\theta=0$, which in their work is to be interpreted as
a surface orthogonal to the limit cycle. The replacements (\ref{eq: HM substituion 1})
and (\ref{eq: HM substituion 2}) would reflect the choice $\phi_{S}\left(\vec{\psi}\right)=0$,
for which the Poincare section is an isochron. One could extend HM's
rigorous work, generalizing it for arbitrary Poincare sections in
moving orthogonal coordinates. We opt instead to present a less formal
derivation for arbitrary sections in the simpler phase-isostable coordinates.

Mirroring HM, we begin with the solution of (\ref{eq: phase averaged vec})
and (\ref{eq: isostable averaged vec}) with the initial condition
restricted to the section: $\vec{\psi}\left(t_{n}\right)=\vec{\psi}_{n}$
and $\phi\left(t_{n}\right)=\vec{m}_{S}^{T}\vec{\psi}_{n}$. The subscript
$n$ reflects that this starting point is the $n^{\textit{th}}$ crossing of
the section. The equation for $\phi$, (\ref{eq: phase averaged vec}),
simply can be integrated, yielding
\begin{equation}
\phi\left(t\right)\sim\mod_{1}\left(\vec{m}_{S}^{T}\vec{\psi}_{n}+t-t_{n}+\sigma\vec{z}_{G}^{T}\left(\vec{W}\left(t\right)-\vec{W}\left(t_{n}\right)\right)\right)\label{eq:phi(t)_approximation}
\end{equation}
to first order in $\sigma$ and $\vec{\psi}$. We rewrite (\ref{eq: isostable averaged vec})
using an integrating factor of $\mathrm{e}^{\mathbf{K}(t-t_{n})}$
and (\ref{eq:phi(t)_approximation}),
\begin{equation}
d\left(\mathrm{e}^{\mathbf{K}(t-t_{n})}\vec{\psi}\right)\sim\sigma\mathrm{e}^{\mathbf{K}\left[(t-t_{n})+1-\mathrm{mod}_{1}\left(t-t_{n}+\mathcal{O}\left(\sigma,\left|\vec{\psi}\right|\right)\right)\right]}\mathbf{Y}_{G}^{T}\,d\vec{W}\left(t\right).\label{eq: psi_vec_integrating_factor}
\end{equation}
Since the first return time to the Poincare section is $t_{n+1}=t_{n}+1+\mathcal{O}\left(\sigma,\left|\vec{\psi}\right|\right)$,
we will integrate (\ref{eq: psi_vec_integrating_factor}) over an
interval $t\in\left(t_{n},t_{n}+1+\mathcal{O}\left(\sigma,\left|\vec{\psi}\right|\right)\right)$.
Note that $\mathrm{mod}_{1}\left(t-t_{n}+\mathcal{O}\left(\sigma,\left|\vec{\psi}\right|\right)\right)$
is simply $t-t_{n}+\mathcal{O}\left(\sigma,\left|\vec{\psi}\right|\right)$
for all but a subset of measure $\mathcal{O}\left(\sigma,\left|\vec{\psi}\right|\right)$
of that interval. Given that (\ref{eq: psi_vec_integrating_factor})
is overall $\mathcal{O}\left(\sigma,\left|\vec{\psi}\right|\right)$,
we may therefore replace the factor $\left[(t-t_{n})+1-\mathrm{mod}_{1}\left(\ldots\right)\right]$
in the exponent on the right hand side of (\ref{eq: psi_vec_integrating_factor})
with $1$ to first order approximation. Integrating then yields
\[
\mathrm{e}^{\mathbf{K}(t-t_{n})}\vec{\psi}\left(t\right)\sim\vec{\psi}_{n}+\sigma\mathrm{e}^{\mathbf{K}}\mathbf{Y}_{G}^{T}\left(\vec{W}\left(t\right)-\vec{W}\left(t_{n}\right)\right),
\]
for $t\in\left(t_{n},t_{n}+1+\mathcal{O}\left(\sigma,\left|\vec{\psi}\right|\right)\right)$.
Once we further evaluate at $t=t_{n+1}$, $\mathrm{e}^{\mathbf{K}(t-t_{n})}=\mathrm{e}^{\mathbf{K}}+\mathcal{O}\left(\sigma,\left|\vec{\psi}\right|\right)$,
and so the Poincare map takes $\vec{\psi}_{n}$ to
\begin{equation}
\vec{\psi}_{n+1}\equiv\vec{\psi}\left(t_{n+1}\right)\sim\mathrm{e}^{-\mathbf{K}}\vec{\psi}_{n}+\sigma\mathbf{Y}_{G}^{T}\Delta\vec{W}_{n},\label{eq: psi_map_step}
\end{equation}
where $\Delta\vec{W}_{n}=\vec{W}\left(t_{n}+1\right)-\vec{W}\left(t_{n}\right)$
is a standard normal vector random variable. Also, from (\ref{eq:phi(t)_approximation})
\begin{equation}
\phi\left(t_{n+1}\right)\sim\vec{m}_{S}^{T}\vec{\psi}_{n}+\sigma\vec{z}_{G}^{T}\Delta\vec{W}_{n}.\label{eq: phi_map_step}
\end{equation}

We now derive the first passage time to first order approximation.
Because the Poincare section is defined via (\ref{eq:phi Poincare section}),
the first return time is the first time $t_{n+1}$ such that 
\begin{equation}
\phi\left(t_{n+1}\right)=\vec{m}_{S}^{T}\vec{\psi}\left(t_{n+1}\right)\label{eq: first_return_condition}
\end{equation}
and $t_{n+1}=t_{n}+1+\mathcal{O}\left(\sigma,\left|\vec{\psi}\right|\right)$.
Estimating the first passage time $\Delta T_{n}^{S}=t_{n+1}-t_{n}$
to first order in $\sigma$ and $\vec{\psi}_{n}$, we write it as
\begin{equation}
\Delta T_{n}^{S}\sim1+\vec{\delta\mu}^{T}\vec{\psi}_{n}+\sigma\zeta_{n}.\label{eq: DeltaT ansatz}
\end{equation}
Inserting $t_{n+1}=t_{n}+\Delta T_{n}^{S}$ into (\ref{eq: first_return_condition})
and expanding it around $t_{n}+1$ using (\ref{eq: psi_map_step})
and (\ref{eq: phi_map_step}) and the facts that $\dot{\phi}=1+\mathcal{O}\left(\sigma,\left|\vec{\psi}\right|\right)$
and $\dot{\vec{\psi}}=\mathcal{O}\left(\sigma,\left|\vec{\psi}\right|\right)$
then gives to first order
\begin{eqnarray}
\phi\left(t_{n}+1\right)+\dot{\phi}\left(t_{n}+1\right)\left(\vec{\delta\mu}^{T}\vec{\psi}_{n}+\sigma\zeta_{n}\right) & \sim & \vec{m}_{S}^{T}\left(\vec{\psi}\left(t_{n}+1\right)+\dot{\vec{\psi}}\left(t_{n}+1\right)\left(\vec{\delta\mu}^{T}\vec{\psi}_{n}+\sigma\zeta_{n}\right)\right)\nonumber \\
\left(\vec{m}_{S}^{T}\vec{\psi}_{n}+\sigma\vec{z}_{G}^{T}\Delta\vec{W}_{n}\right)+\left(\vec{\delta\mu}^{T}\vec{\psi}_{n}+\sigma\zeta_{n}\right) & \sim & \vec{m}_{S}^{T}\left[\mathrm{e}^{-\mathbf{K}}\vec{\psi}_{n}+\sigma\mathbf{Y}_{G}^{T}\Delta\vec{W}_{n}\right].\label{eq: DeltaT match}
\end{eqnarray}
From (\ref{eq: DeltaT match}) we see that
\begin{equation}
\vec{\delta\mu}^{T}\equiv-\vec{m}_{S}^{T}\left(\mathbf{I}-\mathbf{\Lambda}\right),\label{eq: deltaMu}
\end{equation}
where $\mathbf{\Lambda}\equiv\mathrm{e}^{-\mathbf{K}}$, and
\begin{eqnarray}
\zeta_{n} & \equiv & \left(\vec{m}_{S}^{T}\mathbf{Y}_{G}^{T}-\vec{z}_{G}^{T}\right)\Delta\vec{W}_{n}\nonumber \\
 & = & \left(\begin{array}{cc}
-1, & \vec{m}_{S}^{T}\end{array}\mathbf{K}^{-\frac{1}{2}}\right)\mathbf{R}_{G}^{T}\Delta\vec{W}_{n},\label{eq: zeta}
\end{eqnarray}
where $\bm{R}_{G}=\left(\begin{array}{ccccc}
\vec{z}_{G}, & \vec{y}_{G,1}, & \vec{y}_{G,2}, & \cdots, & \vec{y}_{G,d}\end{array}\right)$, as defined in (\ref{eq: Xi factor}). Following (\ref{eq: psi_map_step}),
(\ref{eq: DeltaT ansatz}), and (\ref{eq: zeta}), the covariance
matrix for $\left(\begin{array}{cc}
\Delta T_{n}^{S}; & \vec{\psi}_{n+1}\end{array}\right)$ conditioned on $\vec{\psi}_{n}$ is given by
\begin{eqnarray}
\left(\begin{array}{cc}
\alpha & \vec{\beta}^{T}\\
\vec{\beta} & \mathbf{\Gamma}
\end{array}\right) & \equiv & \left(\begin{array}{cc}
\left(\vec{m}_{S}^{T}\mathbf{Y}_{G}^{T}-\vec{z}_{G}^{T}\right)\left(\mathbf{Y}_{G}\vec{m}_{S}-\vec{z}_{G}\right) & \left(\vec{m}_{S}^{T}\mathbf{Y}_{G}^{T}-\vec{z}_{G}^{T}\right)\mathbf{Y}_{G}\\
\mathbf{Y}_{G}^{T}\left(\mathbf{Y}_{G}\vec{m}_{S}-\vec{z}_{G}\right) & \mathbf{Y}_{G}^{T}\mathbf{Y}_{G}
\end{array}\right)\label{eq: PM_covariance_matrix}\\
 & = & \left(\begin{array}{cc}
-1 & \vec{m}_{S}^{T}\\
\vec{0} & \mathbf{I}
\end{array}\right)\mathbf{\Xi}\left(\begin{array}{cc}
-1 & \vec{0}^{T}\\
\vec{m}_{S} & \mathbf{I}
\end{array}\right),
\end{eqnarray}
where $\mathbf{\Xi}=\mathbf{R}_{G}^{T}\mathbf{R}_{G}$. It will be
useful to rewrite $\alpha$ and $\vec{\beta}$ in terms of 
\begin{equation}
\mathbf{\Gamma}\equiv\mathbf{Y}_{G}^{T}\mathbf{Y}_{G}\label{eq: Gamma in terms of Y}
\end{equation}
as
\begin{eqnarray}
\alpha & = & \vec{m}_{S}^{T}\mathbf{\Gamma}\vec{m}_{S}-2\vec{m}_{S}^{T}\mathbf{Y}_{G}^{T}\vec{z}_{G}+\vec{z}_{G}^{T}\vec{z}_{G}\nonumber \\
\vec{\beta} & = & \mathbf{\Gamma}\vec{m}_{S}-\mathbf{Y}_{G}^{T}\vec{z}_{G}.\label{eq: alpha, beta in terms of Gamma}
\end{eqnarray}

\subsection{Steady-State Probability Densities\label{subsec: SI - Steady-state Densities}}

Computation of the TVGR will require the steady-state distribution
of $\vec{\psi}_{n}$ for $n$ large, (\ref{eq: steady-state distribution}),
the joint distribution of $\vec{\psi}_{n}$ and $\vec{\psi}_{n+m}$
for large $n$, (\ref{eq:mth joint distribution}), and the distribution
of $\Delta T_{n}$ conditioned on $\vec{\psi}_{n}$ and $\vec{\psi}_{n+1}$,
(\ref{eq:inter-step interval dist}) below. These all follow from
the linearized Poincare map and first passage time (\ref{eq: linearized PM}).
Written in terms of a conditional distribution, 
\begin{equation}
\left(\begin{array}{c}
\Delta T_{n}^{S}-1\\
\vec{\psi}_{n+1}
\end{array}\right)\stackrel[\mathrm{given}\,\vec{\psi}_{n}]{dist.}{\sim}\mathcal{N}\left(\left(\begin{array}{c}
\vec{\delta\mu}^{T}\\
\mathbf{\Lambda}
\end{array}\right)\vec{\psi}_{n},\sigma^{2}\left(\begin{array}{cc}
\alpha & \vec{\beta}^{T}\\
\vec{\beta} & \mathbf{\Gamma}
\end{array}\right)\right),\label{eq: FPT_PM dist}
\end{equation}
where $\mathcal{N}\left(\vec{\mu},\mathbf{\Sigma}\right)$ is the
multivariate normal distribution with mean $\vec{\mu}$ and covariance
matrix $\mathbf{\Sigma}$ and $\alpha$, $\vec{\beta}$ and $\mathbf{\Gamma}$
are as defined in Section \ref{subsec: SI - Linearized Map}. The
distribution for $\vec{\psi}_{n}$ is found in Section \ref{subsec:Linearized-Poincare-Map},
and we outline the derivation of the others here.

To start, we cite elementary formulas for connecting conditional and
joint multivariate normal distributions (see e.g. \citep{eaton_multivariate_2007}).
Generally, for two (potentially vector-valued) random variables $y_{1}$
and $y_{2}$ that are jointly normally distributed, i.e.
\begin{equation}
\left(\begin{array}{c}
y_{1}\\
y_{2}
\end{array}\right)\stackrel{dist.}{\sim}\mathcal{N}\left(\left(\begin{array}{c}
\mu_{1}\\
\mu_{2}
\end{array}\right),\left(\begin{array}{cc}
\mathbf{\Sigma}_{11} & \mathbf{\Sigma}_{12}\\
\mathbf{\Sigma}_{12}^{T} & \mathbf{\Sigma}_{22}
\end{array}\right)\right),\label{eq:general joint normal}
\end{equation}
the distribution of $y_{1}$ conditioned on $y_{2}$ is again normally
distributed: 
\begin{equation}
y_{1}\stackrel[\mathrm{given}\,y_{2}]{dist.}{\sim}\mathcal{N}\left(\mu_{1}+\mathbf{\Sigma}_{12}\mathbf{\Sigma}_{22}^{-1}\left(y_{2}-\mu_{2}\right),\,\mathbf{\Sigma}_{11}-\mathbf{\Sigma}_{12}\mathbf{\Sigma}_{22}^{-1}\mathbf{\Sigma}_{12}^{T}\right).\label{eq:general conditional normal}
\end{equation}
Now, using (\ref{eq:general joint normal}) and (\ref{eq:general conditional normal})
we will recover the joint distribution for $\Delta T_{n}^{S}$, $\vec{\psi}_{n}$,
and $\vec{\psi}_{n+1}$ from the conditional distribution for $\Delta T_{n}^{S}$
and $\vec{\psi}_{n+1}$. Taking $y_{1}=\left(\Delta T_{n}^{S}-1;\vec{\psi}_{n+1}\right)$
and $y_{2}=\vec{\psi}_{n}$ and comparing (\ref{eq: FPT_PM dist})
and (\ref{eq:general conditional normal}), we find that
\begin{eqnarray*}
\mathbf{\Sigma}_{12} & = & \left(\begin{array}{c}
\vec{\delta\mu}^{T}\\
\mathbf{\Lambda}
\end{array}\right)\mathbf{\Sigma}_{22}\\
\mathbf{\Sigma}_{11} & = & \sigma^{2}\left(\begin{array}{cc}
\alpha & \vec{\beta}^{T}\\
\vec{\beta} & \mathbf{\Gamma}
\end{array}\right)+\left(\begin{array}{c}
\vec{\delta\mu}^{T}\\
\mathbf{\Lambda}
\end{array}\right)\mathbf{\Sigma}_{22}\left(\begin{array}{cc}
\vec{\delta\mu} & \mathbf{\Lambda}\end{array}\right)\\
\mu_{1} & = & \left(\begin{array}{c}
\vec{\delta\mu}^{T}\\
\mathbf{\Lambda}
\end{array}\right)\mu_{2},
\end{eqnarray*}
where $\mathbf{\Lambda}\equiv\mathrm{e}^{-\mathbf{K}}$. From (\ref{eq: steady-state distribution})
we have $\mu_{2}=\mathrm{E\left\{ \vec{\psi}_{n+1}\right\} =0}$ and
$\mathbf{\Sigma}_{22}=\mathrm{var}\left\{ \vec{\psi}_{n+1}\right\} =\sigma^{2}\mathbf{\Gamma}_{\textit{ss}}$
in steady-state, and (\ref{eq:general joint normal}) becomes

\begin{equation}
\left(\begin{array}{c}
\Delta T_{n}^{S}-1\\
\vec{\psi}_{n+1}\\
\vec{\psi}_{n}
\end{array}\right)\stackrel{dist.}{\sim}\mathcal{N}\left(0,\sigma^{2}\left(\begin{array}{ccc}
\alpha+\vec{\delta\mu}^{T}\mathbf{\Gamma}_{\textit{ss}}\vec{\delta\mu} & \vec{\beta}^{T}+\vec{\delta\mu}^{T}\mathbf{\Gamma}_{\textit{ss}}\mathbf{\Lambda} & \vec{\delta\mu}^{T}\mathbf{\Gamma}_{\textit{ss}}\\
\vec{\beta}+\mathbf{\Lambda}\mathbf{\Gamma}_{\textit{ss}}\vec{\delta\mu} & \mathbf{\Gamma}_{\textit{ss}} & \mathbf{\Lambda}\mathbf{\Gamma}_{\textit{ss}}\\
\mathbf{\Gamma}_{\textit{ss}}\vec{\delta\mu} & \mathbf{\Gamma}_{\textit{ss}}\mathbf{\Lambda} & \mathbf{\Gamma}_{\textit{ss}}
\end{array}\right)\right).\label{eq: joint distribution}
\end{equation}
Similarly, taking $y_{1}=\Delta T_{n}^{S}-1$ and $y_{2}=\vec{\psi}_{n+1}$
and matching (\ref{eq: FPT_PM dist}) and (\ref{eq:general joint normal}),
(\ref{eq:general conditional normal}) gives the conditional distribution
for $\Delta T_{n}^{S}$,
\begin{equation}
\Delta T_{n}^{S}\stackrel[\mathrm{given}\,\vec{\psi}_{n}\mathrm{and}\vec{\psi}_{n+1}]{dist.}{\sim}\mathcal{N}\left(1+\vec{\delta\mu}^{T}\vec{\psi}_{n}+\vec{\beta}^{T}\mathbf{\Gamma}^{-1}\left(\vec{\psi}_{n+1}-\mathbf{\Lambda}\vec{\psi}_{n}\right),\,\sigma^{2}\left[\alpha-\vec{\beta}^{T}\mathbf{\Gamma}^{-1}\vec{\beta}\right]\right).\label{eq:inter-step interval dist}
\end{equation}

We now consider (\ref{eq:mth joint distribution}), the joint distribution
of $\vec{\psi}_{n}$ and $\vec{\psi}_{n+m}$, for $m>0$ and $n\rightarrow\infty$.
Applying (\ref{eq: linearized PM}) recursively and neglecting higher-order
terms, we have
\[
\vec{\psi}_{n+m}=\mathbf{\Lambda}^{m}\vec{\psi}_{n}+\sum_{k=1}^{m}\mathbf{\Lambda}^{k-1}\vec{\eta}_{n+m-k}
\]
Since the $\vec{\eta}_{n+m-k}$ are i.i.d. with mean $0$ and covariance
matrix $\Gamma$, $\mathrm{var}\left\{ \vec{\psi}_{n+m}\left|\vec{\psi}_{n}\right.\right\} =\sigma^{2}\sum_{k=0}^{m-1}\mathbf{\Lambda}^{k}\mathbf{\Gamma}\mathbf{\Lambda}^{k}$
or, using (\ref{eq: Gamma_ss}), $\sigma^{2}\left(\mathbf{\Gamma}_{\textit{ss}}-\mathbf{\Lambda}^{m}\mathbf{\Gamma}_{\textit{ss}}\mathbf{\Lambda}^{m}\right)$.
So 
\begin{equation}
\vec{\psi}_{n+m}\stackrel[\mathrm{given}\,\vec{\psi}_{n}]{dist.}{\sim}\mathcal{N}\left(\mathbf{\Lambda}^{m}\vec{\psi}_{n},\,\sigma^{2}\left(\mathbf{\Gamma}_{\textit{ss}}-\mathbf{\Lambda}^{m}\mathbf{\Gamma}_{\textit{ss}}\mathbf{\Lambda}^{m}\right)\right).\label{eq: psi+m dist}
\end{equation}
Then, as we did above for (\ref{eq: joint distribution}), we recover
the joint density (\ref{eq:mth joint distribution}) using (\ref{eq:general joint normal})
and (\ref{eq:general conditional normal}) and the facts that $E\left\{ \vec{\psi}_{n}\right\} =\mathrm{E\left\{ \vec{\psi}_{n+m}\right\} =0}$
and $\mathrm{var}\left\{ \vec{\psi}_{n}\right\} =\mathrm{var}\left\{ \vec{\psi}_{n+m}\right\} =\sigma^{2}\mathbf{\Gamma}_{\textit{ss}}$.

\subsection{The TVGR Term-by-Term\label{subsec: SI - TVGR,  term-by-term}}

We compute the TVGR for the limit cycle oscillator via (\ref{eq:FanoVar decomposed})
and using the covariance matrix (\ref{eq: PM_covariance_matrix})
as well as the other distributions cited in Sections \ref{subsec:Linearized-Poincare-Map}
and \ref{subsec: SI - Steady-state Densities}. Recall the state $x$
of the Markov renewal process is the position $\vec{\psi}$ on the
Poincare section. In the following, we use $x$ and $\vec{\psi}$
interchangeably.

In order to evaluate $\mathcal{V}_{E}^{\left(t\right)}$, we find
it useful to first compute $\mathcal{CV}_{\left(x\rightarrow x^{\prime}\right)\mapsto h\left(x\right),\left(x\rightarrow x^{\prime}\right)\mapsto x}^{\left(n\right)}$
and $\mathcal{CV}_{\left(x\rightarrow x^{\prime}\right)\mapsto h\left(x\right),\left(x\rightarrow x^{\prime}\right)\mapsto x^{\prime}}^{\left(n\right)}$
for an arbitrary scalar or vector-valued, deterministic function $h$.
Since $\mathrm{E}\left\{ x_{n}\right\} =\mathrm{E}\left\{ \vec{\psi}_{n}\right\} =0$,
(\ref{eq:funcCV sum of variances}) gives
\begin{equation}
\mathcal{CV}_{\left(x\rightarrow x^{\prime}\right)\mapsto h\left(x\right),\left(x\rightarrow x^{\prime}\right)\mapsto x}^{\left(n\right)}=\mathrm{E}\left\{ h\left(x_{n}\right)x_{n}^{T}\right\} +\sum_{k=1}^{\infty}\mathrm{\mathrm{E}}\left\{ h\left(x_{n}\right)x_{n+k}^{T}\right\} +\sum_{k=1}^{\infty}\mathrm{\mathrm{E}}\left\{ h\left(x_{n+k}\right)x_{n}^{T}\right\} .\label{eq: CV_hx,x sum form}
\end{equation}
Recall that we use $N\left(\cdot,\mu,\Sigma\right)$ to denote the
probability density function for the normal distribution with mean
$\mu$ and covariance matrix $\Sigma$. Note then that
\[
\mathrm{\mathrm{E}}\left\{ h\left(x_{n}\right)x_{n+k}^{T}\right\} =\int_{S\times S}h\left(x_{n}\right)x_{n+k}^{T}N\left(\left(\begin{array}{c}
x_{n+k}\\
x_{n}
\end{array}\right);0,\sigma^{2}\mathbf{C}_{k}\right)dx_{n+k}\,dx_{n},
\]
where $C_{k}$ is as defined in (\ref{eq:mth joint distribution}).
We can integrate over $x_{n+k}$ to find for $k\ge0$
\begin{eqnarray*}
\mathrm{\mathrm{E}}\left\{ h\left(x_{n}\right)x_{n+k}^{T}\right\}  & = & \mathrm{\mathrm{E}}\left\{ h\left(x_{n}\right)\mathrm{\mathrm{E}}\left\{ x_{n+k}^{T}\left|x_{n}\right.\right\} \right\} \\
 & = & \mathrm{\mathrm{E}}\left\{ h\left(x_{n}\right)\mathrm{x_{n}^{T}}\mathbf{\Lambda}^{k}\right\} =\mathcal{H}\mathbf{\Lambda}^{k},
\end{eqnarray*}
where from (\ref{eq: psi+m dist}) $\mathrm{\mathrm{E}}\left\{ x_{n+k}^{T}\left|x_{n}\right.\right\} =\mathrm{x_{n}^{T}}\mathbf{\Lambda}^{k}$
and we define 
\begin{equation}
\mathcal{H}\equiv\int_{S}h\left(x\right)x^{T}N\left(x;0,\sigma^{2}\mathbf{\Gamma}_{\textit{ss}}\right)dx.\label{eq: integral h(x) x}
\end{equation}
Analogously, using (\ref{eq:mth joint distribution}) and (\ref{eq:general conditional normal}),
we have $\mathrm{\mathrm{E}}\left\{ x_{n}^{T}\left|x_{n+k}\right.\right\} =x_{n+k}^{T}\mathbf{\Gamma}_{\textit{ss}}^{-1}\mathbf{\Lambda}^{k}\mathbf{\Gamma}_{\textit{ss}}$,
and, thus, for $k\ge0$,
\begin{eqnarray*}
\mathrm{\mathrm{E}}\left\{ h\left(x_{n+k}\right)x_{n}^{T}\right\}  & = & \mathrm{\mathrm{E}}\left\{ h\left(x_{n+k}\right)\mathrm{\mathrm{E}}\left\{ x_{n}^{T}\left|x_{n+k}\right.\right\} \right\} \\
 & = & \mathrm{\mathrm{E}}\left\{ h\left(x_{n+k}\right)x_{n+k}^{T}\mathbf{\Gamma}_{\textit{ss}}^{-1}\mathbf{\Lambda}^{k}\mathbf{\Gamma}_{\textit{ss}}\right\} \\
 & = & \mathcal{H}\mathbf{\Gamma}_{\textit{ss}}^{-1}\mathbf{\Lambda}^{k}\mathbf{\Gamma}_{\textit{ss}}.
\end{eqnarray*}
So the sums in (\ref{eq: CV_hx,x sum form}) can be written in closed-form
as geometric series,
\begin{eqnarray*}
\mathcal{CV}_{\left(x\rightarrow x^{\prime}\right)\mapsto h\left(x\right),\left(x\rightarrow x^{\prime}\right)\mapsto x}^{\left(n\right)} & = & \mathcal{H}\left[\mathbf{I}+\left(\left(\mathbf{I}-\mathbf{\Lambda}\right)^{-1}-\mathbf{I}\right)+\mathbf{\Gamma}_{\textit{ss}}^{-1}\left(\left(\mathbf{I}-\mathbf{\Lambda}\right)^{-1}-\mathbf{I}\right)\mathbf{\Gamma}_{\textit{ss}}\right]\\
 & = & \mathcal{H}\left[-\mathbf{I}+\left(\mathbf{I}-\mathbf{\Lambda}\right)^{-1}+\mathbf{\Gamma}_{\textit{ss}}^{-1}\left(\mathbf{I}-\mathbf{\Lambda}\right)^{-1}\mathbf{\Gamma}_{\textit{ss}}\right].
\end{eqnarray*}
It will be useful to decompose the expression in brackets as
\begin{equation}
\mathcal{CV}_{\left(x\rightarrow x^{\prime}\right)\mapsto h\left(x\right),\left(x\rightarrow x^{\prime}\right)\mapsto x}^{\left(n\right)}=\mathcal{H}\mathbf{\Gamma}_{\textit{ss}}^{-1}\left(\mathbf{I}-\mathbf{\Lambda}\right)^{-1}\mathbf{\Gamma}\left(\mathbf{I}-\mathbf{\Lambda}\right)^{-1},\label{eq: CV_hx,x}
\end{equation}
which can be done via the fact that $\mathbf{\Gamma}_{\textit{ss}}=\mathbf{\Lambda}\mathbf{\Gamma}_{\textit{ss}}\mathbf{\Lambda}+\mathbf{\Gamma}$
(see the discussion preceding (\ref{eq: Gamma_ss})). It can similarly
be shown that
\begin{eqnarray}
\mathcal{CV}_{\left(x\rightarrow x^{\prime}\right)\mapsto h\left(x\right),\left(x\rightarrow x^{\prime}\right)\mapsto x^{\prime}}^{\left(n\right)} & = & \mathcal{H}\left[\mathbf{\Lambda}+\left(\left(\mathbf{I}-\mathbf{\Lambda}\right)^{-1}-\left(\mathbf{I}+\mathbf{\Lambda}\right)\right)+\mathbf{\Gamma}_{\textit{ss}}^{-1}\left(\mathbf{I}-\mathbf{\Lambda}\right)^{-1}\mathbf{\Gamma}_{\textit{ss}}\right]\nonumber \\
 & = & \mathcal{H}\mathbf{\Gamma}_{\textit{ss}}^{-1}\left(\mathbf{I}-\mathbf{\Lambda}\right)^{-1}\mathbf{\Gamma}\left(\mathbf{I}-\mathbf{\Lambda}\right)^{-1}\label{eq: CV_hx,xPrime}
\end{eqnarray}
is precisely the same as $\mathcal{CV}_{\left(x\rightarrow x^{\prime}\right)\mapsto h\left(x\right),\left(x\rightarrow x^{\prime}\right)\mapsto x}^{\left(n\right)}$.

It is also useful to consider the expression $\Delta t\left(x,x^{\prime}\right)$
that appears in the TVGR. Recall from the discussion in Section \ref{subsec:TVGR as SVGR}
and SI Section \ref{sec: SI - Event TVGR Formula} that $\Delta t\left(x,x^{\prime}\right)$
is a random function of $x$ and $x^{\prime}$: even when conditioned
on $x$ and $x^{\prime}$, there may be uncertainty in its value.
For the linearized Poincare map dynamics, $\Delta t\left(\vec{\psi}_{n},\vec{\psi}_{n+1}\right)$
has the same distribution as $\Delta T_{n}^{S}$ conditioned on $\vec{\psi}_{n}$
and $\vec{\psi}_{n+1}$ and can be decomposed as 
\begin{equation}
\Delta t\left(\vec{\psi}_{n},\vec{\psi}_{n+1}\right)=\overline{\Delta t}\left(\vec{\psi}_{n},\vec{\psi}_{n+1}\right)+\delta t_{n},\label{eq: DeltaT decomposed}
\end{equation}
where $\overline{\Delta t}\left(x,x^{\prime}\right)$ is the mean
of $\Delta t\left(x,x^{\prime}\right)$ conditioned on $x$ and $x^{\prime}$
and the $\delta t_{n}$ are zero mean i.i.d. Gaussian random variables
independent of the $\vec{\psi}_{j}$ for any $j$ (see (\ref{eq:inter-step interval dist})).
Since each of the terms constituting the mixed component are covariances
of $\Delta t\left(\vec{\psi}_{n},\vec{\psi}_{n+1}\right)$ with $1_{E}\left(\vec{\psi}_{j}\right)$
(see (\ref{eq:funcCV sum of variances})), the mixed component does
not depend on the $\delta t_{n}$, i.e. it can be rewritten as
\[
\mathcal{CV}_{\left(x\rightarrow x^{\prime}\right)\mapsto1_{E}\left(x\right),\left(x\rightarrow x^{\prime}\right)\mapsto\Delta t\left(x,x^{\prime}\right)}^{\left(n\right)}=\mathcal{CV}_{\left(x\rightarrow x^{\prime}\right)\mapsto1_{E}\left(x\right),\left(x\rightarrow x^{\prime}\right)\mapsto\overline{\Delta t}\left(x,x^{\prime}\right)}^{\left(n\right)}.
\]
While $\delta t_{n}$ likewise does not contribute to the covariances
in the temporal component, $\mathcal{V}_{\left(x\rightarrow x^{\prime}\right)\mapsto\Delta t\left(x,x^{\prime}\right)}^{\left(n\right)}$,
it does affect the first, variance term in (\ref{eq:FV sum of variances})
$\mathrm{var}\left\{ \Delta t\left(\vec{\psi}_{0},\vec{\psi}_{1}\right)\right\} $.
By the law of total variance
\begin{eqnarray}
\mathrm{var}\left\{ \Delta t\left(\vec{\psi}_{0},\vec{\psi}_{1}\right)\right\}  & = & \mathrm{var}\left\{ \mathrm{E}\left\{ \Delta t\left(\vec{\psi}_{0},\vec{\psi}_{1}\right)\left|\vec{\psi}_{0},\vec{\psi}_{1}\right.\right\} \right\} +\mathrm{E}\left\{ \mathrm{var}\left\{ \Delta t\left(\vec{\psi}_{0},\vec{\psi}_{1}\right)\left|\vec{\psi}_{0},\vec{\psi}_{1}\right.\right\} \right\} \nonumber \\
 & = & \mathrm{var}\left\{ \overline{\Delta t}\left(\vec{\psi}_{0},\vec{\psi}_{1}\right)\right\} +\mathrm{var}\left\{ \delta t_{0}\right\} .\label{eq:var Dt decomp}
\end{eqnarray}
The first term of (\ref{eq:var Dt decomp}) is the first term of $\mathcal{V}_{\left(x\rightarrow x^{\prime}\right)\mapsto\overline{\Delta t}\left(x,x^{\prime}\right)}^{\left(n\right)}$.
So, beyond those terms included in $\mathcal{V}_{\left(x\rightarrow x^{\prime}\right)\mapsto\overline{\Delta t}\left(x,x^{\prime}\right)}^{\left(n\right)}$,
$\mathcal{V}_{\left(x\rightarrow x^{\prime}\right)\mapsto\Delta t\left(x,x^{\prime}\right)}^{\left(n\right)}$
has the additional contribution $\mathrm{var}\left\{ \delta t_{0}\right\} $.
Thus,
\[
\mathcal{V}_{E}^{\left(t\right)}=\mathcal{V}_{x\mapsto1_{E}\left(x\right)}^{\left(n\right)}-2\,\mathcal{E}\,\mathcal{CV}_{\left(x\rightarrow x^{\prime}\right)\mapsto1_{E}\left(x\right),\left(x\rightarrow x^{\prime}\right)\mapsto\overline{\Delta t}\left(x,x^{\prime}\right)}^{\left(n\right)}+\mathcal{E}^{2}\mathcal{V}_{\left(x\rightarrow x^{\prime}\right)\mapsto\overline{\Delta t}\left(x,x^{\prime}\right)}^{\left(n\right)}+\mathrm{var}\left\{ \delta t_{0}\right\} .
\]

We now evaluate $\mathcal{V}_{E}^{\left(t\right)}$ term-by-term.
\begin{enumerate}
\item $\mathcal{V}_{x\mapsto1_{E}\left(x\right)}^{\left(n\right)}$\\
This term is as written in (\ref{eq: Markov-only sum}).
\item $\mathcal{CV}_{\left(x\rightarrow x^{\prime}\right)\mapsto1_{E}\left(x\right),\left(x\rightarrow x^{\prime}\right)\mapsto\overline{\Delta t}\left(x,x^{\prime}\right)}^{\left(n\right)}$
\\
From (\ref{eq:inter-step interval dist}) and (\ref{eq: DeltaT decomposed})
we see that $\overline{\Delta t}\left(\vec{\psi}_{n},\vec{\psi}_{n+1}\right)$
is linear in $x_{n}=\vec{\psi}_{n}$ and $x_{n+1}=\vec{\psi}_{n+1}$,
we write
\begin{equation}
\overline{\Delta t}\left(\vec{\psi}_{n},\vec{\psi}_{n+1}\right)=1+\vec{b}_{0}^{T}\vec{\psi}_{n}+\vec{b}_{1}^{T}\vec{\psi}_{n+1},\label{eq: deltaTBar linear}
\end{equation}
where, using (\ref{eq: deltaMu}), (\ref{eq: alpha, beta in terms of Gamma}),
and (\ref{eq:inter-step interval dist}),
\begin{eqnarray*}
\vec{b}_{1} & = & \mathbf{\Gamma}^{-1}\vec{\beta}=\vec{m}_{S}-\mathbf{\Gamma}^{-1}\mathbf{Y}_{G}^{T}\vec{z}_{G}\\
\vec{b}_{0} & = & \vec{\delta\mu}-\mathbf{\Lambda}\vec{b}_{1}=-\vec{m}_{S}+\mathbf{\Lambda}\mathbf{\Gamma}^{-1}\mathbf{Y}_{G}^{T}\vec{z}_{G}.
\end{eqnarray*}
Because it will be useful, we note that their sum is simply
\begin{equation}
\vec{b}_{0}+\vec{b}_{1}=-\left(\mathbf{I}-\mathbf{\Lambda}\right)\mathbf{\Gamma}^{-1}\mathbf{Y}_{G}^{T}\vec{z}_{G}.\label{eq: b0 + b1}
\end{equation}
Then, since $\mathcal{CV}$ is a bilinear functional (see (\ref{eq: funcCV})),
\[
\mathcal{CV}_{\left(x\rightarrow x^{\prime}\right)\mapsto1_{E}\left(x\right),\left(x\rightarrow x^{\prime}\right)\mapsto\overline{\Delta t}\left(x,x^{\prime}\right)}^{\left(n\right)}=\mathcal{CV}_{\left(x\rightarrow x^{\prime}\right)\mapsto1_{E}\left(x\right),\left(x\rightarrow x^{\prime}\right)\mapsto x}\vec{b}_{0}+\mathcal{CV}_{\left(x\rightarrow x^{\prime}\right)\mapsto1_{E}\left(x\right),\left(x\rightarrow x^{\prime}\right)\mapsto x^{\prime}}\vec{b}_{1},
\]
where $x=\vec{\psi}$. Now we can apply (\ref{eq: CV_hx,x}) and (\ref{eq: CV_hx,xPrime})
with $h\left(x\right)=1_{E}\left(x\right)$, yielding
\[
\mathcal{CV}_{\left(x\rightarrow x^{\prime}\right)\mapsto1_{E}\left(x\right),\left(x\rightarrow x^{\prime}\right)\mapsto\overline{\Delta t}\left(x,x^{\prime}\right)}^{\left(n\right)}=\mathcal{X}^{T}\mathbf{\Gamma}_{\textit{ss}}^{-1}\left(\mathbf{I}-\mathbf{\Lambda}\right)^{-1}\mathbf{\Gamma}\left(\mathbf{I}-\mathbf{\Lambda}\right)^{-1}\left(\vec{b}_{0}+\vec{b}_{1}\right),
\]
where, using (\ref{eq: integral h(x) x}),
\[
\mathcal{X}^{T}=\mathrm{E}\left\{ 1_{E}\left(x_{n}\right)x_{n}\right\} =\int_{E}x^{T}N\left(x;0,\sigma^{2}\mathbf{\Gamma}_{\textit{ss}}\right)dx.
\]
Note that $\mathcal{X}=x_{E}\mathcal{E}$, where $x_{E}$ and $\mathcal{E}$
are as defined in Section \ref{subsec:Fano-Variance_linear}. We rewrite
$\mathcal{CV}_{\left(x\rightarrow x^{\prime}\right)\mapsto1_{E}\left(x\right),\left(x\rightarrow x^{\prime}\right)\mapsto\overline{\Delta t}\left(x,x^{\prime}\right)}$
as $-\frac{1}{2}\mathcal{X}^{T}\vec{b}$, where
\begin{equation}
\vec{b}=2\mathbf{\Gamma}_{\textit{ss}}^{-1}\left(\mathbf{I}-\mathbf{\Lambda}\right)^{-1}\mathbf{Y}_{G}^{T}\vec{z}_{G}\label{eq: FanoVariance_bVec}
\end{equation}
follows from (\ref{eq: b0 + b1}) and contains the factors that are
independent of $\sigma$ and $E$.
\item $\mathcal{V}_{\left(x\rightarrow x^{\prime}\right)\mapsto\overline{\Delta t}\left(x,x^{\prime}\right)}^{\left(n\right)}$\\
Again decomposing $\overline{\Delta t}$ via (\ref{eq: deltaTBar linear}),
we have from the linearity of $\mathcal{V}$ and $\mathcal{CV}$ (see
(\ref{eq:FV sum of variances}) and (\ref{eq:funcCV sum of variances})),
\begin{equation}
\mathcal{V}_{\left(x\rightarrow x^{\prime}\right)\mapsto\overline{\Delta t}\left(x,x^{\prime}\right)}^{\left(n\right)}=\left(\begin{array}{cc}
\vec{b}_{0}^{T}, & \vec{b}_{1}^{T}\end{array}\right)\left(\begin{array}{cc}
\mathcal{V}_{\left(x\rightarrow x^{\prime}\right)\mapsto x}^{\left(n\right)} & \mathcal{CV}_{\left(x\rightarrow x^{\prime}\right)\mapsto x,\left(x\rightarrow x^{\prime}\right)\mapsto x^{\prime}}^{\left(n\right)}\\
\mathcal{CV}_{\left(x\rightarrow x^{\prime}\right)\mapsto x^{\prime},\left(x\rightarrow x^{\prime}\right)\mapsto x}^{\left(n\right)} & \mathcal{V}_{\left(x\rightarrow x^{\prime}\right)\mapsto x^{\prime}}^{\left(n\right)}
\end{array}\right)\left(\begin{array}{c}
\vec{b}_{0}\\
\vec{b}_{1}
\end{array}\right).\label{eq: var_deltaTBar decomp}
\end{equation}
Note that $\mathcal{V}_{\left(x\rightarrow x^{\prime}\right)\mapsto x}^{\left(n\right)}=\mathcal{CV}_{\left(x\rightarrow x^{\prime}\right)\mapsto x,\left(x\rightarrow x^{\prime}\right)\mapsto x}^{\left(n\right)}$,
and so we can apply (\ref{eq: CV_hx,x}) and (\ref{eq: CV_hx,xPrime})
for each of the matrix-valued entries in the matrix appearing in (\ref{eq: var_deltaTBar decomp}).
But since (\ref{eq: CV_hx,x}) and (\ref{eq: CV_hx,xPrime}) give
the same result, all of those entries are the same: $\mathcal{V}_{\left(x\rightarrow x^{\prime}\right)\mapsto x}^{\left(n\right)}$.
To compute $\mathcal{V}_{\left(x\rightarrow x^{\prime}\right)\mapsto x}^{\left(n\right)}$,
we take $h$ equal to the identity function. (\ref{eq: integral h(x) x})
gives ${\cal H}=\sigma^{2}\mathbf{\Gamma}_{\textit{ss}}$, and thus
\[
\mathcal{V}_{\left(x\rightarrow x^{\prime}\right)\mapsto x}^{\left(n\right)}=\sigma^{2}\left(\mathbf{I}-\mathbf{\Lambda}\right)^{-1}\mathbf{\Gamma}\left(\mathbf{I}-\mathbf{\Lambda}\right)^{-1}.
\]
Then we have from (\ref{eq: b0 + b1}) and (\ref{eq: var_deltaTBar decomp})
\begin{eqnarray}
\mathcal{V}_{\left(x\rightarrow x^{\prime}\right)\mapsto\overline{\Delta t}\left(x,x^{\prime}\right)}^{\left(n\right)} & = & \left(\vec{b}_{0}+\vec{b}_{1}\right)^{T}\mathcal{V}_{\left(x\rightarrow x^{\prime}\right)\mapsto x}^{\left(n\right)}\left(\vec{b}_{0}+\vec{b}_{1}\right)\nonumber \\
 & = & \sigma^{2}\vec{z}_{G}^{T}\mathbf{Y}_{G}\mathbf{\Gamma}^{-1}\mathbf{Y}_{G}^{T}\vec{z}_{G}\label{eq: VGR conditional mean}
\end{eqnarray}
\item $\mathrm{var}\left\{ \delta t_{0}\right\} $\\
(\ref{eq:inter-step interval dist}) reveals that the conditional
variance of $\Delta t\left(x,x^{\prime}\right)$ is given by $\sigma^{2}\left[\alpha-\vec{\beta}^{T}\mathbf{\Gamma}^{-1}\vec{\beta}\right]$.
Then following (\ref{eq: alpha, beta in terms of Gamma}),
\begin{eqnarray}
\mathrm{var}\left\{ \delta t_{0}\right\}  & = & \sigma^{2}\left[\alpha-\vec{\beta}^{T}\mathbf{\Gamma}^{-1}\vec{\beta}\right]\nonumber \\
 & = & \sigma^{2}\left[\vec{m}_{S}^{T}\mathbf{\Gamma}\vec{m}_{S}-2\vec{m}_{S}^{T}\mathbf{Y}_{G}^{T}\vec{z}_{G}+\vec{z}_{G}^{T}\vec{z}_{G}-\left(\vec{m}_{S}^{T}\mathbf{\Gamma}-\vec{z}_{G}^{T}\mathbf{Y}_{G}\right)\mathbf{\Gamma}^{-1}\left(\mathbf{\Gamma}\vec{m}_{S}-\mathbf{Y}_{G}^{T}\vec{z}_{G}\right)\right]\nonumber \\
 & = & \sigma^{2}\left[\vec{z}_{G}^{T}\vec{z}_{G}-\vec{z}_{G}^{T}\mathbf{Y}_{G}\mathbf{\Gamma}^{-1}\mathbf{Y}_{G}^{T}\vec{z}_{G}\right].\label{eq: conditional var}
\end{eqnarray}
\item $\mathcal{V}_{E}^{\left(t\right)}$\\
Putting all of the above pieces together, we arrive at (\ref{eq:FanoVar Poincare}),
\[
\mathcal{V}_{E}^{\left(t\right)}=\left(\mathcal{E}-\mathcal{E}^{2}\right)+2\sum_{m=1}^{\infty}\left(\mathcal{E}_{m}-\mathcal{E}^{2}\right)+\left(\vec{b}\cdot x_{E}\right)\mathcal{E}^{2}+c\frac{\sigma^{2}}{2}\mathcal{E}^{2},
\]
where $\vec{b}$ is given by (\ref{eq: FanoVariance_bVec}) and $c$
follows from (\ref{eq: VGR conditional mean}) and (\ref{eq: conditional var}):
\begin{eqnarray}
c & = & \frac{2}{\sigma^{2}}\left(\mathcal{V}_{\left(x\rightarrow x^{\prime}\right)\mapsto\overline{\Delta t}\left(x,x^{\prime}\right)}+\mathrm{var}\left\{ \delta t_{0}\right\} \right)\nonumber \\
 & = & 2\vec{z}_{G}^{T}\vec{z}_{G}.\label{eq: FanoVariance_c}
\end{eqnarray}
\end{enumerate}

\section{Bounds on the Markov-only Component\label{sec: SI - Bounds}}

In this SI section we derive (\ref{eq:bounds on indicator funcVar}):
upper and lower bounds on $\mathcal{V}_{x\mapsto1_{E}\left(x\right)}^{\left(n\right)}$,
the Markov-only component of the temporal variance growth rate, for
planar oscillators. For the lower bound, we show that $\mathcal{E}_{m}-\mathcal{E}^{2}>0$
for all finite $m\ge0$. For the upper bound, we bound each term $\left(\mathcal{E}_{m}-\mathcal{E}^{2}\right)$
individually in such a way that the infinite sum can be carried out
as a geometric series. Specifically, we find a $\mathcal{B}$ that
is independent of $\Lambda$ and $m$ such that
\begin{equation}
\mathcal{E}_{m}-\mathcal{E}^{2}\le\Lambda^{m}\mathcal{B},\label{eq: upper bound}
\end{equation}
and therefore
\begin{eqnarray*}
\mathcal{V}_{x\mapsto1_{E}\left(x\right)}^{\left(n\right)} & = & \mathcal{E}\left(1-\mathcal{E}\right)+2\sum_{m=1}^{\infty}\left(\mathcal{E}_{m}-\mathcal{E}^{2}\right)\\
 & \le & \mathcal{B}+2\frac{\Lambda}{1-\Lambda}\mathcal{B}=\frac{1+\Lambda}{1-\Lambda}\mathcal{B}.
\end{eqnarray*}

\begin{figure}
\noindent \begin{centering}
\includegraphics[width=0.75\paperwidth]{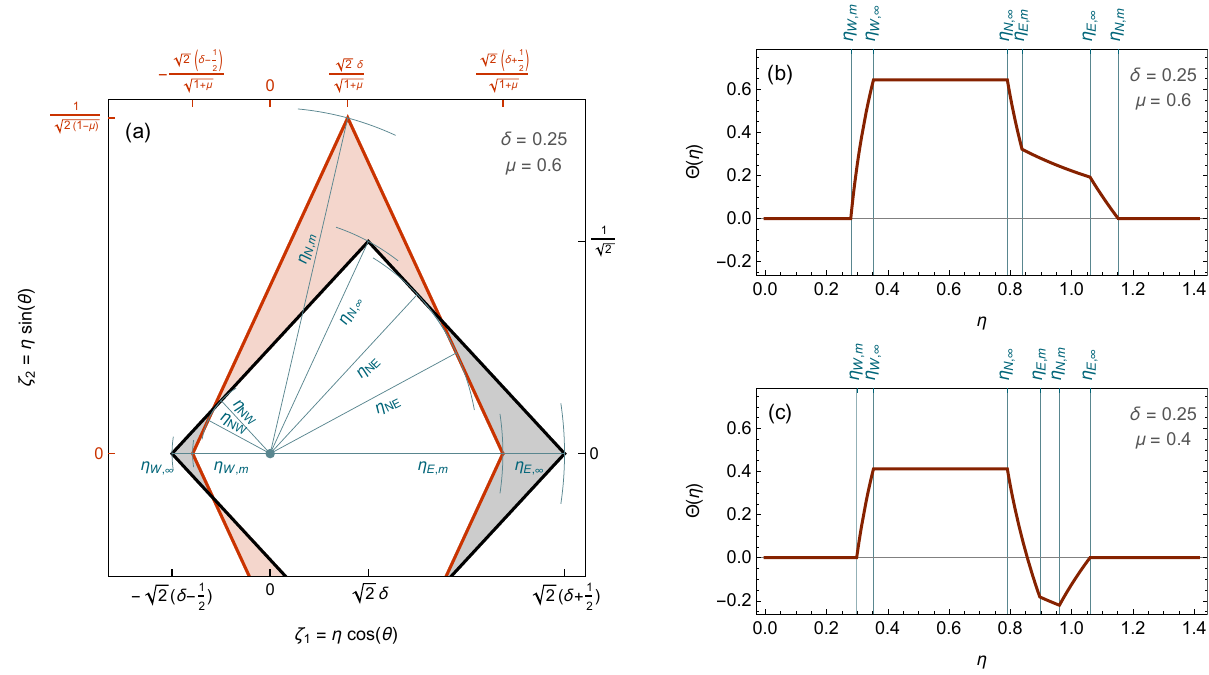}
\par\end{centering}
\caption{(a): Regions of integration for $\mathcal{E}_{m}$ (bounded by the
red lines) and $\mathcal{E}_{\infty}$ (bounded by the black lines)
in the $\zeta$-plane. The red regions count positively towards $\mathcal{E}_{m}-\mathcal{E}_{\infty}$
and the black regions negatively. This picture is to scale for $\delta=0.25$
and $\mu=0.6$, but is qualitatively representative for all $\delta\in\left(0,\frac{1}{2}\right)$
and $\mu\in\left(0,1\right)$. (b) and (c): Radial representation
$\Theta\left(\eta\right)$ of the exact difference in the areas of
the regions. In (a) and (b), $\delta=0.25$ and $\mu=0.6$. In (c),
where $\delta=0.25$ and $\mu=0.4$, $\Theta\left(\eta\right)$ becomes
negative for an interval of $\eta$. \label{fig: sample_zeta_plane + r_plot}}
\end{figure}

Both the lower and upper bounds involve a comparison of 
\[
\mathcal{E}_{m}=\int_{E\times E}N\left(z;0,\sigma^{2}\mathbf{C}_{m}\right)dz
\]
with $\mathcal{E}^{2}$, where
\[
\mathbf{C}_{m}=\left(\begin{array}{cc}
1 & \Lambda^{m}\\
\Lambda^{m} & 1
\end{array}\right).
\]
Note that, since $0\le\Lambda<1$, $C_{\infty}=I$, and so $\mathcal{E}^{2}=\mathcal{E}_{\infty}$.
We start our analysis with a transformation of variables $z$ to $\zeta$
such that the resulting integrands are independent of $m$,
\[
\zeta=\frac{1}{\sqrt{2}}\left(\begin{array}{cc}
\frac{1}{\sqrt{1+\mu}} & \frac{1}{\sqrt{1+\mu}}\\
-\frac{1}{\sqrt{1-\mu}} & \frac{1}{\sqrt{1-\mu}}
\end{array}\right)z,
\]
where $\mu\equiv\Lambda^{m}$. This amounts to a rotation by $-\frac{\pi}{4}$
and a scaling of the two resulting components by $\frac{1}{\sqrt{1+\mu}}$
and $\frac{1}{\sqrt{1-\mu}}$, respectively. Under this transformation,
the integrand becomes the probability density function for a 2D normal
random variable $\zeta$ with uncorrelated components: $N\left(\zeta;0,\sigma^{2}I\right)$.
The cost of this step, of course, is that the domain of integration
$z\in E\times E=\left[\delta-\frac{1}{2},\delta+\frac{1}{2}\right]\times\left[\delta-\frac{1}{2},\delta+\frac{1}{2}\right]$
is no longer rectangular and now depends on $m$. On the other hand,
$\mathcal{E}_{m}$ and $\mathcal{E}_{\infty}$ can now be directly
compared, since they share the same probability density $N\left(\zeta;0,\sigma^{2}I\right)$.
Each term in the infinite sum, $\mathcal{E}_{m}-\mathcal{E}_{\infty}$,
is given by the difference in the total probability mass in a diamond-shaped
region and in a square region in the $\zeta$-plane (red and black
regions, respectively, Figure \ref{fig: sample_zeta_plane + r_plot}a).

Since the probability density has circular symmetry, we analyze the
difference in probability mass in polar coordinates. Progress can
be made by integrating over the angular coordinate, which yields
\begin{equation}
\mathcal{E}_{m}-\mathcal{E}_{\infty}=\frac{1}{2\pi\sigma^{2}}\int_{0}^{\infty}\eta\,\Theta\left(\eta\right)\mathrm{e}^{-\frac{\eta^{2}}{2\sigma^{2}}}d\eta,\label{eq: radial integral}
\end{equation}
where $\eta$ is the radial coordinate and $\Theta\left(\eta\right)$
is the net angular content of the diamond-shaped region less the square
region at a distance of $\eta$ from the origin. Note that for $\mu<1$
(i.e. $\Lambda<1$ and $m>0$) and for $\eta$ sufficiently small,
$\Theta\left(\eta\right)=0$, since the diamond and square overlap
at small $\eta$. $\Theta\left(\eta\right)$ is also $0$ for sufficiently
large $\eta$ since the diamond and square are finite for $\mu<1$.
These facts will act as ``boundary conditions'' in our analysis,
since we characterize $\Theta\left(\eta\right)$ in terms of its derivative
in the following.

Between its initial and final values, $\Theta\left(\eta\right)$ is
a piecewise smooth function that depends parametrically on $\delta$
and $\mu$. We expect it might become non-smooth at radii corresponding
to 1) the distance between the origin and the corners of the square
and diamond and 2) the minimum distance between the origin and the
sides of the square and diamonds (Figure \ref{fig: sample_zeta_plane + r_plot}a).
As it turns out, though, the distance to the sides of the square and
to the sides of the diamond are equal; those radii / distances are
invariant in $\mu$. At those values of $\eta$, $\Theta\left(\eta\right)$
is smooth, since the contributions from the square and diamond cancel.
There are $6$ remaining, unique ``non-smooth radii'', taking the
symmetry about the $\zeta_{1}$-axis into account. We label them via
cardinal directions (north, east, and west; south is equivalent to
north) and an index of $m$ if that radius is specific to the diamond
and of $\infty$ if it is specific to the square (Figure \ref{fig: sample_zeta_plane + r_plot}a):
\begin{equation}
\eta_{N,m}=\sqrt{\frac{1}{2\left(1-\mu\right)}+\frac{2\delta^{2}}{1+\mu}};\;\eta_{E,m}=\sqrt{\frac{2}{1+\mu}}\left(\frac{1}{2}+\delta\right);\;\eta_{W,m}=\sqrt{\frac{2}{1+\mu}}\left(\frac{1}{2}-\delta\right)\,.\label{eq: eta non-smooth}
\end{equation}
Here the $m=\infty$ counterparts are given by replacing $\mu=\Lambda^{m}$
with $0$. Note that $\eta_{E,m}$ and $\eta_{W,m}$ are strictly
decreasing as functions of $\mu$, while $\eta_{N,m}$ is strictly
increasing for $\mu\in\left[0,1\right)$ when $\delta\in\left[0,\frac{1}{2}\right)$.

\begin{figure}[h]
\noindent \begin{centering}
\includegraphics[width=0.75\paperwidth]{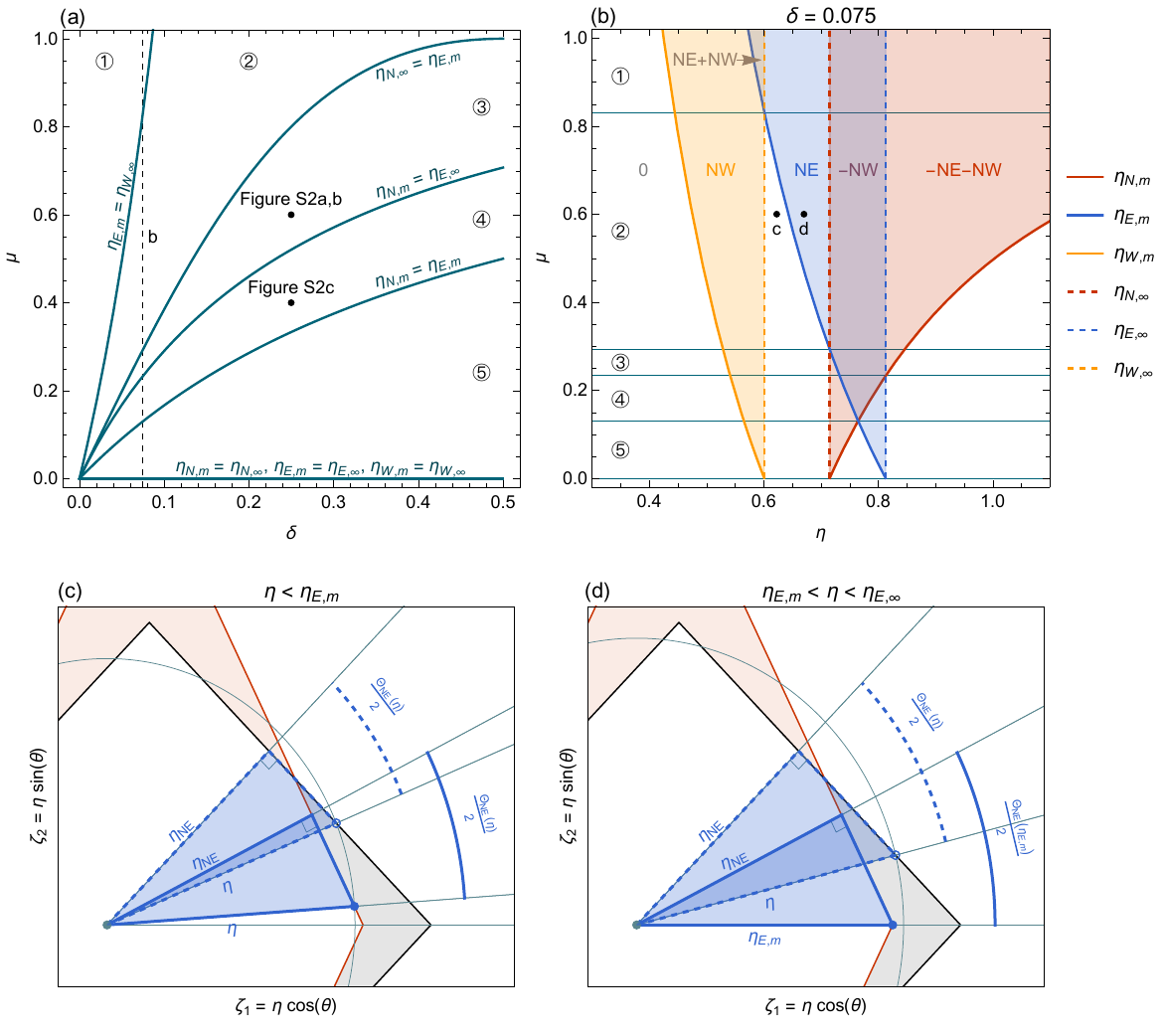}
\par\end{centering}
\caption{(a): Curves in the $\delta$-$\mu$ plane on which two of the non-smooth
radii are degenerately equal. In the numbered regions between the
curves, the radii have a particular ordering and $\Theta\left(\eta\right)$
a particular piecewise behavior. (b): Curves in the $\eta$-$\mu$
plane that separate the different smooth segments of $\Theta\left(\eta\right)$.
While (b) is to-scale only for $\delta=0.075$ and corresponds to
the vertical dashed line in (a), each of the numbered horizontal strips
is representative of the corresponding region in (a). They indicate
the ordering of non-smooth radii and smooth segments, which is invariant
within each region. The shading in (b) indicates the smooth behavior
that $\Theta\left(\eta\right)$ takes on in that interval of $\eta$.
$0$ means $\Theta$ is constant, $\mathrm{NE}$ stands in to mean
$\Theta^{\prime}\left(\eta\right)=\Theta_{\textit{NE}}^{\prime}\left(\eta\right)$,
$-\mathrm{NE}-\mathrm{NW}$ for $\Theta^{\prime}\left(\eta\right)=-\Theta_{\textit{NE}}^{\prime}\left(\eta\right)-\Theta_{\textit{NW}}^{\prime}\left(\eta\right)$,
etc. The points marked c and d correspond to the situations shown
in subfigures (c) and (d). (c) and (d) give a geometric interpretation
of $\Theta_{\textit{NE}}\left(\eta\right)$ in the $\zeta$-plane, as an example:
$\frac{\Theta_{\textit{NE}}\left(\eta\right)}{2}$ is one of the acute angles
in right triangles with adjacent side length $\eta_{\textit{NE}}$ and hypotenuse
$\eta$.\label{fig: radii orderings}}
\end{figure}

The behavior of the function $\Theta\left(\eta\right)$ between those
$6$ radii can be fully specified by the ordering of the radii. For
example, if $\eta_{E,m}>\eta_{N,\infty}$ (as in the two cases shown
in Figure \ref{fig: sample_zeta_plane + r_plot}), then $\Theta$
is constant between $\eta=\eta_{W,\infty}$ and $\eta=\eta_{N,\infty}$.
$5$ different orderings of the non-smooth radii and therefore $5$
different qualitative behaviors of $\Theta\left(\eta\right)$ appear
non-degenerately for $\delta\in\left(0,\frac{1}{2}\right)$ and $\mu\in\left(0,1\right)$
(Figure \ref{fig: radii orderings}a,b). Between subsequent pairs
of the non-smooth radii, $\Theta\left(\eta\right)$ takes on one of
only six potential functional forms up to an additive constant. In
particular, each piecewise segment of the derivative $\Theta^{\prime}\left(\eta\right)$
is one of $0$, $\Theta_{\textit{NE}}^{\prime}\left(\eta\right)$, $\pm\Theta_{\textit{NW}}^{\prime}\left(\eta\right)$,
or $\pm\left(\Theta_{\textit{NE}}^{\prime}\left(\eta\right)+\Theta_{\textit{NW}}^{\prime}\left(\eta\right)\right)$,
with
\begin{equation}
\Theta_{Nd}\left(\eta;\delta\right)\equiv2\,\mathrm{arcsec}\left(\frac{\eta}{\eta_{Nd}\left(\delta\right)}\right)\label{eq: Theta_Nd}
\end{equation}
for $d=E,W$, where $\eta_{\textit{NE}}$ and $\eta_{\textit{NW}}$ are the distances
from the origin to the sides of diamond/square (Figure \ref{fig: sample_zeta_plane + r_plot}a),
\begin{equation}
\eta_{\textit{NE}}=\frac{1}{2}+\delta;\;\eta_{\textit{NW}}=\frac{1}{2}-\delta.\label{eq: eta sides}
\end{equation}
The functional form of $\Theta_{Nd}\left(\eta;\delta\right)$ follows
from elementary geometry along with the ``initial condition'' of
$\Theta=0$ for small $\eta$. Since $\Theta$ is $0$ initially,
we see the parts of the circle of radius $\eta$ that protrude \emph{outside}
of the square (diamond) as contributing positively (negatively) towards
$\Theta$. For $d=E$, for example, $\frac{\Theta_{\textit{NE}}\left(\eta\right)}{2}$
is the angle subtended by each the northeast and southeast side of
the square or diamond when restricted to the \emph{exterior} of a
circle of radius $\eta$ (Figure \ref{fig: radii orderings}c,d).
Note that while $\eta<\eta_{E,m}<\eta_{E,\infty}$, i.e. before the
circle passes through the eastern corner of the diamond, the net contribution
from this eastern quadrant is $0$ (in Figure \ref{fig: radii orderings}c,
the dashed and solid arcs are equal in length). It is only when $\eta_{E,m}<\eta<\eta_{E,\infty}$,
i.e. when the circle intersects the square and not the diamond on
the northeast and southeast sides, that the net angle accumulates
at a rate of $\Theta_{\textit{NE}}^{\prime}\left(\eta\right)$ (in Figure \ref{fig: radii orderings}d,
$\eta$-dependent dashed arc is longer than the fixed solid arc).
The accumulation is due to the increasing part of the circle falling
outside the square and stops once $\eta>\eta_{E,\infty}$ (note how
the blue shading ends at $\eta=\eta_{E,\infty}$ in Figure \ref{fig: radii orderings}b).
Note that $\Theta_{Nd}^{\prime}\left(\eta\right)>0$ and therefore
the appearance of a minus sign in some of preceding functional forms
indicates that $\Theta\left(\eta\right)$ is decreasing, as it does
in the northern quadrant. Generally, $\Theta_{Nd}^{\prime}$ makes
a positive contribution to $\Theta$ when $\eta_{d,m}<\eta<\eta_{d,\infty}$
and a negative contribution when $\eta_{N,\infty}<\eta<\eta_{N,m}$
(Figure \ref{fig: radii orderings}b). 

With the above, common analysis set-up, we now treat the lower and
upper bounds separately.

\subsection{The Lower Bound}

We aim to show that (\ref{eq: radial integral}) can be bounded as
\begin{equation}
\mathcal{E}_{m}-\mathcal{E}_{\infty}\propto\int_{0}^{\infty}\eta\,\Theta\left(\eta\right)g\left(\eta;\sigma\right)d\eta\ge0,\label{eq: lower bound condition}
\end{equation}
where $g\left(\eta;\sigma\right)\equiv\mathrm{e}^{-\frac{\eta^{2}}{2\sigma^{2}}}$.
If $\Theta\left(\eta\right)$ is non-negative, the above inequality
will hold trivially. This is the case for values of $\delta$ and
$\mu$ corresponding to regions 1-3 of Figure \ref{fig: radii orderings}a.
For $\mu<1$ (i.e. $\Lambda<1$ and $m>0$), $\Theta\left(\eta\right)$
must eventually be $0$ at large $\eta$, since both the diamond and
square extend only to finite $\eta$. In regions 1-3, as $\eta$ increases,
$\Theta\left(\eta\right)$ is first (non-strictly) increasing from
$0$, and then decreases monotonically to $0$, remaining $0$ for
$\eta>\eta_{N,m}$ (consider the ``boundary conditions'' $\Theta\left(0\right)=\Theta\left(\infty\right)=0$
and the signs of $\Theta^{\prime}\left(\eta\right)$ associated with
the shading in Figure \ref{fig: radii orderings}b). This implies
that $\Theta\left(\eta\right)$ remains non-negative for all $\eta$.

In regions 4 and 5, the argument is more subtle. Here $\Theta\left(\eta\right)$
initially increases (non-strictly) for $\eta<\eta_{N,\infty}$, then
decreases (again, non-strictly) for $\eta\in\left(\eta_{N,\infty},\eta_{N,m}\right)$,
and then again increases until $\eta=\eta_{E,\infty}$, where $\Theta=0$.
Therefore $\Theta\left(\eta\right)$ is initially $0$, becomes positive
for some interval of $\eta$, and then crosses $0$, becoming negative
before returning to $0$ (as in e.g. Figure \ref{fig: sample_zeta_plane + r_plot}c).
The same is true of $\eta\,\Theta\left(\eta\right)$. Most importantly,
the interval where $\Theta\left(\eta\right)$ and $\eta\,\Theta\left(\eta\right)$
are negative is on the rightmost end of their support. At the same
time, for any $0<\eta_{1}<\eta_{2}$, $\nicefrac{g\left(\eta_{2};\sigma\right)}{g\left(\eta_{1};\sigma\right)}$
is strictly increasing as a function $\sigma$. This means that, in
the integral in (\ref{eq: lower bound condition}), the negative part
of $\eta\,\Theta\left(\eta\right)$ has the greatest relative contribution
in the limit $\sigma\rightarrow\infty$. It is therefore sufficient
that $\int_{0}^{\infty}\eta\,\Theta\left(\eta\right)d\eta\ge0$ in
order to show that (\ref{eq: lower bound condition}) holds. But the
integral $\int_{0}^{\infty}\eta\,\Theta\left(\eta\right)d\eta$ is
just the difference in area between the diamond and the square, $\frac{1}{\sqrt{1-\mu^{2}}}-1$,
which is positive for $0<\mu<1$.

We conclude that $\mathcal{E}_{m}-\mathcal{E}_{\infty}>0$ for any
$\Lambda\in\left(0,1\right)$, $m>0$, and $\delta\in\left(0,\frac{1}{2}\right)$.

\subsection{The Upper Bound}

We recall our goal, a bound $\mathcal{B}$ independent of $\mu=\Lambda^{m}$
such that (cf. (\ref{eq: radial integral}) and (\ref{eq: upper bound}))
\[
\frac{\mathcal{E}_{m}-\mathcal{E}_{\infty}}{\mu}=\frac{1}{2\pi\sigma^{2}\mu}\int_{0}^{\infty}\eta\,\Theta\left(\eta;\delta,\mu\right)g\left(\eta;\sigma\right)d\eta\le\mathcal{B}\left(\sigma;\delta\right),
\]
and we therefore aim to select an appropriate upper bound of $\frac{1}{\mu}\Theta\left(\eta;\delta,\mu\right)$
uniform in $\mu$. However, many of the obvious (e.g. piecewise constant)
upper bounds on $\Theta$ result in divergences in the above integral
as $\mu\rightarrow1$. The difficulty is that $\eta_{N,m}\rightarrow\infty$
as $\mu\rightarrow1$. Indeed, the limiting shape of the diamond region
is the infinite strip $\left[\delta-\frac{1}{2},\delta+\frac{1}{2}\right]\times\left(-\infty,\infty\right)$,
and so $\Theta$ remains nonzero for arbitrarily large values of $\eta$.
We find it necessary to incorporate the behavior in the limit as $\mu\rightarrow1$
into our bound. In particular, we will demonstrate in the following
that
\begin{equation}
\frac{1}{\mu}\Theta\left(\eta;\delta,\mu\right)\le\Theta\left(\eta;\delta,\mu=1\right).\label{eq: upperBound Theta}
\end{equation}
(\ref{eq: upperBound Theta}) in turn gives $\mathcal{B}=\mathcal{E}\left(1-\mathcal{E}\right)$,
since $\mathcal{E}_{\infty}=\mathcal{E}^{2}$ and $\mathcal{E}_{m}\rightarrow\mathcal{E}$
as $\mu\rightarrow1$.

To demonstrate that our bound (\ref{eq: upperBound Theta}) holds
true, we first recall that $\Theta\left(\eta;\delta,\mu\right)$ is
either $0$ or negative for $\eta>\eta_{N,m}\left(\delta,\mu\right)$
and $\eta_{N,m}$ increases as a function of $\mu$ (Figure \ref{fig: radii orderings}b).
Since $\Theta\left(\eta;\delta,\mu=1\right)$ is non-negative for
all $\eta$, the bound holds trivially for $\eta>\eta_{N,m}\left(\delta,\mu\right)$.
For $\eta\le\eta_{N,m}$, consider the decomposition $\Theta=\Theta_{\textit{ENE}}+\Theta_{\textit{NNE}}+\Theta_{\textit{NNW}}+\Theta_{\textit{WNW}}$,
where for $d=E,W$
\begin{eqnarray}
\Theta_{NNd}\left(\eta;\delta,\mu\right) & \equiv & \begin{cases}
0, & \eta\le\eta_{N,\infty}\\
-\int_{\eta_{N,\infty}}^{\eta}\Theta_{Nd}^{\prime}\left(\eta\right)d\eta, & \eta_{N,\infty}<\eta\le\eta_{N,m}\\
\Theta_{Nd}\left(\eta_{N,\infty}\right)-\Theta_{Nd}\left(\eta_{N,m}\right), & \eta>\eta_{N,m}
\end{cases}\nonumber \\
 & = & \Theta_{Nd}\left(\min\left(\eta,\eta_{N,\infty}\left(\delta\right)\right);\delta\right)-\Theta_{Nd}\left(\min\left(\eta,\eta_{N,m}\left(\delta,\mu\right)\right);\delta\right)\label{eq: Theta NNd}
\end{eqnarray}
and $\Theta_{\textit{NNE}}+\Theta_{\textit{NNW}}$ reflects the ``$-\mathrm{NE}-\mathrm{NW}$''
contribution that, as a function of $\eta$, monotonically decreases
in the red region in Figure \ref{fig: radii orderings}b and otherwise
remains constant. Similarly,
\begin{equation}
\Theta_{dNd}\left(\eta;\delta,\mu\right)\equiv\Theta_{Nd}\left(\min\left(\eta,\eta_{d,\infty}\left(\delta\right)\right);\delta\right)-\Theta_{Nd}\left(\min\left(\eta,\eta_{d,m}\left(\delta,\mu\right)\right);\delta\right)\label{eq: Theta dNd}
\end{equation}
reflects for $d=E$ the blue ``$\mathrm{NE}$'' and for $d=W$ the
yellow ``$\mathrm{NW}$'' contributions. It is sufficient to show
that $\frac{1}{\mu}\Theta_{NNd}\left(\eta;\delta,\mu\right)\le\Theta_{NNd}\left(\eta;\delta,1\right)$
and $\frac{1}{\mu}\Theta_{dNd}\left(\eta;\delta,\mu\right)\le\Theta_{dNd}\left(\eta;\delta,1\right)$
for $d=E,W$ and $\eta\le\eta_{N,m}\left(\delta,\mu\right)$ to validate
(\ref{eq: upperBound Theta}). The first of these is trivial, since
$\Theta_{NNd}\left(\eta;\delta,\mu\right)$ is independent of $\mu$
and non-positive for $\eta\le\eta_{N,m}\left(\delta,\mu\right)$,
and thus $\frac{1}{\mu}\Theta_{NNd}\left(\eta;\delta,\mu\right)\le\Theta_{NNd}\left(\eta;\delta,\mu\right)=\Theta_{NNd}\left(\eta;\delta,1\right)$.
The latter follows if $h\left(x;\mu\right)\le h\left(x;1\right)$
for all $x\ge0$ and $\mu\in\left(0,1\right)$, where (cf. (\ref{eq: eta non-smooth}),
(\ref{eq: Theta_Nd}), (\ref{eq: eta sides}), and (\ref{eq: Theta dNd}))
\[
h\left(x;\mu\right)\equiv\frac{1}{\mu}\left[\mathrm{arcsec}\left(\min\left(x,\sqrt{2}\right)\right)-\mathrm{arcsec}\left(\min\left(x,\sqrt{\frac{2}{1+\mu}}\right)\right)\right].
\]
Note that $h^{\prime}\left(x;\mu\right)$ is non-negative and only
depends on $\mu$ by a factor of $\frac{1}{\mu}$. So, for any given
value of $x$, it is minimized as $\mu\rightarrow1$. It is therefore
sufficient to verify the inequality $h\left(x;\mu\right)\le h\left(x;1\right)$
in the limit as $x\rightarrow\infty$; if $h\left(\infty;\mu\right)\le h\left(\infty;1\right)$,
\begin{eqnarray*}
h\left(x;\mu\right) & = & h\left(\infty;\mu\right)-\int_{x}^{\infty}h^{\prime}\left(u;\mu\right)du\\
 & \le & h\left(\infty;1\right)-\int_{x_{0}}^{\infty}h^{\prime}\left(u;1\right)du\\
 & = & h\left(x;1\right)
\end{eqnarray*}
 for all $x\ge0$. So we verify that
\[
h\left(\infty;\mu\right)=\frac{1}{\mu}\left[\frac{\pi}{4}-\mathrm{arcsec}\left(\sqrt{\frac{2}{1+\mu}}\right)\right]\le\frac{\pi}{4}=h\left(\infty;1\right).
\]
This is equivalent to
\[
\sin\left(\frac{\pi}{4}\left(1-\mu\right)\right)^{2}\le\frac{1-\mu}{2},
\]
which is satisfied for $\mu\in\left(0,1\right)$, since equality is
obtained at $\mu=0,1$, $\frac{1-\mu}{2}$ is linear, and $\sin\left(\frac{\pi}{4}\left(1-\mu\right)\right)^{2}$
is a concave-up function of $\mu$ on the open interval.

\section{Dependence on Event Interval Width\label{sec: SI - w-dependence}}

\begin{figure}
\noindent \begin{centering}
\includegraphics[width=0.5\textwidth]{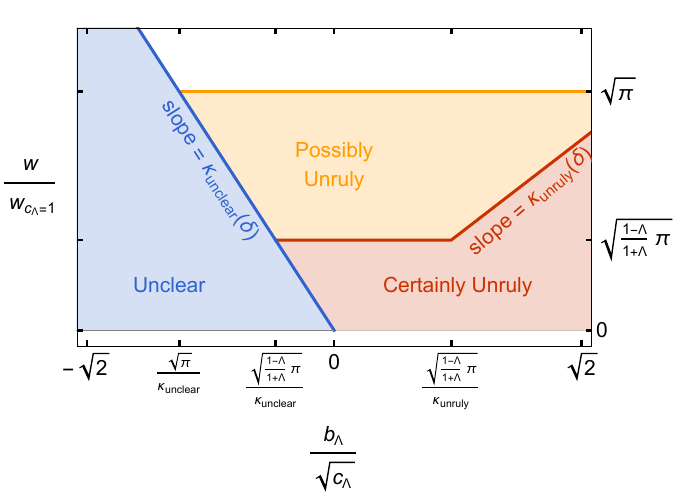}
\par\end{centering}
\caption{An alternative representation of the phase diagram, Figure \ref{fig:phaseDia w-delta}a,
showing the dependence of unruliness on the event interval width $w$.\label{fig:DphaseDia w}}
\end{figure}

In this SI section, we derive the relationship between the event interval
width $\left|E\right|=w$ and unruliness in planar oscillators. From
Figure \ref{fig:phaseDia w-delta}a, we see that changing $w$ with
a fixed value of $\frac{b_{\Lambda}}{\sqrt{c_{\Lambda}}}$ moves the
oscillator between the regions of possible unruliness, certain unruliness
and uncertainty. We therefore first translate the $b_{\Lambda}$-$c_{\Lambda}$
phase diagram (Figure \ref{fig:phaseDia w-delta}a) into a phase diagram
over $\frac{b_{\Lambda}}{\sqrt{c_{\Lambda}}}$ and $w$ (Figure \ref{fig:DphaseDia w}).
It is useful to normalize $w$ by a characteristic width, the value
of $w$ that is required to fix $c_{\Lambda}=1$: from (\ref{eq:c_general})
and (\ref{eq:bLambda_cLambda}),
\[
w_{c_{\Lambda}=1}=\sqrt{\frac{1+\Lambda}{1-\Lambda}\frac{\Gamma_{\textit{ss}}}{c}}=\frac{1}{\sqrt{2}\left(1-\Lambda\right)}\frac{\left\Vert \vec{y}_{G}\right\Vert }{\left\Vert \vec{z}_{G}\right\Vert }.
\]
It then turns out that the linear boundaries from the $b_{\Lambda}$-$c_{\Lambda}$
phase diagram remain linear and with the same slope in the $\frac{b_{\Lambda}}{\sqrt{c_{\Lambda}}}$-$\frac{w}{w_{c_{\Lambda}=1}}$
phase diagram. Referring to Figure \ref{fig:DphaseDia w}, we see
the following:
\begin{enumerate}
\item[i)] If $\frac{b_{\Lambda}}{\sqrt{c_{\Lambda}}}>0$, the TVGR is not unruly
for $w>w_{\textit{possible}}^{*}$, is possibly unruly for $w\in\left(w_{\textit{certain}}^{*},w_{\textit{possible}}^{*}\right)$,
and is certainly unruly for $w\in\left(0,w_{\textit{certain}}^{*}\right)$,
where
\[
w_{\textit{possible}}^{*}=\sqrt{\pi}w_{c_{\Lambda}=1}=\sqrt{\frac{\pi}{2}}\frac{1}{\left(1-\Lambda\right)}\frac{\left\Vert \vec{y}_{G}\right\Vert }{\left\Vert \vec{z}_{G}\right\Vert }
\]
ands
\begin{eqnarray*}
w_{\textit{certain}}^{*} & = & \max\left(\sqrt{\frac{1-\Lambda}{1+\Lambda}\pi},\kappa_{\textit{unruly}}\left(\delta\right)\frac{b_{\Lambda}}{\sqrt{c_{\Lambda}}}\right)w_{c_{\Lambda}=1}\\
 & = & \max\left(\sqrt{\frac{\pi}{2\left(1-\Lambda^{2}\right)}}\frac{\left\Vert \vec{y}_{G}\right\Vert }{\left\Vert \vec{z}_{G}\right\Vert },\frac{\kappa_{\textit{unruly}}\left(\delta\right)}{1-\Lambda}\frac{\vec{z}_{G}^{T}\vec{y}_{G}}{\left\Vert \vec{z}_{G}\right\Vert ^{2}}\right).
\end{eqnarray*}
\item[ii)] If $\frac{1}{\kappa_{\textit{unclear}}\left(\delta\right)}\sqrt{\frac{1-\Lambda}{1+\Lambda}\pi}<\frac{b_{\Lambda}}{\sqrt{c_{\Lambda}}}<0$,
the TVGR is possibly unruly for $w\in\left(w_{\textit{certain}}^{*},w_{\textit{possible}}^{*}\right)$,
is certainly unruly for $w\in\left(w_{\textit{unclear}}^{*},w_{\textit{certain}}^{*}\right)$,
where
\[
w_{\textit{unclear}}^{*}=\kappa_{\textit{unclear}}\left(\delta\right)\frac{b_{\Lambda}}{\sqrt{c_{\Lambda}}}w_{c_{\Lambda}=1}=\frac{\kappa_{\textit{unclear}}\left(\delta\right)}{1-\Lambda}\frac{\vec{z}_{G}^{T}\vec{y}_{G}}{\left\Vert \vec{z}_{G}\right\Vert ^{2}}.
\]
\item[iii)] If $\frac{1}{\kappa_{\textit{unclear}}\left(\delta\right)}\sqrt{\pi}<\frac{b_{\Lambda}}{\sqrt{c_{\Lambda}}}<\frac{1}{\kappa_{\textit{unclear}}\left(\delta\right)}\sqrt{\frac{1-\Lambda}{1+\Lambda}\pi}$,
the TVGR is possibly unruly for $w\in\left(w_{\textit{unclear}}^{*},w_{\textit{possible}}^{*}\right)$,
but the situation is unclear for $w<w_{\textit{unclear}}^{*}$.
\item[iv)] If $\frac{b_{\Lambda}}{\sqrt{c_{\Lambda}}}<\frac{1}{\kappa_{\textit{unclear}}\left(\delta\right)}\sqrt{\pi}$,
the TVGR is not unruly for $w>w_{\textit{unclear}}^{*}$ and the situation
is unclear for $w<w_{\textit{unclear}}^{*}$.
\end{enumerate}

\end{document}